# Power System Stability Analysis using Neural Network

by

## Md. Rayid Hasan Mojumder

A thesis submitted in partial fulfillment of the requirements for the degree of Master of Science in Engineering in the Department of Electrical and Electronic Engineering

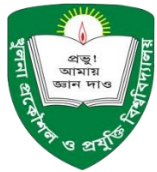

Khulna University of Engineering & Technology

Khulna 9203, Bangladesh

**November, 2022**

# Declaration

This is to certify that the thesis work entitled "Power System Stability Analysis using Neural Network" has been carried out by Md. Rayid Hasan Mojumder in the Department of Electrical and Electronic Engineering, Khulna University of Engineering & Technology, Khulna 9203, Bangladesh. The above thesis work or any part of this has not been submitted anywhere for the award of any degree or diploma.

Signature of Supervisor                    Signature of Candidate

# Approval

This is to certify that the thesis work submitted by Md. Rayid Hasan Mojumder entitled "Power System Stability Analysis using Neural Network" has been approved by the board of examiners for the partial fulfillment of the requirements for the degree of Master of Science in Engineering in the Department of Electrical and Electronic Engineering, Khulna University of Engineering & Technology, Khulna 9203, Bangladesh in November 2022.

## BOARD OF EXAMINERS

1. 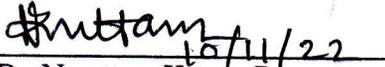
   Dr. Naruttam Kumar Roy
   Professor, Department of Electrical and Electronic Engineering
   Khulna University of Engineering & Technology
   
   Chairman
   (Supervisor)

2. 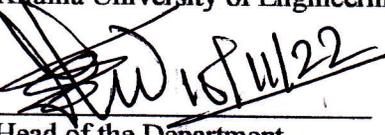
   Head of the Department
   Department of Electrical and Electronic Engineering
   Khulna University of Engineering & Technology
   
   Member

3. 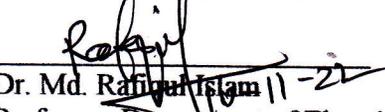
   Dr. Md. Rafiqul Islam
   Professor, Department of Electrical and Electronic Engineering
   Khulna University of Engineering & Technology
   
   Member

4. 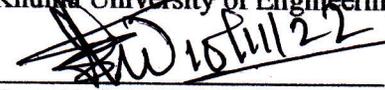
   Dr. Md. Salah Uddin Yusuf
   Professor, Department of Electrical and Electronic Engineering
   Khulna University of Engineering & Technology
   
   Member

5. 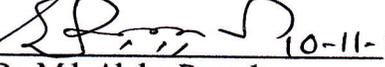
   Dr. Md. Abdur Razzak
   Professor, Department of Electrical & Electronic Engineering
   Independent University, Bangladesh
   
   Member
   (External)

# Acknowledgment

I would like to express my deep gratitude to the supervisor of this thesis work, Dr. Naruttam Kumar Roy, Professor, Department of Electrical and Electronic Engineering, Khulna University of Engineering & Technology, Khulna 9203, Bangladesh, for his constant support, constructive feedback, continuous inspiration, and validation throughout the research work. During this difficult time of Covid 19, his cooperation made it possible for me to stay foot in the research direction. His creative insight, expertise, and knowledge of power system control and stability analysis helped me to address many of the research difficulties. I wish him all the best in the world.

I would also like to pay my tribute to Dr. Md. Salah Uddin Yusuf, Professor & Head of the Department of Electrical and Electronic Engineering, and to Dr. Md. Rafiqul Islam and Dr. Mohammad Shaifur Rahman, Professor & Former Head of the Department of Electrical and Electronic Engineering; they made it easier for me to continue at the department, to learn, and to complete the courses with all the required facilities, at the best possible ways. Thank you very much for your continuous support and guidance.

I also acknowledge all the honorable faculty members and staffs of the Department of Electrical and Electronic Engineering, Khulna University of Engineering & Technology, Khulna 9203, Bangladesh, for their direct and indirect support during this thesis work and for establishing an outstanding educational platform along with an excellent research-intensive collaborative culture.

In particular, I want to acknowledge the works of Prof. Luigi Vanfretti and ALSETLab at RPI, Troy, NY, for contributing to OpenModelica and Dymola tools for power system stability classification and multi-physics simulations.

Finally, I would like to express my love and gratitude to my family members for their sacrifices over the years. Without them, the completion of my study would not be possible.

# Abstract

This work focuses on the design of modern power system controllers for automatic voltage regulators (AVR) and the applications of machine learning (ML) algorithms to correctly classify the stability of the IEEE 14 bus system. The performance of the PID, LQR, and LQG controllers directly depends on the AVR sub-system gain parameters and time constants. The LQG controller performs the best time domain characteristics compared to PID and LQG, while the sensor and amplifier gain is changed in a dynamic passion. After that, the IEEE 14 bus system is modeled, and contingency scenarios are simulated in the System Modelica – Dymola environment. Application of Monte Carlo principle with modified Poissons probability distribution principle is reviewed from literature that reduces the total contingency from 1000k to 20k. The damping ratio of the contingency is then extracted, pre-processed, and fed to ML algorithms, such as logistic regression, support vector machine, decision trees, random forest, Naïve Bayes, and k-nearest neighbor. A neural network (NN) of one, two, three, five, seven, and ten hidden layers with 25%, 50%, 75%, and 100% data size is considered to observe and compare the prediction time, accuracy, precision, and recall value. At 100% data size, the accuracy of Naïve Bayes, k-nearest neighbor, support vector machine, logistic regression, and decision trees result in 97.08%, 99.90%, 71.79%, 78.04%, and 98.83%. With a neural network of one, two, and three hidden layers, the accuracy becomes 98.38%, 98.53%, and 98.53%. When the data size is reduced to lower points, at 75%, the time of prediction reduces for all the algorithms; however, the accuracy drops to nearly 66% and 77% for support vector machine and logistic regression, respectively. At 25% dataset, Naïve Bayes, k-nearest neighbor, and decision trees, although could still maintain ~97% efficiency, the overall F1 score, considering precision and accuracy, respectively, results in 91.1%, 99.2%, and 95.8%. The F1 score for a neural network with two hidden layers, at 75% data, is 92.81%. However, when the time of prediction is also considered, the k-nearest neighbor takes the highest time of 6.5 seconds, compared to only 0.04 s using neural networks. At lower data size, 25%, in the neural network with two-hidden layers and a single hidden layer, the accuracy becomes 95.70% and 97.38%, respectively. Increasing the hidden layer of NN beyond a second does not increase the overall score and takes a much longer prediction time, thus could be discarded for similar analysis. Moreover, when five, seven, and ten hidden layers are used, the F1 score reduces. However, in practical scenarios, where the data set contains more features and a variety of classes, a higher data size is required for NN for proper training. This research will provide more insight into the damping ratio-based system stability prediction with traditional ML algorithms and neural networks.

# Contents

# LIST OF TABLES

# LIST OF FIGURES

© Md. Rayid Hasan Mojumder - 2022

**Nomenclature**

| | |
|---|---|
| $V_{ref}(s)$ | Reference Voltage |
| $V_s(s)$ | Sensor Output Voltage |
| $E(s)$ | Error Signal |
| $V_R(s)$ | Exciter Input Voltage |
| $V_F(s)$ | Field Excitation Voltage |
| $V_T(s)$ | Generator Terminal Voltage |
| $r(t)$ | Reference Set Point Value |
| $y(t)$ | Measured Value at the Output |
| $e(t)$ | Error Value |
| $\sigma$ | Real Part of Eigen Value |
| $\omega$ | Real-valued Frequency Variable |
| $\lambda$ | Complex Eigen Value |
| $\xi$ | Damping Ratio |
| $\omega_{syn}$ | Synchronous Speed |
| $\delta$ | Load Angle |
| H | Normalized Inertia Component |
| $P_m$ | Mechanical Power |
| $P_e$ | Electrical Power |
| $P_a$ | Accelerating Power |
| $P$ | Real Power |
| $Q$ | Reactive Power |
| $v$ | Voltage |
| $x$ | State Variable |
| $y$ | Output Variable |
| $u$ | Input Variable |
| $P(X)$ | Poisson's Distribution |
| $\sigma^2$ | Variance |
| $P$ | Probability |
| $N$ | Sample Number |
| $Q$ | Unavailability or Failure Probability |
| $\alpha$ | Variation Coefficient |
| $t$ | Computational Time |
| $G_{AVR}(s)$ | Transfer Function of AVR Model |
| $A, B, C, D$ | Input, Output, and Feed-Forward matrices, respectively |
| $J$ | Quadratic Objective Function |
| $R$ | Positive Control Weighing Matrix |
| $Q$ | Non-negative State Weighning Matrix |
| $P$ | Solution of Algebraic Riccati Equation |
| $L$ | Kalman Estimator Gain Matrix |
| $K_k$ | Kalman Estimator Gain |
| $n$ | Number of Total Branches in IEEE 14 Bus System |
| $k$ | Number of Simultaneous Contingencies |
| $T$ | Total Number of Possible Contingencies |
| $p(k)$ | Modified Poisson Distribution |
| $Logit(P_i)$ | Log Transformation of Dependent Variable $P_i$ |
| $W$ | Feature Weighing Coefficient |
| $b$ | Bias |
| $\phi$ | Transformation Function |
| $d(x, y)$ | Distance Between Points |
| $Entropy(S)$ | The entropy of the Dataset $(S)$ |
| $c$ | Classes in a Dataset |
| $P(x_i|y)$ | Bayesian Conditional Probability |

## Abbreviations

| | |
|---|---|
| FACTS | Flexible AC Transmission System |
| LFC | Load Frequency Control |
| PSS | Power System Stabilizer |
| SG | Synchronous Generator |
| LQR | Linear Quadratic Regulator |
| LQG | Linear Quadratic Gaussian |
| MPC | Model Predictive Control |
| ML | Machine Learning |
| SCADA | Supervisory Control and Data Acquisition |
| PMU | Phasor Measurement Unit |
| SE | State Estimation |
| MLT | Machine Learning Technique |
| AI | Artificial Intelligence |
| AVR | Automatic Voltage Regulator |
| PID | Proportional Integral Derivative |
| PIDA | Proportional Integral Derivative Acceleration |
| FOPID | Fraction Orders Proportional Integral Derivative |
| SFL | Sugeno Fuzzy Logic |
| ZN | Ziegler Nichols |
| CC | Cohen Coon |
| PSO | Particle Swarm Optimization |
| ABC | Artificial Bee Colony |
| BAT | Bat Search Technique |
| ACO | Ant Colony Optimization |
| CS | Cuckoo Search |
| MOL | Many Optimizing Liaisons |
| GSA | Gravitational Search Algorithm |
| TLBO | Teaching Learning Based Optimization |
| HSA | Harmony Search Algorithm |
| GA | Genetic Algorithm |
| LUS | Local Unimodal Sampling |
| PQD | Power Quality Disturbance |
| SVM | Support Vector Machine |
| DT | Decision Trees |
| NN | Neural Network |
| ANN | Artificial Neural Network |
| CNN | Convolutional Neural Network |
| PNN | Probabilistic Neural Network |
| VSMI | Voltage Stability Margin Index |
| k-NN | k-nearest neighbor |
| HVDC | High Voltage Direct Current |
| PEC | Power Electronic Converter |
| DG | Distributed Generation |
| SG | Smart Grid |
| MG | Micro Grid |
| PCC | Point of Common Coupling |
| CIG | Converter Interfaced Generation |
| SSR | Subsynchronous Resonance |
| SSO | Subsynchronous Oscillations |
| DDSSO | Device Dependent Subsynchronous Oscillations |
| PLL | Phase Locked Loop |
| VSC | Voltage Source Converter |
| TSVM | Transductive Support Vector Machine |
| ASI | Artificial Super Intelligence |
| RNN | Recurrent Neural Network |
| DNN | Deep Neural Network |
| ARE | Algebric Riccati Equation |
| OpenIPSL | Open Instance Power System Library |
| PCF | Phase Cross-over Frequency |
| GCF | Gain Cross-over Frequency |
| GM | Gain Margin |
| PM | Phase Margin |
| MSE | Mean Squared Error |

© Md. Rayid Hasan Mojumder - 2022

# CHAPTER I

# Introduction

## 1.1 Background

Power system analysis is one of the core parts of modern grid control and optimization. In the electrical power system, uninterrupted power flow is expected from the generation end to the distribution point with maximum efficiency and reliability. The reliability of power flow depends on the ability of the system to maintain stability at all dynamic or transient disturbances and contingencies. This could be achieved through an appropriate measure to continuously monitor the state of the power system model, and dispatching required control strategies. The usual practice to assess contingencies lies in contingency identification and ranking in terms of severity. The primary focus of power system stability analysis and control is to facilitate the identification of system states and damping ratio and to initiate required changes in the system state to maintain stability. Traditionally, the practice of assessing the reliability of power systems consisted of the tedious numerical solution of the load flow models. The investigation involved a range of time domains containing the steady-state, transient, and sub-transient portions of operation and control. The numerical solution strategy was, however, very much time-consuming and became an unfeasible option to consider with the practical large-scale power grid network.

In conventional coal, hydro, and combined-cycle power plants, the use of a synchronous generator (SG) is common place (Fig. 1.1). A turbine and a speed governor are embedded to properly drive the SG for electricity generation. The specified terminal voltage induced on the SG directly depends on the excitation at the rotor side and the mechanical standing and design of the core windings, and the number of poles. When an SG is loaded for large reactive power demand, the line loss increases, and this leads to an alteration in the rated voltage output of the SG [1]. These voltage fluctuations also affect the total connected load to the system. As a result, significant system instability follows and leads to higher operating and maintenance costs. The use of an automatic voltage regulator (AVR) system maintains a static terminal voltage for the SG and helps to abate system stability. When contingency arises and the SG's terminal voltage deviates, the AVR accordingly alters the exciter output level and forces the terminal voltage toward the nominal voltage [2]. While reacting to maintain nominal voltage at the SG's terminal, the operation of the AVR needs to be fast and dynamic, which also leads to lower voltage instability. The performance of the AVR system could be boosted by integrating automatic close-loop controllers such as PID, LQG, SFL, and MPC into the system dynamics. Among these, the structure and operation of a PID controller are very simple but very susceptible to load fluctuation and non-linearity of the load [3].

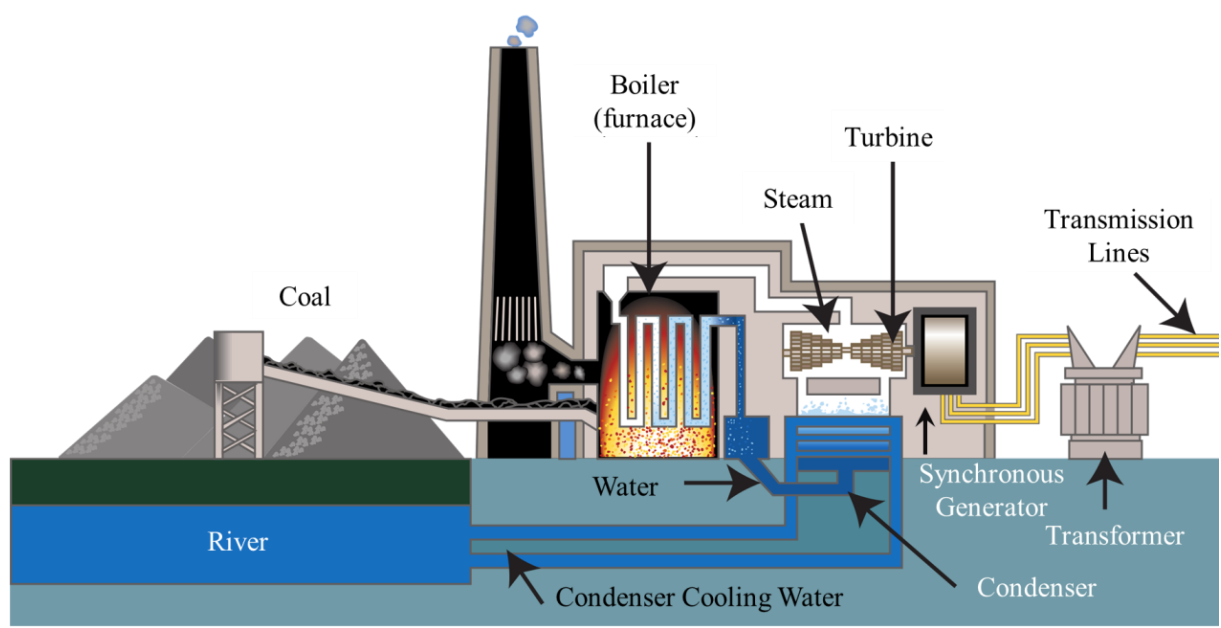

Figure 1.1: Coal-fired power plant diagram [4].

The power system study has faced a few turns in the last decade. Researchers traditionally used dynamic control systems to plot eigenvalues and observe system behavior. With the advancement of power electronics-based devices and high processing capabilities, adaptive controllers became a core part of the system. Two primary control loops used in controlling the frequency and voltage output of a generating station or synchronous generator were the turbine-governor loop control and automatic voltage regulator, respectively. Soon, these two loops became insufficient to provide power system protection against the diversely and randomly occurring contingencies. Thus, additional controlling devices were added to boost the dynamic performance and system stability margin. The controlling devices were classified into (i) normal or preventive control: discrete or continuous automatic control strategy for continuous power system operation, such as flexible AC transmission systems (FACTs), power system stabilizers (PSSs), and load frequency control (LFC); and (ii) emergency controls: used for contingency situation, such as connecting or disconnecting system load, tripping of generators, controlling of exciter units in SG. Fig. 1.2 shows a basic synchronous generator with LFC and AVR control loop. When normal control fails to mitigate a contingency profile, the emergency controls come into play and try to isolate the disturbance. When used together, the strategies help maintain the reliability of the power system and ensure first-swing stability by reducing the damping oscillation and stabilizing the system voltage. However, even such control strategies had their limitation in maintaining the power system at its stability margin when a large contingency occurs.

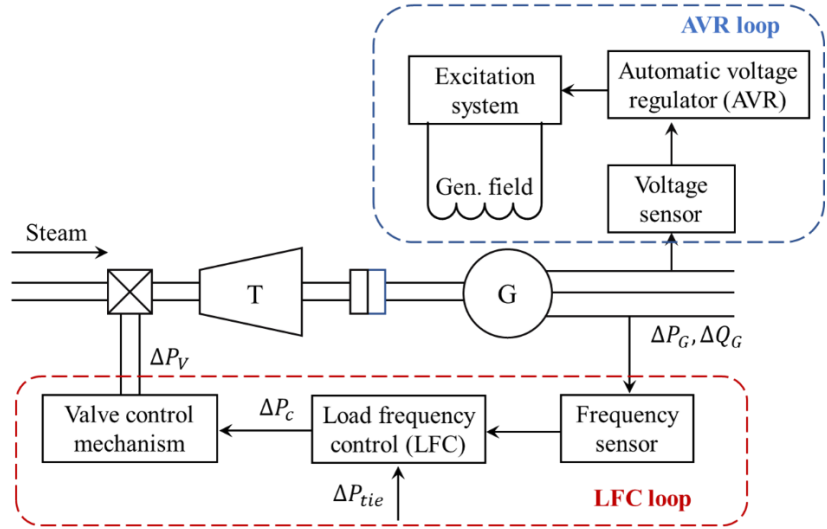

Figure 1.2: LFC and AVR in a synchronous generator model [5].

In current literature, researchers have constantly been trying to apply various techniques to address significant control problems. The initial way was the application of control theory and linear optimal control technique. The methods were formulated, focusing only on the linear dynamics of the power system for simplicity. For a power system operating at the nominal condition and having nominal parameters, the linearized control technique could show optimal performance. However, with a large power system with complex system dynamics, uncertainties, and disturbances, the models fail to provide sufficient performance. Controllers such as proportional integral derivative (PID), linear quadratic regulator (LQR), linear gaussian controller (LQG), and model predictive control (MPC) were embedded with advanced optimization procedures to control the system dynamics properly. The consideration of modern control techniques, such as adaptive control, linear quadratic control, model predictive control, $H_\infty$ control, etc., became a more efficient way to handle the limitations provided by the linear optimal control (conventional controller). Modern controllers also showed higher performance and facilitated swift change in controller parameters to revert the system to stability in a post-contingency situation. However, these modern controllers require very complex mathematical formulations, higher implementation costs, and require a high-performing computing platform for swift parameter identification, optimization, and updating of the system states.

AVR system alters the exciter voltage and tries to address the variation of voltage at the SG terminal due to dynamic load fluctuations. The step-response of a PID-based AVR system with various optimization algorithms demonstrates that embedding PID-controller keeps the maximum peak overshoot within 1.0 to 1.3 per unit (p.u.) of the system's nominal value [6]. Also, the time to reach the peak response stays near ~ 0.35 sec with the majority of the PID-optimization algorithms. With the fractal search algorithm, the rise time and settling time obtained are 0.103 sec and 0.584 sec [7]. The quality and performance of the optimization procedures largely depend on the setting of the objective and cost functions being used, and in current literature vastly differ. Heuristical algorithms are a good choice to reduce the computation burden while tuning and estimating the gain parameters of the PID block of an AVR system [8]. However, the recent trend is to utilize modern controllers such as LQR, LQG, and MPC in applications across power systems, vehicle designs, robotics and automation, and aerospace [9]–[11]. Substituting PID controllers with LQR or LQG could lead to more robustness of the system with lower susceptance to load fluctuation.

In [12], the DC control strategy of an AVR system is realized with an LQG controller. The considered single-area grid isolated system shows sufficient improvement in terms of overshoot reduction and improving damping efficiency. When compared to PID, LQR, and LQG controllers, the resultant LQG-DC system supersedes the rest in terms of efficient operation. A hybrid AC/DC microgrid is examined in [13] for damping calculation and to track the subgrid voltage level with the help of an integral-LQG stage. For practical simulation, diverse and undefined loads are connected to observe the voltage fluctuation at the subgrid level. The model, when simulated, shows excellent robustness. Even in the LFC strategy, the extensive use of LQR and LQG controllers is present. In [12], the use of an LQR controller reduced the settling time of an LFC-based control scheme from 8 sec to less than 5 sec. In LQR, the unique minimum of the cost function is obtained by evaluating the algebraic Riccati equation. The parameters Q and R hold the penalty for state variables and control signals and are used to weight them for proper tuning of the plant operation. A larger value of R and Q places higher control weight and lower required change in the state variables. Usually for large control signals, the R matrix value is kept in check so that no saturaion arises from the large control signals at the actuator or other stages in terms of noise. In LQG, Gaussian distribution of noise is placed. A state estimator (usually used one is Kalman filter) is required with since the system state, unlike LQR, could not be directly observable in the presence of randomness of the variables and noises from the controller gains and system.

Investigation of power system stability is one of the fundamental aspects of maintaining and controlling electrical grid parameters. The electrical grid, comprising the generation, transmission, and distribution sides, is basically an interconnected system that in-takes power from multiple distributed generating stations located sporadically or structurally throughout the entire region of the power network and transfers the power towards the load center in a real-time passion. Any alteration in the load scheduling, values of the grid parameters, and control signals could pave adverse results of diverse magnitude. As far as the definition of the IEEE/CIGRE, the ability of an electrical to regain its stable operating condition in the post-contingency situation, having the possible variables retained within the stability boundary, could be termed as the stability in the power system [14]. In general, a stable power system follows where the restoring forces in response to any disturbing forces are equal or even greater to such a degree that the system states could maintain an equilibrium profile.

A study on power system stability originated in 2004 [14]. The generator units back then were primarily responsible for reactive power (voltage) and active power (frequency) control during normal operation and ride-through for both voltage and frequency disturbances. The lack of vast application of fast power electronic devices constricted the stability analysis to focus on fairly slow electromechanical phenomena of synchronous generators for dynamic control and performance investigation [15]. In such a scenario, the electrical system modeling could be simplified by assuming a dominance of fundamental voltage and current harmonics. The consideration of steady-state voltage and current phasor results in a quasi-static phasor modeling approach. However, in recent times, the application of power electronic converters (PECs), FACTS, and high voltage direct current (HVDC) is prevalent. Moreover, the inclusion of newer power delivery topologies, such as distributed generations (DGs), smart grids (SGs), and micro grids (MGs), made it possible to incorporate diverse renewables (solar, wind, hydro, biomass, etc.) generation through converter interfaced generation technologies (CIGs) to the point of common coupling (PCC). Even the loads and transmission system nowadays enjoys converters to improve power factor and efficiency. Oftentimes, the electromagnetic transients coming from the power electronics circuitry fall beyond the range of the traditional scope of stability investigation. In Fig. 1.3, the time scale associated with diverse power system dynamic responses is attached. It could be observed that the time range for CIGs control is quite small ($\mu s$ to $ms$) compared to the synchronous machine dynamics ($ms$ to $min$) [16]. This is because, inherently, the CIGs provide no short circuit current during fault/disturbance and have a very small inertia footprint. Such vast utilization of fast-responsive power converters has made the dynamic stability analysis of power systems very much difficult and complex [15]. The stability issues that arise in the power transmission due to the loading of distributed resources at the distribution end are also important to consider.

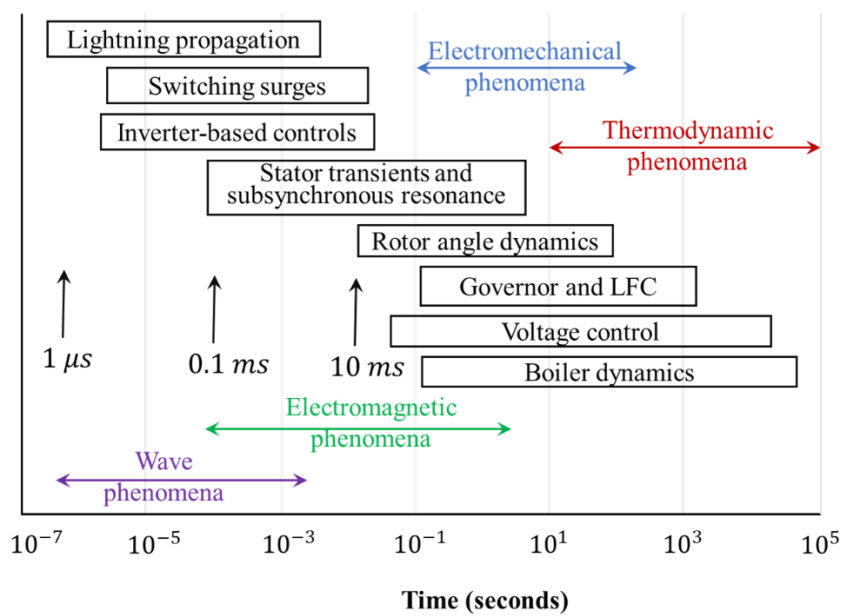

Figure 1.3: Time scales for various power system dynamics and control [16].

There has been extensive use of automatic controllers embedded in the simulation and monitoring hardware in recent times. Automation has become a solid root for extracting comprehensive real-time data of the power grid system, the big data. Big data in other sectors, especially in data science and statistics, has shown significant competence in predicting and indicating a target variable's behavior. With big data and high computing power on hand, incorporating machine learning (ML) and artificial intelligence (AI) models in power systems is also economically feasible [17]. Fig. 1.4 represents the taxonomy of transient stability analysis of power systems using traditional and ML methods [18].

The traditional hard-coded system control programs are still present in analyzing a known power system's behavior. However, network topology alters when distributed renewable resources with power electronics stages are embedded in the system [19]. Furthermore, additional load, like the charging pattern of electric vehicle batteries and vehicle-to-grid technology, tremendously curves the conventional load pattern [20]. Variation in load behavior places extra strain on the transmission line and power system components. It could push the power system beyond its stability margin, resulting in instabilities, adversaries, and disturbances [21]. With the current power system architecture, any adversary on the network nodes can alter the control commands and cause a blackout and, eventually, significant financial loss. In addition to that, the incorporation of modern tools such as supervisory control and data acquisition (SCADA), phasor measurement unit (PMU), and state estimation (SE) also exposes the prevalent power system to cyber vulnerabilities [22]. However, the biggest crisis lies in the slow computing process during an electrical component or fault outage, which leads to difficult post-contingency situations. As a result, the fast computational power of machine learning techniques (MLTs) is required to classify contingency appropriately and identify stability boundaries [23].

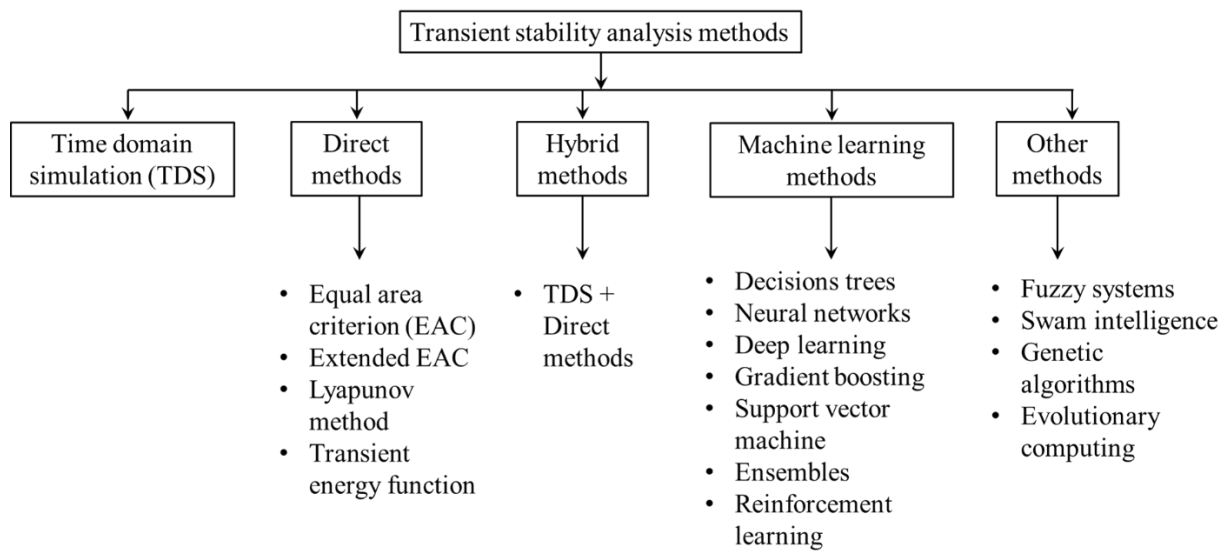

Figure 1.4: Power system transient stability analysis methods [18].

The efficiency of the ML model never reaches the perfect level. However, it was concluded in previous research that with the neural network, the efficiency could be improved [24]. By embedding a fast-performing ML model, the end entities could make faster decisions. The concept is becoming more critical when the energy economics and load dispatch are primarily shifted to a fully autonomous profile. In the coming revolution of the decentralized grid systems, bidirectional power flow between the consumer and grid is becoming a pivotal concept; ML would become the most crucial tool to model rapid load scheduling, system control, and stability analysis. Power systems analysis with ML and artificial intelligence (AI) is still a growing field. Over the years, researchers from industry and academia have come forward to reduce the complexity of ML/AI models while improving their fault detection and stability analysis performance.

## 1.2 Research Motivation

In recent investigations, MLTs primarily focus on power system stability analysis, disturbance classification, and security improvement [25]. According to Swarma *et al.*, MLTs are classified into supervised, unsupervised, and reinforcement learning [26]. Researchers use a simulation platform (Matlab/Simulink/Modelica/PSS/E) and produce massive feature data from IEEE standard power systems [27]. The large-scale data is then preprocessed and filtered to apply MLTs. In addition, support vector machine (SVM), neural network, k-nearest neighbor (kNN), and other algorithms are often incorporated into the disturbance classification [28]. In [25], a comparative performance analysis of the recently proposed MLTs for power quality disturbance classification has been presented. In [28]–[30], it is stated that for the feature extraction process, consideration of dominant features from either time (using empirical mode decomposition) or frequency (using Fourier transform) domain is unfeasible for dynamic system models, and use of wavelet transform is a good solution. Although various combinations of feature extraction techniques and classification algorithms are used to improve ML performances, it was observed that the handling of massive data set becomes computationally unfeasible. Moreover, it is difficult to map the possible contingency since the nodes are pretty large. In [25]-[30], it is observed that the proposed MLTs fail to provide a fast way to incorporate possible disturbances to the dataset without creating a massive data burden.

In this work, the neural network architecture and a few conventional MLTs are investigated to predict small signal stability using IEEE-14 bus system's eigen value dataset. In addition, the static dynamic of AVR model with close-loop controllers are also observed. The eigen value dataset is produced through applying Monte Carlo-based contigency sampling technique of IEEE-14 bus system. Moreover, a comparison with classical ML models such as SVM, kNN, logistic regression, Naïve-Byes, decision trees, and random forest, performed on the same filtered dataset, is provided to outline a comparative performance evaluation. The Monte Carlo operation reduces the extraneous contingency scenarios from the dataset, thus reducing the data burden significantly. It is expected that ML scheme would be very much faster than the conventional power system stability classification method. With a high-performance score and faster speed, the provided method would become very useful in fast power system stability analysis with different types of load, renewable/non-renewable generating units, and failure of grid equipment.

## 1.3 Research Objectives

In this work, eigen value dataset generated through using Monte Carlo sampling technique is used to demonstrate efficient power system stability classification using neural network techniques. The objectives of the proposed research are to-

1. Apply neural network to observe the stability profile of IEEE test systems,
2. Study the Monte Carlo sampling technique for contingency reduction,
3. Design a few physics-based traditional ML algorithms (logistic regression, kNN, SVM, etc.), and
4. Formulate a generic process of the proposed model for data-driven applications, such as load scheduling, energy economics, and big data.

## 1.4 Dissertation Organization

The rest of this dissertation is structured in the following manner:

CHAPTER II provides the core theoretical information and background for investigating the power system stability with neural networks.

CHAPTER III introduces the mathematical modeling, computation steps, and tools for power system data generation, processing, and stability analysis in a machine learning platform.

CHAPTER IV depicts the results obtained during the stability analysis of IEEE busbar systems with various traditional and proposed machine learning and neural network classifiers, along with their comparison with previous relevant research works with brief explanations.

CHAPTER V wraps up the whole investigation, enlists the outcomes of the thesis, and paves possibilities and impact of the thesis with future research directions.

# CHAPTER II

# Literature Review

## 2.1 Introduction

Power system control and stability are two major aspects of maintaining the real-time power flow within the safety limit of connected switchgear, control gear, transmission line, busbar, and communication relays. The power system control using automatic voltage regulator with proportional integral derivative, linear quadric gaussian, and linear quadric regulator controllers are excellent combinations to investigate the effect of systems' gain parameter variation. The unit response of automatic voltage regulator blocks subdivided into generator, sensor, exciter, and amplifier stage change could be considered as a reference for future investigation with AVR with the model predictive controller or fuzzy-neural controllers. The power system stability considers the system states and their pole-zero plots on the complex plane to observe the overall stable nature of the system during the contingency and under normal operating conditions. The usual way to observe whether a system is stable or not is to plot the real and imaginary parts of the complex damping ratio of the system. Machine learning and neural network exercise the path to stability consideration in a faster manner. The use of Poisons distribution and the Monte-Carlo optimization technique are fantastic ways to weigh only the contingency combination that are the most likely to happen. The IEEE-14 bus is the main bus configuration throughout this analysis.

## 2.2 Power System Control

### 2.2.1 Automatic Voltage Regulator System

An AVR system addresses the voltage instability issue of the power system by employing excitation control. At various operation conditions, the AVR system retains the SG terminal voltage close to the nominal level. The field exciter is economically operated to control the SG voltage. The overall AVR system primarily consists of an amplifier, exciter, generator, and sensor subsystems [31]. Usually, the non-linearity and saturation phenomenon is overlooked in the simplistic analysis of the control strategies. The simplified transfer function of those subsystems is then considered with explicitly defined subsystem time constants and gain parameters.

In Fig. 2.1, the subsystems are denoted to properly realize the AVR function. The $V_{ref}(s)$, $V_s(s)$, $E(s)$, $V_R(s)$, $V_F(s)$, and $V_T(s)$ refers to the reference voltage, sensor output voltage, error signal, exciter input voltage, field excitation voltage, and generator terminal voltage [32]. The combined subsystem provides the AVR model, which, when provided with step response, behaves very poorly. Figure 2.2 demonstrates the resultant time to rise to the peak, settling time, percent overshoot, and steady-state error values [32]. It is thus necessary to embed the AVR model with dynamic controller blocks to make it more robust and stable for diverse system operation

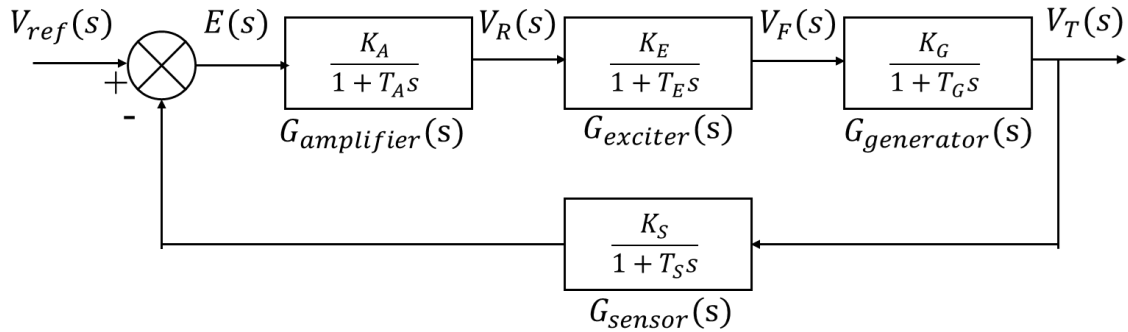

Figure 2.1: Block diagram of an AVR system [32].

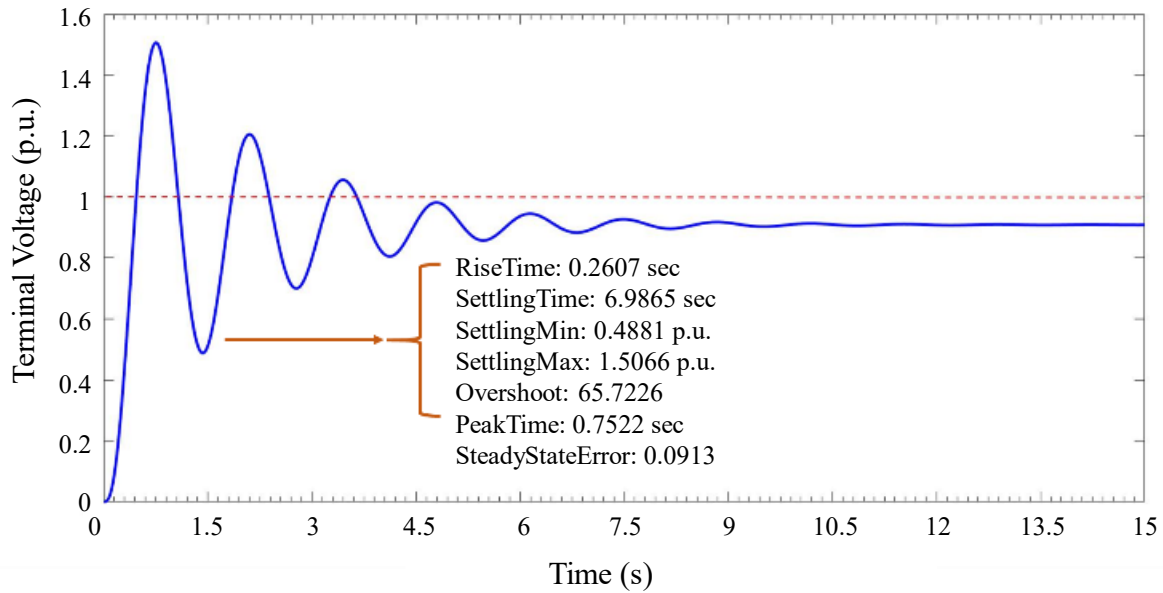

Figure 2.2: Terminal voltage variation of an AVR system [32].

### 2.2.2 Automatic Voltage Regulator with Modern Controllers

The automatic voltage regulator (AVR) of a synchronous generator (SG) is essentially used in power system utilities to improve the voltage stability of the system. Any drop in the generator terminal voltage can result in increased line losses, voltage fluctuations, damage to the loads, and causes financial issues. Thus, it is required that the output voltage is efficiently controlled. The AVR governs the terminal voltage of an SG to keep it to the nominal rating under various system phenomena (no-load, full-load, half-load, disturbance, and others) by altering the exciter voltage of the generator [33]. In a generator, the large inductance value of the field windings and rapid load fluctuations reduces the dynamic response of the AVR [34]. Therefore, improving the dynamic response of the AVR is a must. In this regard, diverse control techniques comprising Proportional Integral Derivative (PID) [35], Proportional Integral Derivative Acceleration (PIDA) [36], Fraction Orders PID (FOPID) [37], Sugeno Fuzzy Logic (SFL) [38] controllers have been investigated by researchers to address and to improve the AVR system dynamics. However, a PID controller is the most preferable due to its simple design structure and robustness in operation [39].

Though the augmentation of the controller segment can somewhat improve the AVR system dynamics, the controller can not maintain stability during variable operating points, nonlinear loads, and time delays. The drawbacks are addressed by tuning the controller gain parameters in dynamic system presets. In the case of the PID controller, the gain parameter values can be enumerated by considering trial and error [40], conventional Ziegler Nichols (ZN), or Cohen Coon (CC) methods ([41]). The trial-and-error process requires a tremendous amount of time to converge and seldom yields any optimal result. The traditional ZN and CC method adjusts the gain parameters by considering linear system modeling for an operating point. Neither of the two ways can efficiently withstand system non-linearity during discrepancy and often results in undesirable overshoots along with long-term oscillations in the system.

Moreover, both methods require a higher load of numerical analysis to extract the optimal parameters of the PID controllers and, thus, are inefficient. Optimization techniques are introduced to overcome the problem of PID gain parameters tuning. Intelligent optimization techniques can adjust the system in varying system conditions. Artificial intelligence optimization techniques fall in the category of neural networks and fuzzy logic [42]. In an artificial neural network, the training process considers a significant amount of data and higher convergence. In the fuzzy logic system, the system efficiency depends on the analysis of data, tuning of a model, and the designer's competency during the creation of fuzzy membership functions [43]–[45]. Thus, both of the methods tune differently based on the device, amount of available data, computation complexity, and mindset of the operator.

Apart from the PID controller, the use of LQR, LQG, model predictive control (MPC), and $H_\infty$ controller in power system control has been very popular [46]–[48]. In an isolated power system with AVR, the LQG controller performs better than the LQR or PID controllers [32]. The performance of the AVR system could be improved by implementing a damping compensator with the LQG (DC-LQG) controller [49]. DC-LQG system helps in improving the overshoot damping efficiency [12]. In [50], the LQR control strategy was used with a free intelligent model to design a power system stabilizer that facilitates close-loop system identification. The model could be expanded for a multi-machine system [51]. When used in a decentralized model, the LQR controller helps in efficient dynamic state estimation to control the power system oscillations [52]. For applications in intermitted renewable energy integration, the LQG controller is often used. The LQG controller is used to ensure optimal variable speed control of wind farms [53]. Moreover, the LQG controller with loop transfer algorithm helps in damping out the inter-area oscillations in the power system [54]. LQG is also considered for thyristor-controlled series capacitors; it dampens the inter-area modes while retaining the local modes unaltered [55]. When the power system itself is unregulated, and disturbance follows, the use of $H_\infty$ controller applied in LFC ensures a few lower-order controllers which could work in the nonlinear system condition [56]. The use of a coordinated distributed MPC with LFC provides a feasible way to incorporate intermitted wind energy into the traditional system [57]. MPC control could also lead to optimal power sharing between generating station and retaining DC bus voltage unaltered [58]. Oftentimes, distributed MPC is used, where each power system subsystem is embedded with an MPC unit, for coordinate controlling and boosting the operation of automatic generation control [59]. With times, more and more control system are being proposed by the research to properly control the power system dynamics in a more efficient way with reduced computational burden and complexity.

In recent times, meta-heuristic optimization algorithms have been getting more attention due to their tunability of controller parameters in a much simpler and easier way without requiring any information gradient. Such optimization algorithms can address system fluctuations in a dynamic environment with significant efficacy. The heuristic algorithms can primarily be categorized under four different groups: swarm-based algorithms, physics-based algorithms, human-based algorithms, and evolutionary algorithms. Researchers have introduced different types of swarm-based algorithms that include Particle Swarm Optimization (PSO) [60], Artificial Bee Colony (ABC) [61], Bat Search (BAT) [62], Ant Colony Optimization (ACO) [63], Cuckoo Search (CS) [64], Many Optimizing Liaisons (MOL) [65]. Physics-based algorithms group is comprised of Gravitational Search Algorithm (GSA) [66] and Bio-geography-Based Optimization (BBO) [67]. Under the human-based algorithms, there are Teaching Learned Based Optimization (TLBO) [68] and Harmony Search Algorithm (HSA) [69]. The corresponding authors have tried to investigate the transient response, root locus, and bode diagrams for the optimal PID-AVR system design using the algorithm. Moreover, the robustness of the proposed models is depicted by changing the exciter and generator model's transfer function gains as a dummy for system disturbance.

The practice of artificial intelligence techniques, such as a neural network for power system controlling and monitoring, has become a trend of research interest. When added to the conventional mathematical approaches, this machine learning technique could facilitate even more effective ways to find solutions to complex load flow equations and could help the precise tuning in automatic controllers to dispatch required system control to minimize system oscillation and to keep the system operation stable.

### 2.2.3 Controller Design

The PID, LQR, and LQG controllers are considered in this thesis to observe how they could improve the response of an AVR system. This section provides the structural configuration of the controllers. The dependency of the considered three controllers on time constant and gained parameters of the subsystems are also observed.

### 2.2.3.1 Design of PID controller

The most simplistic controller is the PID controller, which operates by the weightage placed on its proportional, integral, or derivative gain parameter value. The basic function of the PID lies in considering the errors between the estimated and actual parameter values. In Fig. 2.3, the $r(t)$ and $y(t)$ respectively refer to the reference set point and the measured value for the PID controller [32]. The error between the reference and measured, $e(t) = r(t) - y(t)$, is considered and addressed by changing the gain of the PID block. Each gain parameter is responsible for certain system behavior, and the weightage average tries to reduce the value of $e(t)$ over time.

The proportional part (P) determines the error between the measured value and the desired value and reacts proportionately to minimize the error. If there presents no error, the proportional part becomes zero. The past values of the error signal over time are integrated and taken into consideration by the integral part (I). The integral part tries to minimize any residual error remaining after the system has been dealt with by the proportional part. The derivate (D) unit functions in predicting the future change in the error signal. The decision is obtained by measuring the current slope of the error signal $e(t)$. Due to the derivate nature of control, the higher the rate of change of error is, the higher would be the damping control.

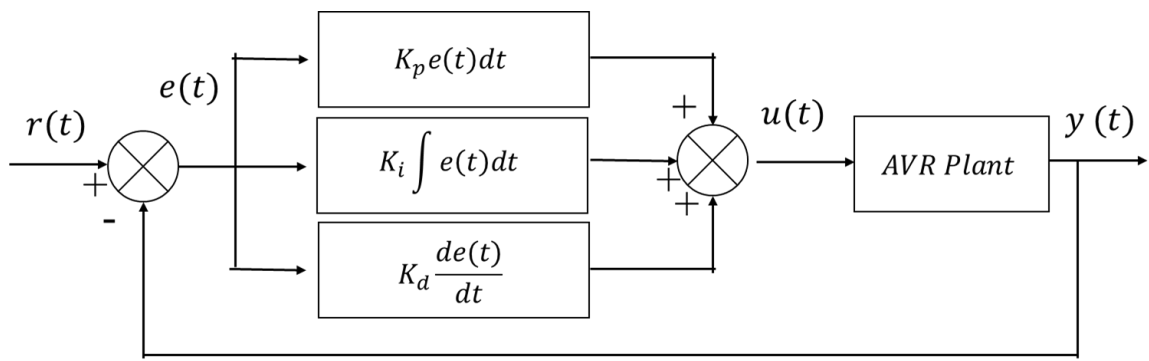

Figure 2.3: Block diagram of a PID feedback controller [32].

The efficient operation of a controller is usually assessed in terms of performance, such as rise time, steady-state error, and static or dynamic robustness. A higher proportional gain would place more contribution to reducing the rise time of the system but would also be improving overshoot. The integral gain minimizes the steady-state error with the penalty of increased settling time. One way to reduce overshoot and minimize settling time is to increase the derivative gain. Usually, a combination of these three gains is used with the gain value adjusted accordingly in a way that varies from application to application, ensuring the desired control characteristics.

### 2.2.3.2 Design of LQR controller

The LQR controller enhances the system operation by ensuring an optimal control regulator system with a feedback gain. A quadratic cost function is minimized to observe the proper control flow throughout the process. The block diagram of the LQR block can be expressed in Fig. 2.4 [32].

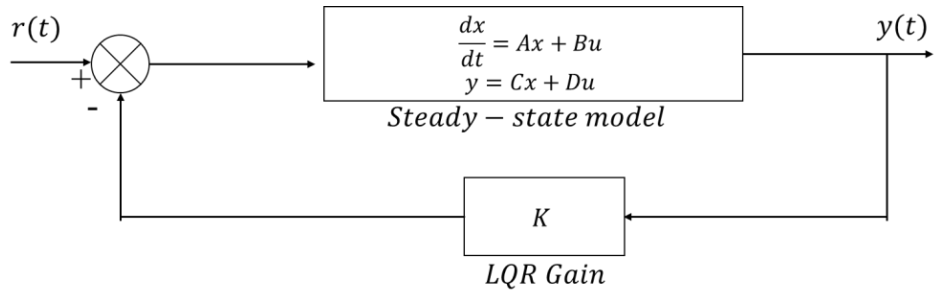

Figure 2.4: Block diagram of an LQR feedback controller [32].

### 2.2.3.3 Design of LQG controller

The LQG controller operates on the principle of the LQR with an additional state estimator unit. LQG provides robust control of the system over a wide variant of system dynamics, uncertainty, and disturbances. During the modeling stage, Gaussian white noise is forcefully entered into the dynamics to simulate real-life disturbances. This white noise distorts the state signal and makes the state information incomplete. As a result, at the summation node of the control strategy, the error is misinformed. In LQG, an additional state estimator like the Kalman filter is added that takes distorted state variables as input and, using an optimal recursive data processing technique, renders state value with preciseness. In this way, the error between the controller output and command signal is optimized for further improvement. The simplified layout of the LQG system is provided in Fig. 2.5 [32]. The design steps of the Kalman filter and the recursive data processing methodologies can be obtained in ref. [70].

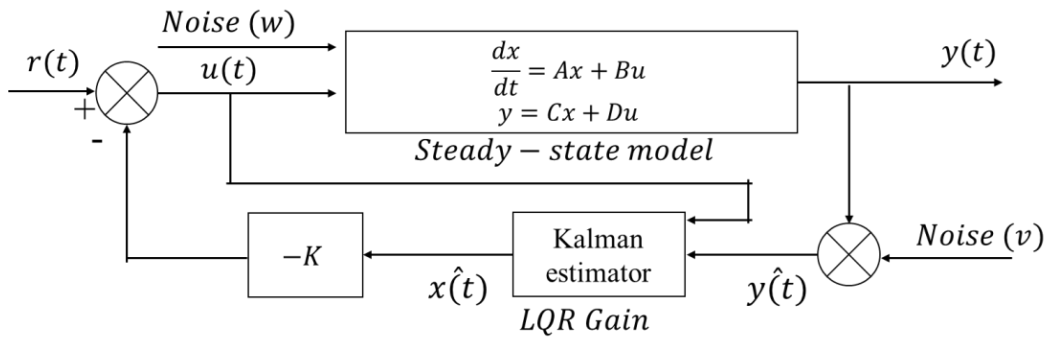

Figure 2.5: Block diagram of an LQG feedback controller [32].

### 2.3 Power System Stability

#### 2.3.1 Power System Stability and Types

In 2004, a generalized stability classification was proposed by omitting the sub-synchronous resonance because, at that time, the presence of sub-synchronous resonance was insignificant. The proposed variable-wise stability types consisted of rotor angle, frequency, and voltage stability. The prevalent use of power electronics converter stages has been placing disturbances to the system dynamics at the sub-synchronous level. Thus, called to include the electromagnetic transient to the other parts of the stability to encircle the total stability profile. In 2016, a report published by IEEE PES included resonance stability and converter-driven stability to the original stability types produced in 2004 [16].

Traditionally, the variable-wise power system stability classification included – rotor angle stability, voltage stability, and frequency stability. In this classification proposed in 2004, the sub-synchronous resonance was not incorporated. However, the recent involvement of CIGs and PECs in the power networks requires further including electromagnetic transient for proper stability analysis. In 2016, a reform was dispatched to mitigate such a hurdle, which concluded in a report published in IEEE PES that added two more parts to the original stability classification: resonance stability and converter-driven stability [16], [71]. Fig. 2.6 demonstrates the extended classification of power system stability [16].

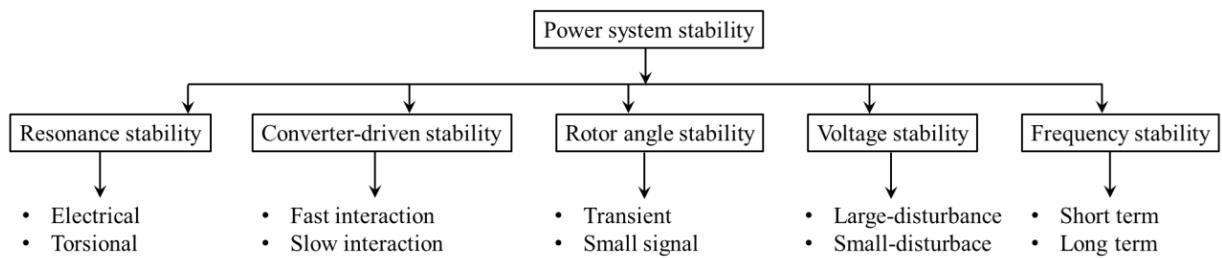

Figure 2.6: Classifications of power system stability [16].

#### 2.3.1.1 Rotor Angle Stability

The rotor angle stability, when viewed from an interconnected power system analysis, is defined as the inherent ability of a synchronous machine to revert to synchronism after it is subjected to noise or disturbance. The synchronism happens for a machine when the electromagnetic torque is equal to the prime mover's mechanical torque. The higher the ability of the synchronous machine to maintain the equilibrium between these two opposite torques, the higher the stability of the power system follows. The instability from the rotor angle primarily happens due to the lack of efficient damping or synchronizing torques [72]. The earlier leads to oscillatory instability, and the latter results in non-oscillatory instability. When in a system, the negative damping torque becomes absent, the effect of small disturbance sustains and leads to small-disturbance oscillatory stability. The oscillatory disturbance happens in either the intra-area mode, inter-area mode, control mode, or torsion mode [73]. Usually, non-oscillatory or aperiodic transients result from inefficient or negative synchronizing torques. The instability that follows aperiodic transient involves larger rotor angle excursions, usually investigated through numerical integration. In current literature, the rotor angle stability is sometimes also divided into small signal stability and transient stability. The former is disrupted due to small noises and disturbances in the system, and the transient stability depends on the presence of large disturbances and major faults in the system [74]–[76].

The small signal stability of a linearized power model could be investigated by determining the eigenvalue of the characteristics matrix A. Depending upon whether the eigenvalue is a real or imaginary, $\lambda = \sigma + j\omega$, a damping ratio ($\xi$) equation (in %) is formulated. The damping ratio provides the most important information regarding the system's level of stability and is defined as [77]:

$$\xi = 100 \times \frac{-\sigma}{\sqrt{\sigma^2 + \omega^2}}, \qquad (2.1)$$

where $\sigma$ and $\omega$ are real values, a negative value of $\xi$ refers to an unstable system. It usually requires nearly 3% to 5% of $\xi$ to make a system stable in terms of performance [77].

In the case of transient stability, the non-linear swing equation is considered to observe the acceleration or deceleration of the rotor with a change in current flow in the stator circuit at dynamic load conditions [34], [35]. From ref. [35], the swing equation could be formulated as follows:

$$\frac{2H}{\omega_{syn}} \frac{d^2\delta(t)}{dt^2} = P_m(t) - P_e(t) = P_a, \qquad (2.2)$$

where, H, $\omega_{syn}$, $\delta(t)$, $P_m(t)$, $P_e(t)$, and $P_a$ are the normalized inertia constant, load angle, mechanical power, electrical power, and accelerating power, respectively. After a change in system topology or in the presence of system disturbance, the eigenvalues of the state matrix of a linearized system are changed from the left half plane to the right half plane [78], [79]. The left plane of a complex plot is defined to be a stable region, whereas the right half plane refers to an unstable condition. The presence of a complex-conjugate pair of eigenvalues demonstrates the oscillatory nature of the system [80].

Different methods have been proposed in recent times to address transient stability, which include:

1. Direct or conventional methods, where the stability criteria are fulfilled at the position when the accelerating area of the synchronous machine equals the decelerating area. Some procedures used to determine such are equal area criterion, single machine equivalent method, direct Lyapunov's theorem, potential energy boundary, boundary controlling unstable, shadowing method, and comprehensive method [72].

2. Integration methods occur on the basis of finding the time response of a rotor angle by integrating the rotor momentum waveform. Procedures used for integration include Runge-Kutta methods, implicit trapezoidal rule, and mixed Adams-backward differentiation formulae.

### 2.3.1.2 Voltage Stability

Under the voltage stability condition, the power system can retain prescribed voltage labels at all of its buses even after being subjected to an external disturbance. The voltage stability is directly related to the security of a power system. On a daily basis, the reactive power demand from the load center significantly varies and causes a power imbalance between the generation and demand which leads to voltage instability [74]. The effect of voltage instability is realized in the loss of load area, unnecessary tripping of distribution centers, and damage and malfunction of connected network components and protective devices. A large magnitude of voltage instability could even result in cascading failure and push the rest generating stations to go out of synchronism. The change in field current flow at the SG may damage the mechanical parts or could increase loss [16]. Cascading failure leads to a collapse in national grid operation and could create black-out conditions. Factors such as the transmission line strength, magnitude of power being transferred from generation to the load side, dynamic load variation pattern, reactive power flow level, types of power and voltage compensators used, operation and coordination between controllers and protective equipment, intermittent penetration of changing generating units like renewables could contribute to the system collapse due to voltage instability.

The analysis of voltage stability is divided into short-term and long-term bases. Short-term voltage stability concerns electrical components with fast load dynamics, such as induction motors, electronic loads, high-voltage DC links, and inverter-based generators. The short-term stability assesses the system stability, such as rotor angle stability or converter-driven stability, only within a several seconds period [78]. Such slow interaction requires the knowledge of dynamic load modeling and probable short circuit faults in the main system. On the other hand, long-term voltage stability considers the dynamics of sluggishly acting components, which involves tap-changing transformers, theoretically controlled loads, and current limiters of generators. The long-term stability usually appears in the form of a progressive reduction of the node voltages. During the process of instability, after the field or armature current is overloaded at the SGs, the maximum power transfer and voltage support become limited. The long-term stability study period usually could extend to several minutes and requires extensive, lengthy simulations to properly analyze the dynamic performance of the system. Generally, this type of stability does not depend on the presence of a fault in the system but rather on the presence of an outage of transmission or generating equipment after the fault is cleared. When the system load dynamics try to restore the power consumption limit, even hitting a limit beyond the maximum power transfer level causes a loss of long-term equilibrium condition, and this directly leads to long-term instability. Apart from that, instability may arise in a failed attempt to restore a stable post-disturbance equilibrium. It is crucial to observe the disturbance trend; for instance, disturbance leading to instability could be just a progressing accumulation of load like the morning load increase.

In simpler terms, the concept of the steady-state voltage stability principle is generally explained with a 2-bus system, as shown in Fig. 2.7 [81].

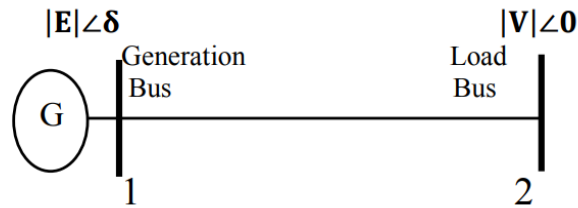

Figure 2.7: Two bus power systems [81].

The real power ($P$) and reactive power ($Q$) transfer from bus 1 to bus 2 follows the following relations [82]:

$$P = \frac{EV}{X} \sin \delta \qquad (2.3)$$

$$Q = \frac{V^2}{X} + \frac{EV}{X} \cos \delta \qquad (2.4)$$

The normalized value of equations 3 and 4 leads to real positive solutions of the voltage as [82]:

$$v = \sqrt{\left( \frac{1}{2} - q \pm \sqrt{\left( \frac{1}{4} - p^2 - q \right)} \right)} \qquad (2.5)$$

In the current research, various techniques and tools are exploited to properly assess the instability of voltage, that includes the p-v curve technique, v-q curve technique, reactive power assessment, analysis of nodes of collapse through finding the singularity of the Jacobian matrix, continuation power flow technique [80], [82].

### 2.3.1.3 Frequency Stability

When a large disturbance happens in the system, the balance between the generation and load shifts significantly. Frequency stability refers to the ability of the system to revert to stable behavior following a significant disturbance. In a large interconnected grid system, when the large system is split into the small islanded network, frequency instability could cause unnecessary tripping in generating units and loads. The increased penetration of renewable energy sources has recently been an agenda to maintain frequency stability [83]. In traditional power plants, load frequency control (LFC) is exploited to enhance control of frequency mismatch. LFC consists of primary, secondary, tertiary, and time control. Moreover, measures for emergency conditions are embedded during LFC design.

### 2.3.1.4 Resonance Stability

In a power system, resonance refers to the periodical oscillatory energy exchange between inductance and capacitance. This oscillation is usually dissipated by the path resistance. However, when the dissipated energy is insufficient to maintain the resonance, the oscillation grows and could boost voltage/current/torque magnitudes beyond the prescribed level. Generally, the subsynchronous resonance (SSR) is encompassed by the resonance, although resonance could actually be associated with the electromechanical or entirely electrical phenomenon. The original definition of the SSR refers to two possible forms of resonance [84]; the first occurs between the electrical series compensation and mechanical torsional frequencies of the turbine-generator shaft, and the second form comprises only electrical resonance as an induction generator effect [85]. The first form of SSR, the resonance between the series compensated line(s) and turbine-generator shaft directly related to the traditional SG operation and has been well investigated in the literature [84], [86], [87].

In [47], the IEEE working group has classified the subsynchronous oscillations (SSO) into SSR and device-dependent subsynchronous oscillations (DDSSO). According to the report, SSR involves a significant amount of energy exchange between the power network and the natural sub-synchronous torsional modes of the turbo-generator mechanical shaft [86]. The torsional oscillations happening in SSR could range from damped, undamped, and negative damped and could be growing in nature; this directly burdens the mechanical stability of the turbo-generator shaft. DDSSO results due to the fast interaction of control components, such as static var compensators, static synchronous compensators, PSS, and other power electronics controller stages with the mechanical torsional modes of the turbo-generators in the vicinity [88]. In a doubly-fed induction generator circuit, the possible resonance appears between the generator and series compensators [89]. At times, self-excitation type SSR appears due to the interaction of the series capacitance with the inductance present in the generators at the sub-synchronous frequency and results in the cumulative circuit resistance is negative.

### 2.3.1.5 Converter-driven Stability

In the present power system, the vast presence of power electronics converter stages has become a common place situation. The use of a voltage source converter (VSC) circuit to interface with the utility grid has made CIG quite distinct from the traditional SGs [70]. The proper operation strategy of the CIG depends on the control loops and control algorithms to provide fast responses in real-time. The use of phase-locked loops (PLL) and inner-current control loops are two of the convention. Control of CIGs requires lengthier time lengths, and as a result, the control signals makes interaction with both the electromechanical dynamics and electromagnetic transients. This makes the power system move towards an unstable region, followed by unnecessary oscillations of a wide range of frequencies [90]. Instability oscillation produced by the CIGs control loop could be of low frequencies (~ less than 10 Hz) and higher frequencies (tens of Hz to a few kHz), the first term is slow-interaction converter-driven stability, and the second one is fast-interaction converter-driven stability.

### 2.3.2 Small Signal Stability Analysis

In the power system, the small signal stability refers to the ability of a power system to retain synchronism in the presence of a small contingency. Any non-linear relationship, when considered from a tiny time scale, could be expressed through a linear-like relationship. A linear system could be expressed with output variable $y$, input variable $u$, and state variable $x$. System variables that change with time are considered in the state matrix. Thus, the system could be expressed through the following functional relationships:

$$y = f(x, u) \tag{2.6}$$
$$x' = g(x, u), \tag{2.7}$$

where $x'$ is the rate of change of state variables as time passes. The relation among the changed state, current state, input vectors, and output vectors are included in the state and output equations. For a linear system, the equations are:

$$x' = Ax + Bu \tag{2.8}$$
$$y = Cx + Du, \tag{2.9}$$

where A, B, C, and D are, respectively, the system matrix, control matrix, output matrix, and feed-forward matrix. When the change in the state becomes independent of the current state (input), values matrix A (matrix C) become zero. Similarly, when the system output is rendered independently of the system state (input) values, matrix C (matrix D) is null. In the case of a basic feedback system, the value of D is usually forced to zero.

### 2.3.3 Power System Stability Analysis with Machine Learning

The power system needs to be secured and stable to maintain a steady electricity supply to the consumers. The stability of a power system after being exposed to an external disturbance needs to be properly ensured. The stability of a power system usually refers to the stable operation of the system in normal conditions, maintaining equilibrium, and the ability of the system to revert to the stability margin in the post-contingency situation. The disturbance of the system could be small or large. Small disturbances usually appear due to step changes in loads or generation. Large disturbances are associated with major faults, such as faults in a transmission line or switchgear equipment, drastic loss of a huge chunk of load, or generating plants. Due to these disturbances, a variable degree of oscillations appears in the system parameters, voltage, current, frequency, and phase angles. This oscillation makes the restraining torques and the forward propagating torques at the rotor part become unequal to each other, leading to major maloperation in the power system.

The stability of the system stability is usually defined in terms of the rotor angle stability and voltage stability. The formal is exploited to detail the profile of all the rotors connected to the system and their degree of being in synchronization with each other following a disturbance. The voltage stability usually explains and cares about the voltage profile at all the system buses after a disturbance been occurred. The reactive power balance and the rotor torque balance are considered within the voltage and rotor angle stability study, respectively. Moreover, the magnitude of an external or internal disturbance occurring for a defined time initiates the concepts of transient and small signal stability. Small signal stability is studied in terms of steady-state and dynamic stability. Steady-state stability refers to the ability of the system to remain in synchronism when a gradual change of load has occurred, whereas dynamic stability concerns small disturbances or faults in the power system.

Due to small disturbances, oscillations appear in the system. When these oscillations are damped in a fast manner, they place a negligible burden on the system. However, with the uncontrolled growth of these small oscillations, the system becomes dynamically unstable. Dynamic stability response is studied for 5 to 10 sec, but it could take nearly 30 sec. When the fault is of a lower degree and places little burden on the equilibrium, easy to solve, it falls under the small signal stability. In comparison, the sudden placement of large faults is studied within the transient study. Large faults need to be detailed with the complex non-linear equation, which makes the computation trickier. The transient stability response is usually computed within 1 sec. For simple cases, transient stability analysis could be performed by manual computation, but the majority of the time, computer simulation is required. Traditionally, graphical approaches such as the computation of equal area criterion had been considered to determine the stability profile of the system following a contingency. In [91], the effect of the model complexity in a multi-machine system with external noises and disturbances is provided that directly changes the computational requirement and process time. Conventionally, load flow analysis, the study of equal area criterion, and the single machine to infinite bus bar system were the basis of studying the power system stability with traditional SG-based generation. Researchers have continuously been trying to produce a mathematical model with practical applications that could reduce the computational burden while improving system state determination and ensuring the stability of the power system.

The use of state-space equations for synchronous machine modeling has been recognized [92]. In a traditional power system, rotor angle-based stability analysis was commonplace. Analysis with the Jacobian singular matrix, Liapunov function, and Zubov's method has facilitated the analysis of the non-linear dynamics of the power system [93]. The parameter of the system could be identified using the Kalman filter for dynamic stability analysis [94]. The use of field-programmable gate arrays for precise d-q axis control of the rotor angle was investigated for stability analysis [95]. In [96], a delay differential-algebraic equation was proposed with its effect on small signal stability patterns in IEEE 14 bus system. Moreover, it is observed that the induction generator run by a prime mover could help in improving the transient stability in an SG-based power system [97]. In some cases, advanced models of SG are considered to improve the stability margin of the system [98].

The recent incorporation of renewables and converter stages introduces newer challenges in controlling the frequency limit and reactive power demand [99]. In [100], a probabilistic power flow method has been carried out to track the higher instantaneous power penetration to the utility grid from a wind energy source. The transient stability-driven wind power penetration profile has been investigated in ref. [101]. The converter's maximum power handling capacity and the grid voltage stability analysis in a weak grid has also been investigated [102]. The intermittent nature of the renewables requires increased use of storage devices. However, this also results in a change in system frequency dynamics [103]. The electromechanical oscillations produced by the use of the converter stage need to be properly handled. In a low-inertia system, the penetration of wind and solar power systems alters the power oscillation damping and changes the system eigenvalues [104], [105]. In [106] synchronous power controllers, the increased small signal and transient stability of large penetration PV cell is investigated. When a large number of non-synchronous machines, such as a commutator converter or voltage source converter, are connected to the system, they alter the voltage stability [107]. The problem of using a large number of converter blocks could be discarded by the introduction of synchronous motor-generator pair (MGP). It is observed that for the same inertia mass, MGP provides more damping than a single generator, which helps maintain the stability of the system [108].

The complex and vastly spread modern-day power system is becoming a big challenge for the conventional method of protection. Any maloperation in reactive power management and voltage stability could lead to global stability issues and blackouts of various magnitudes [109]. Traditionally, P-V and Q-V curves of the load buses are considered to be analyzed with load flow models, which lack the compatibility of the model to work at the dynamic condition at a fast speed [110]. The traditional method requires tedious computation even when the system dynamics are simplified, discarding even some crucial events. Many of the shortcomings in the stability classification with traditional stability computation techniques could be overcome by the use of ML techniques [111].

The use of machine learning and deep learning techniques is becoming very much lucrative to address the majority of the issues that are faced today. The facility of having high performance with high accuracy of prediction in a shorter time makes ML algorithm to be widely considered. The application of ML in power systems incorporates system monitoring, detection of intrusion or power theft, prediction of fault, load forecasting, and classification of stability. Supervised ML techniques such as regression, classification, support vector machine, and decision trees are some of the popular choices. Among the unsupervised ML algorithms, principle component analysis and k-means clustering are well known. In a real power system, the accurate classification of specific events is very difficult since the sources and types of disturbances and contingencies are vast [112].

The power quality disturbance (PQD) could be classified using machine learning and deep learning techniques such as support vector machine (SVM), decision tress (DT), K-means, artificial neural network (ANN), convolutional neural network (CNN), etc. [113]–[116]. However, the use of support vector machines and probabilistic neural network (PNN) are the most widely used due to their suitability in PQD. For instance, PNN is very suitable for high-accuracy data classification that involves signal outliers and works without needing any initial weight settings as other NN [117]. The use of SVM for voltage stability margin is getting interesting. It is observed that compared to the traditional optimized SVM model, the optimized SVM model with a neural network provides much better results [118]. The use of artificial NN costs much lower computation time and provides outstanding voltage stability margin indexing (VSMI) accuracy [119]. But the ANN requires excessive training, and the parameter tuning of the model is cumbersome oftentimes to model predicts substandard performance [120]. A viable option proposed was the use of feed-forward neural network topology like extreme learning machines (ELM), which results in faster and more accurate VSMI [121]. In a recent study with short-term voltage instability, the use of a neural network regressive model (NNRM) resulted in better accuracy than the traditional NN. The parameter tuning process of NNRM is very efficient, which makes it well-suited for VSMI [111]. The use of NN for power system classification thus has had hype in recent years. It is required to work on increasing the stability classification accuracy while reducing the computational process and training time required.

## 2.4 Monte Carlo Optimization

Two main types of system evaluation techniques involve analytical and simulational procedures. The former represents the system by analytical models, which are then evaluated through various mathematical solutions. In the latter one, the model is first built block by block and then simulated to observe the input and output of each stage. The Monte Carlo simulation method estimates the indices of a system through dispatching simulation of the actual system behavior embedded with random events. This method treats the overall system dynamics by first slicing it into a series of events or experiments. Both the analytical and simulation method has their merits and demits, and their application preference largely depends on the circumstance and the response time requirements. Generally, in a system where complex conditions of operation are aborted and the component failure probability is insignificant, a reliable system and analytical techniques are preferable. When the complex system dynamics need to be taken into consideration, and the severe fault or contingency events required to be considered, Monte Carlo methods are more efficient.

The concept of Monte Carlo was first realized in the 18th century by Scientist Buffon when he tried to enumerate the $\pi$ through the famous needle throw test method. According to the needle throw test, when a random throw of a needle of length $d$ and width $a$ is placed on a plane surface where a parallel line is drawn, the probability of a needle hitting a drawn line becomes $P = 2d/\pi a$, provided that $d > a$. As the probability of random events could be determined, this equation could help to calculate the value of $\pi$. The needle-throw test demonstrated the probable powerful use of the Monte Carlo method in both stochastic and deterministic situations. At present, the Monte Carlo technique is widely used in mathematical calculation, determination of medical statistics, analysis of engineering process flow, and power system reliability determination.

In power system stability, the mathematical expectation of the reliability index is considered to properly classify the reliability profile. The Monte Carlo technique works on this expectation feature. Considering the unavailability of a system or the failure of probability being $Q$ and $X_i$ Be a zero-one indicator and expressed as:

$$X_i = \begin{cases} 0; & \text{if the system is in the up state} \\ 1; & \text{if the system is in the down state} \end{cases} \tag{2.10}$$

The system instability, $\bar{Q}$, and the unbiased sample variance, $V$, could be expressed in terms of the number of state samples $N$ as follow [122]:

$$\bar{Q} = \frac{1}{N} \sum_{i=1}^{N} x_i \tag{2.11}$$

$$V(x) = \frac{1}{N-1} \sum_{i=1}^{N} (x_i - \bar{Q})^2 \tag{2.12}$$

Since the value of $X_i$ only results in 1 or 0, the variance becomes [122]:

$$\sum_{i=1}^{N} x_i^2 = \sum_{i=1}^{N} x_i \tag{2.13}$$

$$\begin{aligned} V(x) &= \frac{1}{N} \sum_{i=1}^{N} x_i^2 - \frac{1}{N} \sum_{i=1}^{N} 2x_i \bar{Q} + \frac{1}{N} \sum_{i=1}^{N} \bar{Q}^2 \\ &= \bar{Q} - 2\bar{Q}^2 + \bar{Q}^2 \\ &= \bar{Q} - \bar{Q}^2 \end{aligned} \tag{2.14}$$

The system unavailability estimation is provided in equation (6). If the uncertainty in the vicinity of the estimate needs to be obtained, the variance of the expected estimation needs to be employed as [122]:

$$V(\bar{Q}) = \frac{1}{N} V(x)$$
$$= \frac{1}{N} (\bar{Q} - \bar{Q}^2) \qquad (2.15)$$

Usually, a coefficient of variation, $\alpha$, is used to observe the accuracy of the Monte Carlo simulation and is mathematically related to the system unavailability and uncertainty in estimation by [123]:

$$\alpha = \sqrt{V(\bar{Q})}/\bar{Q} \qquad (2.16)$$

$$\alpha = \sqrt{\frac{1-\bar{Q}}{N\bar{Q}}} \qquad (2.17)$$

$$N = \frac{1-\bar{Q}}{\alpha^2 \bar{Q}} \qquad (2.18)$$

Eqn. (13) refers to the independency of required sample numbers $N$ on the size of the system; it only depends on the system's unavailability $Q$ at a desired accuracy level $\alpha$. This also shows the compatibility of the use of the Monte Carlo technique for large-scale stability evaluation and comparative advantage over the analytical procedures. For the practical reliability evaluation setup, the value of failure of probability or unavailability is very tiny and smaller than unity. Thus,

$$N \approx \frac{1}{\alpha^2 \bar{Q}} \qquad (2.19)$$

This denotes the inversely proportional relationship between the number of samples and the unavailability. This demonstrates the requirement for a very reliable system; the sample size needs to be larger.

A random problem could be solved through a variety of Monte Carlo techniques. Among the techniques includes the generation of random numbers, different sampling approaches or variance reduction techniques, and many more. The comparative efficiency of the type of Monte Carlo technique used is thus required to be compared to other Monte Carlo methods. Considering two Monte Carlo methods being used to evaluate the same system. Also, the reliability index expectation estimates are kept the same. If $t_1$ and $t_2$ refers to the computing time, $\sigma_1^2$ and $\sigma_2^2$ are the variances of reliability index being observed for the two systems respectively, the following relation could be used to compare both models [123]:

$$\eta = \frac{t_1 \sigma_1^2}{t_2 \sigma_2^2} \qquad (2.20)$$

If $\eta < 1$, the first Monte Carlo method supersedes the second one in terms of efficiency, and vice versa. The efficiency of a Monte Carlo system is the algebraic multiplication of the computational time and variance of the estimation, and thus, the sample size is not the only parameter to consider.

## 2.5 Poisson's Probability Distribution

Poisson's probability distribution is a discrete probabilistic model used to determine the number of events occurring within a given interval (usually). The occurrence of the events is considered to be mutually explicit, and knowledge of the probability of an event happening can not be used to predict the occurrence probability of other events. The distribution is based on a few presets. First, only the occurrence of an event could be determined. Second, the events occurring randomly throughout the interval could be divided into subintervals. Third, only one event could occur within a subinterval. Fourth, the length of the subinterval is proportional to the probability of a random event occurring in that subinterval. Fifth, the event occurring in a subinterval is directly independent of the event occurring in the other subinterval. When a given experiment fulfills the five criteria, then the experimental process could be outlined as to be Poisson's process. In a series of randomly placed discrete events, the average time between the occurrence of two random events at two subintervals is known, but the determination of the exact timing of the events is impossible to track. The length of the subinterval varies randomly, and so does the probability of an event occurring.

Consider a $\lambda\,(> 0)$ number of an average event occurring in an interval; the actual number of events occurring at the same interval could be found from the Poisson distribution. $X$ being a random number and carries the value of the actual events occurring; the Poisson distribution could be expressed as [124]:

$$P(X = r) = \mathrm{e}^{-\lambda} \frac{\lambda^r}{r!} \qquad (2.21)$$

Since $\lambda$ holds the number of average events or successes in a set of the interval where the Poissons distribution is taking place, both the mean and variance of the distribution are equal to $\lambda$. This is expressed by the following relationships [125]:

$$E(X) = \lambda \qquad (2.22)$$
$$V(X) = \sigma^2 = \lambda \qquad (2.23)$$

Poisson distribution could be applied to applications such as investigating traffic flow, the probability of defects occurring during the manufacturing process, and fault occurrence in electric cables. In this research, Poissons distribution would primarily be used to find the probability that the number of lines of the IEEE-14 bus system could be faulty at a given period of time. Knowing the faulty lines would help in generating a faulty dataset to investigate the power system stability using machine learning algorithms.

## 2.6 Machine Learning Algorithms

Machine learning (ML) corresponds to the scientific solution of computer-based models without requiring an explicit program. The model could be an algorithm or any other statistical model. ML instructs a computer or machine to process a specific data handling task efficiently and in a fast manner. The data problem requires algorithms to demonstrate the actual relationship. The type of algorithm is chosen based on the nature of the problem on hand, the type of model, the number of variables, and the magnitude of its complexity. The prevalent ML algorithms are demonstrated in Fig. 2.8 [126].

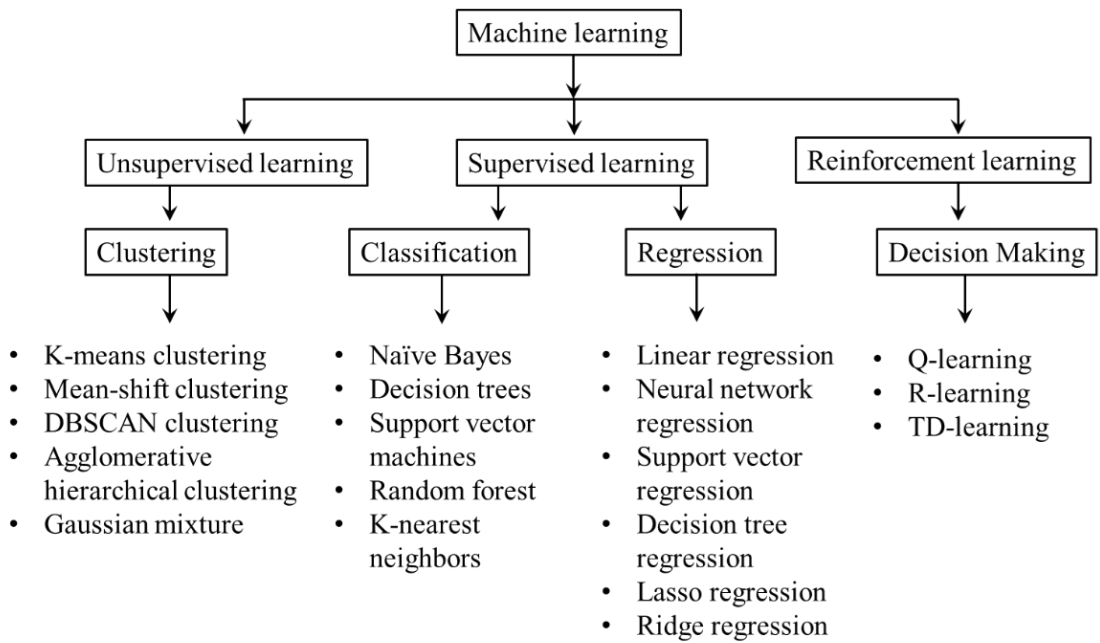

Figure 2.8: Commonly performed ML algorithms [126].

### 2.6.1 Supervised Learning

In supervised learning, the relation and mapping between the input and output are learned based on the previously provided example input-output pairs after the entire dataset is required to be divided into training and testing datasets. It exploits the labeled training data examples to train the model for prediction and, thus, requires external assistance. During the training, the algorithms help to map a relationship between the input variables and the output for predicting or classifying. The basic work steps of supervised learning are provided in Fig. 2.9 [127]. Some of the commonly used supervised learning algorithms for classification purpose includes the Naïve Bayes, support vector machine, k-nearest neighbors, decision trees, and random forest.

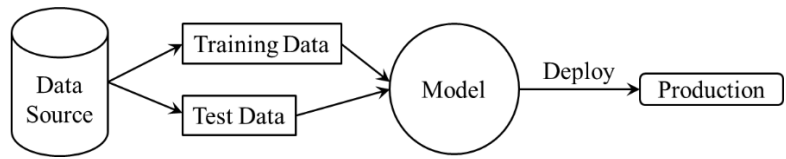

Figure 2.9: Supervised learning workflow [127].

### 2.6.1.1 Decision Tree

The decision tree works on the basic choice phenomenon. The ultimate result to be determined is found through a graph of choices, and the result of choice forms a tree, as shown in Fig. 2.10 [128]. According to the figure, each node corresponds to a choice made, and each branch refers to the value of the node.

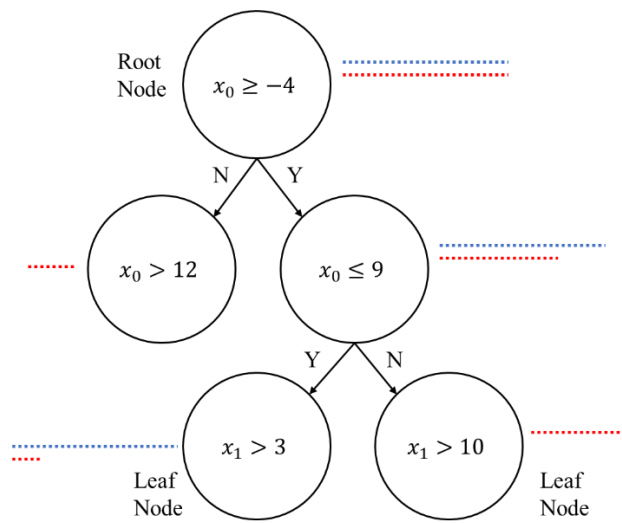

Figure 2.10: Sample decision-making process of a decision tree algorithm [128].

### 2.6.1.2 Naïve Bayes

The naïve Bayes theorem is formulated upon the Bayes theorem, which assumes an independent relationship between the predictors of an event. The salient presence of a particular feature happens in an unrelated manner to the other features. The algorithms are vastly exploited for the text classification industry, where classification and clustering technique with conditional probability of happening is required. The basic functional relation of the prediction principle with Naïve Bayes is provided in Fig. 2.11 [129].

### 2.6.1.3 Support Vector Machine

In the present day, the support vector machine (SVM) is one of the vastly used state-of-art ML algorithms. SVM could be applied for both linear and non-linear classification. It exploits the kernel trick by implicitly routing the inputs to a high-dimensional feature space, with margins between each class. The margins are outlined by keeping the classes at their maximum; thus, the classification error or misclassification is reduced (Fig. 2.12) [130].

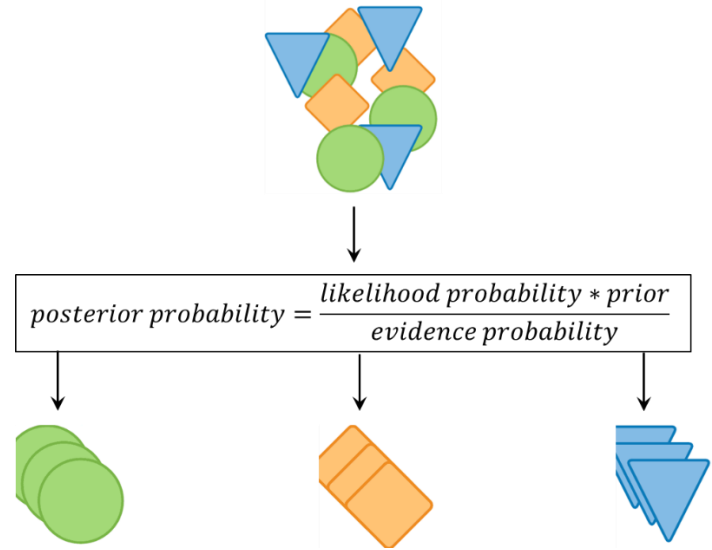

$$posterior\ probability = \frac{likelihood\ probability * prior}{evidence\ probability}$$

Figure 2.11: Probabilistic prediction relation of Naïve Bayes algorithm [129].

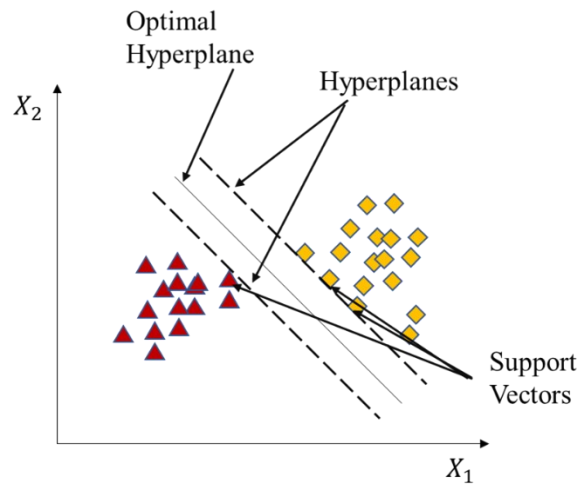

Figure 2.12: Classification plane for support vector machine algorithm [130].

### 2.6.2 Unsupervised Learning

In unsupervised learning, the algorithms are left on their own to recognize any interesting structure from the provided unlabeled dataset. Most times, the unsupervised algorithms can grasp more than one feature across the dataset. There presents no defined correct answer in this learning method, and it is mostly used for applications where clustering and feature reduction are required. Among the most well know unsupervised learning methods are principal component analysis and k-means clustering. The basic flow diagram of unsupervised learning can be expressed in Fig. 2.13 [25].

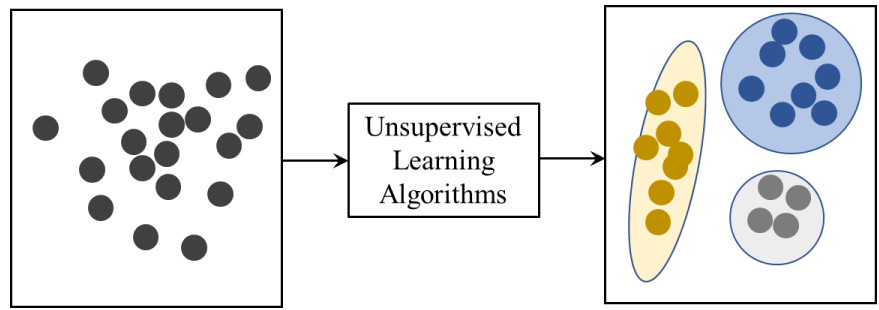

Figure 2.13: Flow diagram of unsupervised machine learning technique [25].

#### 2.6.2.1 Principal Component Analysis

In a principal component analysis algorithm, statistical methods are dispatched on a dataset where different features are correlated. Then orthogonal transformation is applied to convert that data into a set of linearly independent variables, known as to be principal components of the given dataset. This dimension reduction procedure reduces computational time significantly. Moreover, with the help of linear combinatorics, the variance-covariance relationship among the variables could be explained through these algorithmic procedures. The basic structure of dimensionality-reduction using the principal component analysis is given in Fig. 2.14 [25].

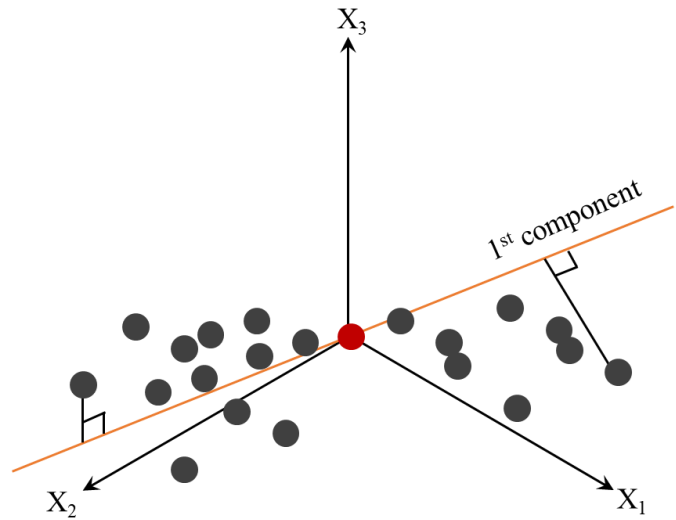

Figure 2.14: Basic structure of dimensionality-reduction using the principal component analysis [25].

### 2.6.2.2 K-Means Clustering

K-means is a well-known unsupervised ML algorithm mostly applied for data clustering practices. It follows a simplistic procedure to classify data from the dataset into clearly isolated clusters of similar features. A k-center is defined for each cluster so that each cluster stays with a clearly large margin than the neighboring cluster (Fig. 2.15) [25]. The data point from a query dataset is then considered, and the algorithms try to associate with the nearest center. This completes the first step. After all the data points are assigned to a cluster, the model validates the accuracy and tries to create a new k number of centroids using the information from the previous step. These k-centroids act as the barycenter of the clusters, and the process is repeated to get the best possible outcomes, and data is classified with preciseness.

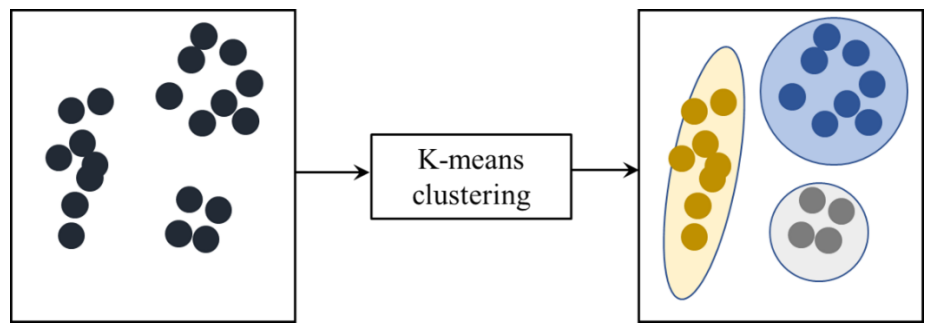

Figure 2.15: Neighboring clustering using a k-means clustering algorithm [25].

### 2.6.3  Semi-Supervised Learning

Semi-supervised learning could be considered as a combination of supervised and un-supervised learning techniques. It is mostly used with unlabelled data, which is very difficult to label without prior realizing the feature relations, or the process itself is tedious. Among the semi-supervised learning methods, the most common ones are generative models, self-training, and transductive SVM.

### 2.6.3.1  Generative Models

In generative models, the complete dataset is used to generate data points, $P(x, y)$, by using a probability distribution, given that $x$ to be input feature and $y$ is the outcomes. The mixture of different features could be easily realized by only one labeled example per component.

### 2.6.3.2  Self-Training

In self-training, the semisupervised classifier is fed with a little chunk of labeled data. Then, the unlabeled data is fed, and the classifier tries to predict the label for these data. The unlabeled data and the predicted label together are combined to create the training data. The procedure is repeated a few times because, in each iteration, the model trains itself and, while doing so, makes a more exact labeled dataset.

### 2.6.3.3  Transductive SVM

The partially labeled data could also be treated by a transductive SVM (TSVM) algorithm. The TSVM model tries to cluster the label and unlabeled data such that a clear margin between the two is present. The foundation of the TSVM method is somewhat difficult to realize, and it is very difficult to produce an exact solution using TSVM.

### 2.6.4  Reinforcement Learning

Reinforcement learning is a reward-based learning technique concerned with software agents. It determines how agents should react in an environment in order to maximize the reward returned from the environment. After each action is placed in the environment of the software agent, the state and reward associated with the effectiveness of the action taken given to agent.

The process is repeated for precise outcomes. The basic work loop of reinforcement learning is provided in Fig. 2.16 [131].

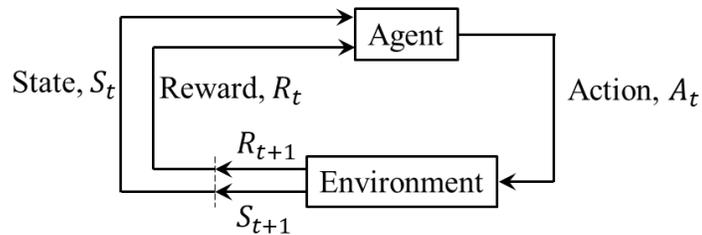

Figure 2.16: Reinforcement learning work-loop [131].

### 2.6.5  Multitask Learning

This is a sub-field of ML which concerns solving multiple tasks parallelly by exploiting the advent of similarities between the tasks. This boosts the learning curves and helps in finding dependency amongst the features under consideration. Mathematically, for n tasks, the tasks or their sub-set being similar but might not be the same, multitask learning method utilizes the feature or patterns contained in each of the similar sub-blocks of data and results in a much better prediction.

### 2.6.6  Ensemble Learning

Ensemble learning methods provide the opportunity to utilize a combination of various models in a strategically structured way to provide the solution to a difficult statistical problem. The performance of the overall model is improved in ensemble learning since the misclassification or prediction of one individual classifier or expert is buttressed by the others. Ensemble learning is mostly applied in precise optimal feature selection, error-correcting, and data fusion. The most important types of ensemble learning algorithms are boosting and bagging.

### 2.6.6.1  Boosting

In traditional ML algorithms, two major issues are bias and variance. A weak algorithm provides a large bias and variance during the learning and prediction. Boosting is an ensemble learning algorithm that boosts the performance of a weak classifier to become stronger. A set of weak learners could lead to a much stronger learner. Through an iterative process, the performance of the week classifier in terms of true classification is well-improved.

### 2.6.6.2 Bagging

Sometimes the stability and accuracy of machine learning algorithms become insufficient for proper prediction and practical uses. The bagging algorithms help to decrease variance and abate the overfitting tendency of a regression or classification algorithm and help in better prediction.

### 2.6.7 Neural Networks

Neural networks (NN) mimic the process of human training in predicting a feature. NN is composed of a set of algorithms that, when exposed to a dataset, try to figure out the implicit relationship between the features. The system of neurons in NN could be seen as organic or artificial. One of the most advantageous features of the NN is that it can adapt very well to dynamic and changing input. Once the criteria for output are set, the algorithms work independently of the data as long as the underlying relationship between the data is kept undistorted. The fundamental structures of NN are shown in Fig. 2.17 [132]. ANN usually comes in three basic layers, the input layer, the hidden layer, and the output layer. The data is provided to the input layer, processed by NN in the hidden layer, and the output is extracted from the output layer. The NN is directly relevant to AI and artificial super-intelligence (ASI), which are becoming very important in this $21^{st}$-century trading system.

Depending on the density of the layers, the structure of the neurons, and data flow depth, the NN or ANN is classified into three fundamental categories: CNN, recurrent NN (RNN), and deep NN (DNN). Sometimes, both CNN and RNN are studied within the DNN since the fundamental difference between the three is very abstract. The CNN only got one hidden layer; when the no of the hidden layer increase by more than one, it could be termed to be a DNN. Both the RNN and DNN are very suited to processing temporal data.

CNN is an artificial neural network algorithm that is specifically exploited in image recognition and image processing applications. The NN of CNN mimics the frontal lobe of the human brain that is uniquely structured to respond to visual stimuli. However, unlike humans, the CNN network requires vast time to learn and extensive data sets to train for good efficiency. Thus, CNN is usually applied for low-resolution images due to scalability issues. In present times, muti-layered NN has been introduced, which helped to apply CNN for other image processing applications much faster than before. A CNN with three hidden layers is shown in Fig. 2.18 [133].

RNN is a type of ANN that is well-developed for processing temporal data like time-series data. RNN is mostly used in applications where continuous data is present, like speech recognition, financial data, DNA sequence, or language translation. The NN nodes in RNN are connected in a temporal sequence and form a directed or undirected graph. Each neuron shows an inherent memory to remember the previous data flown through it and could predict the next expected data, and this gives the unique power of RNN for handling sequential data flow. Fig. 2.19 provides the basic flow diagram of an RNN [132].

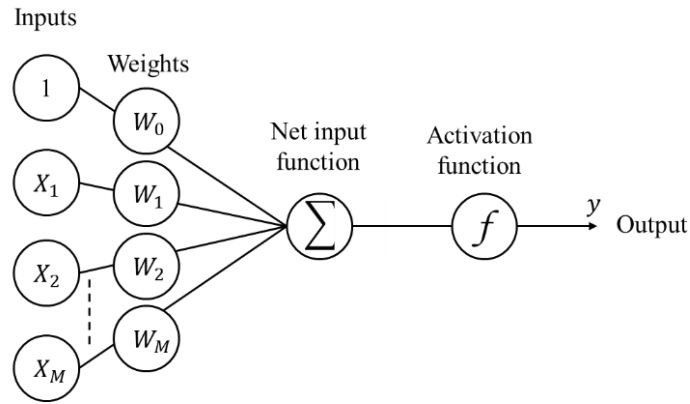

Figure 2.17: Fundamental structure of neural networks [132].

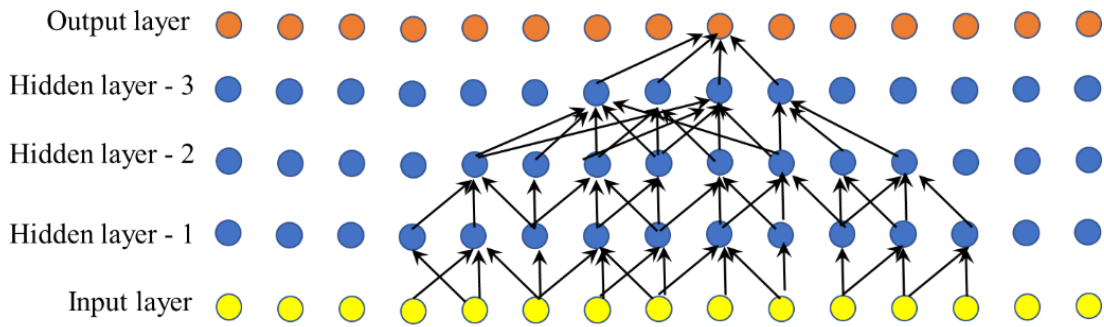

Figure 2.18: Fundamental structure of a convolutional neutral network [133].

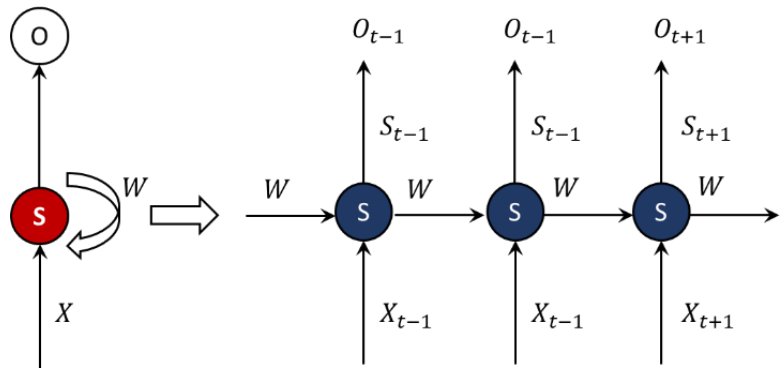

Figure 2.19: Basic flow diagram of a recurrent neural network [132].

DNN is an ANN with multiple hidden. The hidden layer could be formed of CNN or RNN, depending upon the application. Because of the chance of using CNN or RNN, DNN could be applied to vast practical application fields, including sound and image recognition, creative thinking, and the formation of ASI. DNN learning technique is a repetitive process, and it learns by each sample it handles. Thus the more and wide variety of the samples, the more sophisticated the DNN is. Another important feature of the DNN is that it is capable of self-learning and self-recognition of a non-linear relationship between the input and output vectors. The data provided to the input layer travels through the successive hidden layer till an accurate prediction and classification are determined. The DNN then dispatches a feedback loop to observe the strength of its learning curve and works on improving itself. The NN for DNN is shown in Fig. 2.20 [134]. The NN or ANN is generally grouped under supervised NN, unsupervised NN, and reinforced NN.

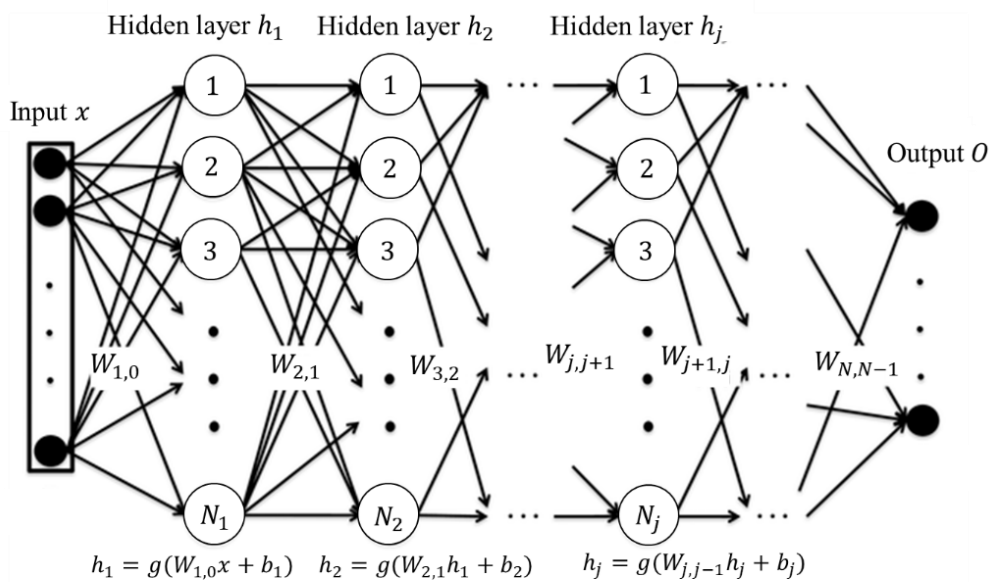

Figure 2.20: Fundamental structure of a deep neural network [134].

### 2.6.7.1 Supervised Neural Network

In a supervised neural network, the NN process resembles the basic structure of supervised learning. The output of the prediction or the goal is kept predetermined and used to check the predicted outcomes from the NN. A feed-forward loop is used to improve the prediction of the NN. While doing so, the NN parameters are tuned to the model and are improved. The basic feed-forward supervised NN is shown in Fig. 2.21 [135].

### 2.6.7.2 Unsupervised Neural Network

In an unsupervised neural network, the outcomes are not known beforehand. The NN tries to categorize the data based on similar features. The model keeps doing that till a clearly defined feature cluster is observed for the unlabelled data in an unsupervised fashion. The process flow of an unsupervised neural network can be presented in Fig. 2.22 [135].

### 2.6.7.3 Reinforced Neural Network

In reinforced NN, the performance of a software agent using the NN is improvised in a reinforced pattern. An objective function is first set, which helps to track the improvement of the NN prediction performance. The correct decisions led by the agent are incentivized, whereas the misclassifications are penalized, which helps in swift efficiency improvement. The metaphorical structure of the reinforced neural network is given in Fig. 2.23 [135].

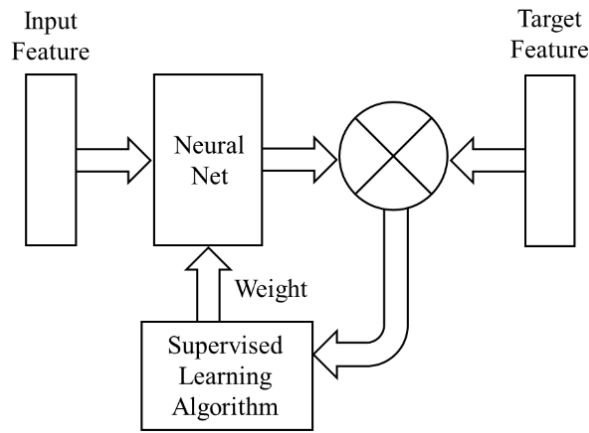

Figure 2.21: Fundamental structure of supervised neural network [135].

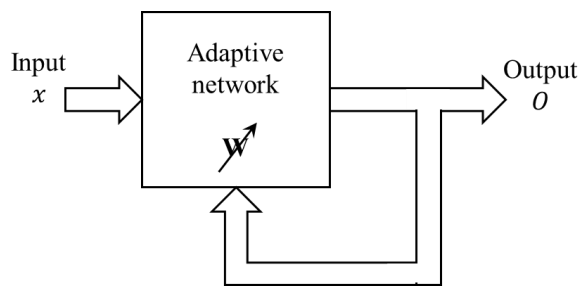

Figure 2.22: Fundamental structure of an unsupervised neural network *[135]*.

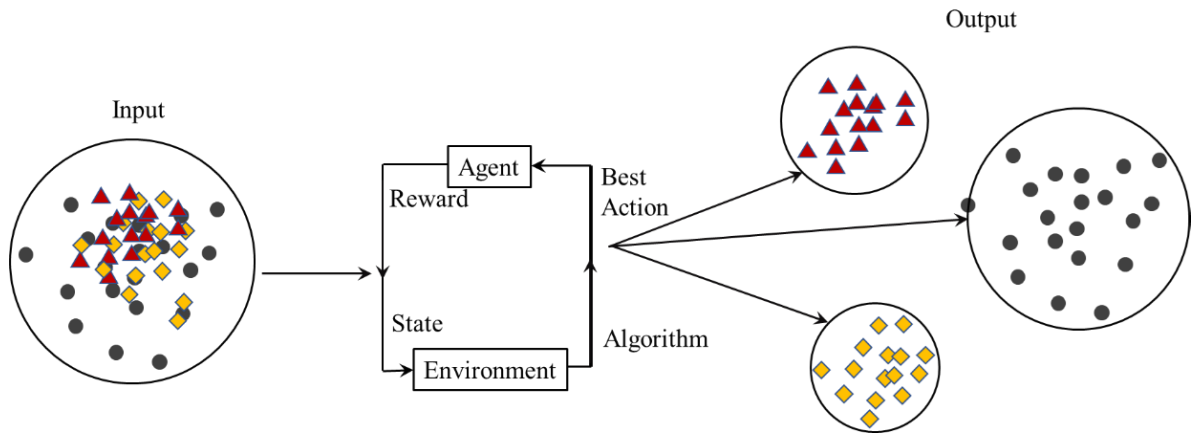

Figure 2.23: Metaphorical structure of the reinforced neural network [135].

### 2.6.8 Instance-Based Learning

The instance-based learning is a family of ML techniques that are uniquely designed to predict the outcomes of an event from a query based on both the chunk of data where the query is made and the data chunk nearest to that within the training set. Unlike decision trees and NN, the instance-based learning methods do not generate any abstraction from random unique instances. This helps to store all the data in the dataset. During a query, the instance-based technique makes use of the answer found by similar queries in all the nearest neighbors. Among the instance-based algorithms, the most popular one is the k-nearest neighbor.

### 2.6.8.1 K-Nearest Neighbor

A specific type of instance-based learning method where when a query is made, a k number of nearest data parts are also used to provide the query result is commonly known to be the k-NN algorithm. CNN is a very simple and supervised ML algorithm specifically well-suited for regression and classification tasks. However, one major problem of kNN is that it becomes very sluggish and slow to respond when provided with a large dataset; this is because, like other instance-based learning methods, it uses large data to make a decision.

### 2.7 Summary

This chapter describes the theoretical background and mathematical modeling related to important power system control strategies and controllers. Moreover, the concept of power system stability and types of stability are detailed in length. The well-known Monte Carlo optimization technique for stability analysis is outlined. The machine learning techniques and their conceptual works are described.

# CHAPTER III

# Methodology

## 3.1 Introduction

Investigation of power system stability comprises a variety of procedures and techniques. Concepts of control system engineering help to investigate real-life system stability margins and propose prospective behavior of the system in the presence of contingencies. Root-locus, pole-placement, or use of controllers could come handly to retain the system within its stability region. The recent development of data-driven investigation techniques, such as machine learning or artificial intelligence, has made it possible to monitor system status much faster and to dispatch control signals more efficiently. The choice of freedom to select certain types of stability controller or ML technique depend on the complexity of the system, the number of states, the computational burden, and the economy. In this research, the required dataset for the stability analysis, the damping ratio of the IEEE-14 bus under system fault, is extracted via the Python-Dymola interface. Different fault conditions were simulated in the Dymola platform, and the simulated data were extracted. The dataset is then utilized for training the neural network model to properly distinguish the unstable system from stable ones.

## 3.2 Modeling of AVR System with PID, LQR, and LQG Controllers

A simplified and linearized AVR system model with being considered in this section to observe the parameter-dependent response of PID, LQR, and LQG controllers. The step response in the time domain and the response in the frequency domain would be analyzed in this aspect. The linearized closed-loop model of the AVR system could be expressed as [32]:

$$G_{AVR}(s) = \frac{V_T(s)}{V_{ref}(s)} = \frac{K_A K_E K_G (1 + T_S s)}{(1 + T_A s)(1 + T_E s)(1 + T_G s)(1 + T_s s) + K_A K_E K_G K_S} \qquad (24)$$

Where the notations $K_A, K_E, K_G, K_S$, corresponds to the gain parameter and $T_A, T_E, T_G, T_S$, to the time constant of the amplifier, exciter, generator, and sensor sub-blocks, respectively. Usually, for an isolated system, the gain parameters and time constants are maintained within a standard range (given in Table 3.1) to preserve convenient operating conditions. Moreover, the value of the constant parameters considered herein in the model is also attached to the table.

The PID controller weights the controlled variable $u(t)$ by changing the P-I-D gains. The measured value of the variable is compared to the reference value to calculate the error. Then depending on the level of error and response characteristics of the system, the PID gains are adjusted, and the $u(t)$ value is updated. The PID controller is mathematically expressed through the following relationship [32]:

$$u(t) = K_p e(t) + K_i \int_0^T e(t)dt + K_d \frac{de(t)}{dt} \tag{25}$$

Where $K_p$ is the proportional gain, $K_i$ is the integral gain, and $K_d$ refers to the derivative gain of the PID controller. The LQR plant is described as follows:

$$\frac{dx(t)}{dt} = Ax(t) + Bu(t) \tag{26}$$

$$y(t) = Cx(t) + Du(t) \tag{27}$$

Where, $\frac{dx(t)}{dt}$ is the first derivative of the state vector, $x(t)$, with $x(t) \in R^n$, $u(t)$ is the input vector with $u(t) \in R^p$, and $y(t)$ is the output vector with $y(t) \in R^m$. The state matrix is represented by the $n \times n$ order matrix $A$, the input matrix of $n \times p$ order is $B$, $C$, and $D$ are output and feed-forward matrices with order $q \times n$ and $q \times p$, respectively. The feed-forward matrix becomes null when the system does not have a feed-forward path value. The quadrature objective function is a tool to maintain the system states to the stable range and helps to properly tune the PID gains; it is defined as [32]:

$$J = \int_0^\infty (x(t)Qx(t) + u^T(t)Ru(t))dt \tag{28}$$

In the above relation, the positive control weighing matrix and non-negative state weighing matrix are represented by the $R(R = R^T)$ and $Q$. The larger weight on the $R$ and $Q$ matrices produces a higher penalty on the control signal and the lowest distortion or change of the control variables. The control strategy also becomes non-economical. The algebraic Riccati equation (ARE) could be used to find the optimal minima of the cost function shown in Equation (5). It is a common practice to make a trade-off between the $Q$ and $R$ values to maintain system robustness while keeping the control expense and complexity within the tolerable limit. The state feedback method, $u(t) = -Kx(t)$, is used to help find the solution of the cost function ($J$), where $K$ refers to the controller gain and depends on the solutions ($P$) of the ARE, which are positive in nature. The solutions ($P$) could be enumerated from the following relationship [32]:

$$K = R^{-1}B^T P \tag{29}$$

$$A^T P + PA - PBR^{-1}B^T P + Q = 0 \tag{30}$$

In LQG, the LQR model is used with added Gaussian noises; this makes it harder to get system states at the summing nodes and requires to use of a Kalman state estimator. The state-space system model with added noise and the associated cost function is presented as [32]:

$$\frac{dx(t)}{dt} = Ax(t) + Bu(t) + D_1 w_{pn} \tag{31}$$

$$y(t) = Cx(t) + Du(t) + D_2 w_{mn} \tag{32}$$

$$J = \lim_{T \to +\infty} E \frac{1}{T} \int_0^T (x(t)Qx(t) + u^T(t)Ru(t))dt \qquad (330)$$

The variables A, B, C, D, $x(t)$, $u(t)$, and $y(t)$ are similar for both LQR and LQG models. The added system noise matrices are referred by $D_1$ and $D_2$. Moreover, the Gaussian process and measurement noises are denoted by $w_{pn}$ and $w_{mn}$, respectively. The expected value is $E$, and R is the non-negative state weighing matrix. The LQG controller could be modeled as [32]:

$$\dot{\hat{x}}(t) = A\hat{x}(t) + Bu(t) + K_k\big(y(t) - \hat{y}(t)\big) \qquad (341)$$
$$u(t) = -L\hat{x}(t) \qquad (352)$$

where, $\dot{\hat{x}}(t)$ and $x(t)$, $\hat{y}(t)$ and $y(t)$ represent the estimated and actual input vectors and output vectors, respectively. The primary task of the Kalman estimator is to pull out the system states from the physically non-measure noisy state. The Kalman estimator got observed the gain matrix, $L$, and Kalman gain, $K_k$. The value of the PID, LQR, and LQG gain parameters are then optimized for the considered linear AVR system model.

Table 3.1: The standard and considered time constant and gain parameter values for a linearized AVR model [32].

|  | Gain parameter | The time constant (s) | Value considered |
|---|---|---|---|
| Amplifier | $10 \leq K_A \leq 40$ | $0.02 \leq T_A \leq 0.1$ | $K_A = 10$, $T_A = 0.1$ s |
| Exciter | $1.0 \leq K_E \leq 10$ | $0.5 \leq T_E \leq 1.0$ | $K_E = 1.0$, $T_E = 0.5$ s |
| Generator | $0.7 \leq K_G \leq 1.0$ | $1.0 \leq T_G \leq 2.0$ | $K_G = 1.0$, $T_G = 1.0$ s |
| Sensor | $1.0 \leq K_S \leq 2.0$ | $0.001 \leq T_S \leq 0.06$ | $K_S = 1.0$, $T_S = 0.01$ s |

### 3.3 Data Generation for ML models

The dataset required for the power system stability analysis is created in the System Modelica – open-instance power system library (OpenIPSL) and Dymola environment, as proposed in current literature [96]. The IEEE 14 bus system is considered for data generation purposes. The IEEE-14 bus comes with five generator units, sixteen lines, four transformers between buses, and a total of twenty branches (Fig. 3.1) [96]. The detailed bus configurations for IEEE 14 bus system are provided in Table 3.2 [136]. The well-known open-source OpenIPSL simulation tool would be used for phasor time-domain contingency simulation of the IEEE bus. Instead of actually discarding any network part, which would change the system states (A matrix), the computational complexity would be reduced by making the line impedance very high (~$10^{12}$ magnitude); this would denote the contingency instead. Only the line branches and transformer branches are to be considered for contingency analysis; generator branches are not considered.

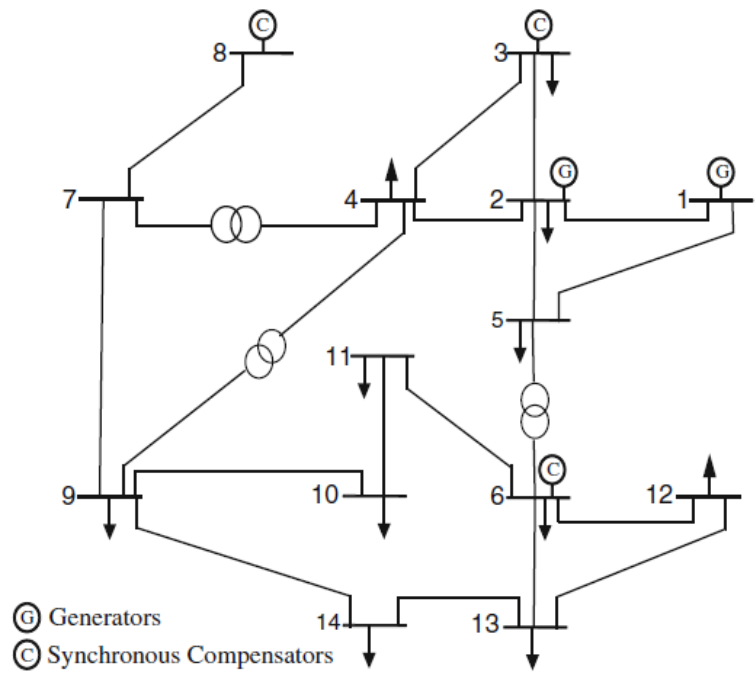

Figure 3.1: Single-line diagram of the IEEE-14 bus system [96].

Table 3.2: The IEEE-14 bus system terminal conditions in 100 MVA base [136].

| Bus number | Bus voltage, $V$ (kV) | Rotor Angle, $\delta$ (degree) | Real power, (p.u.) | Reactive power, (p.u.) |
|---|---|---|---|---|
| 1 (slack bus) | 146.28 | 0.0000 | 2.3239 | -0.1655 |
| 2 | 144.21 | -4.9826 | 0.4000 | 0.4356 |
| 3 | 139.38 | -12.7250 | 0.0000 | 0.2508 |
| 6 | 147.66 | -14.2209 | 0.0000 | 0.1273 |
| 8 | 150.42 | -13.3596 | 0.0000 | 0.1762 |

Table 3.3: The IEEE-14 bus system transmission line characteristics [136].

| Line | | Resistance, $R\left[\frac{pu}{m}\right]\times 10^{-7}$ | Reactance, $X\left[\frac{pu}{m}\right]\times 10^{-6}$ | Susceptance, $B\left[\frac{pu}{m}\right]\times 10^{-7}$ |
|---|---|---|---|---|
| From Bus | To Bus | | | |
| 1 | 2 | 1.94 | 0.59 | 5.28 |
| 1 | 5 | 5.40 | 2.23 | 4.92 |
| 2 | 3 | 4.70 | 1.98 | 4.38 |
| 2 | 4 | 5.81 | 1.76 | 3.40 |
| 2 | 5 | 5.70 | 1.74 | 3.46 |
| 3 | 4 | 6.70 | 1.71 | 1.28 |

| 4 | 5 | 1.34 | 0.42 | 0.01 |
|---|---|------|------|------|
| 6 | 11 | 9.50 | 1.99 | 0.01 |
| 6 | 12 | 12.3 | 2.56 | 0.01 |
| 6 | 13 | 6.62 | 1.30 | 0.01 |
| 7 | 8 | 0.01 | 1.76 | 0.01 |
| 7 | 9 | 0.01 | 1.10 | 0.01 |
| 9 | 10 | 3.18 | 0.85 | 0.01 |
| 9 | 14 | 12.7 | 2.70 | 0.01 |
| 10 | 11 | 8.21 | 1.92 | 0.01 |
| 12 | 13 | 22.1 | 2.00 | 0.01 |
| 13 | 14 | 17.1 | 3.48 | 0.01 |

If $n = 20$ is the total number of branches, and $k = 1$ is the number of simultaneous disconnections, then according to combinatorics, the total number of possible contingencies is [137]:

$$S_{n,k} = \frac{n!}{(n-k)! \, k!} \qquad (363)$$

But, in practice, the minimum and maximum number of simultaneous trippings could be,

$$k_{max} = n - 1 = 19 \qquad (37)$$
$$k_{min} = 1 \qquad (38)$$

Thus, the total number of possible scenarios is [137]:

$$T = \sum\nolimits_{k_{min}}^{k_{max}} S_{n,k} = 1{,}048{,}574 \qquad (396)$$

This is the measure that the traditional dataset considers for their analysis, which is quite large and places a computational burden. This research, however, would reduce the number using a two-stage Monte Carlo contingency selection method. Firstly, modified Poisson distribution is considered to select the number of contingency events (like line outages) that are the most likely to happen,

$$p(k) = \frac{1}{k! \sum_{k_{min}}^{k_{max}} \frac{1}{n!}} \approx 20{,}000 \qquad (407)$$

Secondly, with the likely contingency on hand, a combination of these faults is enumerated. This consists of a combination of branches with branch numbers that are open. Now, from this poll of contingency, random events are selected as our target system disturbance, for which the system stability issue would be analyzed using a neural network [138].

In the python Dymola data-generation interface, the selected scenarios are modeled by placing high transverse impedance to the associated branches [136]. The built-in linearization model in Dymola helps to run an analytical Jacobian of the designed power system model. The numerical process results in the system matrix, A. System matrix is then used to get the eigenvalues of the system. The eigenvalue is used to calculate the damping ratio of the system.

Linearized system equation:

$$\Delta \dot{x} = A\Delta x + B\Delta u \tag{418}$$
$$\Delta y = C\Delta x + D\Delta u \tag{42}$$

From system matrix A, the eigenvalues are calculated by:

$$\lambda_i = \{eig(A)\} \tag{430}$$

The damping Ratio becomes:

$$\zeta_i = \frac{-Re\{\lambda_i\}}{\sqrt{(Re\{\lambda_i\})^2 + (Im\{\lambda_i\})^2}} \tag{44}$$

The damping ratio is used as the discriminative scalar metric to identify the system stability: whether the system is stable/unstable/marginally stable at the present contingency, as shown in Fig. 3.2 [137]. The real and imaginary parts of the damping ratio also help to analyze the stability limit faster than other methods [137].

### 3.4 Data Processing for ML models

The eigenvalues obtained from the previous steps could be plotted in Fig. 3.3 to understand the initial eigenvalue distribution [137]. Sometimes there present outliers, such as the one indicated in the figure that spreads the system dynamics. Moreover, if there is more than a cluster of one imaginary component and there are spread near the stability limit, indicated in Fig. 3.3(a), the precision of the ML classifier suffers. Thus, the obtained data would have to be normalized. Similar to Fig. 3.3(b), a selected normalization could be incorporated that would pull every eigenvalue located outside of the unit circle to the unit circle's perimeter, keeping the angular displacement from the stability boundary unaltered. If data is not normalized, the network would process the data points spread across the input space with different biases (larger values would take a larger effect). This is feasible because the damping ratio of the system only depends on the angle of the eigenvalues with the imaginary axis.

After normalization, the search for any skewed dataset would follow. Since a very high volume of contingency scenarios would be entered into the dataset, there is the probability that some of the events' appearance could be very higher compared to the rest. This unbalanced eigenvalue dataset could result in a biased classifier, which is undesirable. The unbalanced category issues would be reduced if the same number of events are selected for each category. The libraries are imported to Python programming, and then the complex damping ratio dataset is converted to real and imaginary data, and then the data preprocessing follows. In [137], Monte-Carlo based optimization has been applied to IEEE-14 bus system and the reduced contigency scenarios are simulated in Modelica to prepare the eigenvalue dataset for small-signal stability analysis, which has directly been used for this investigation.

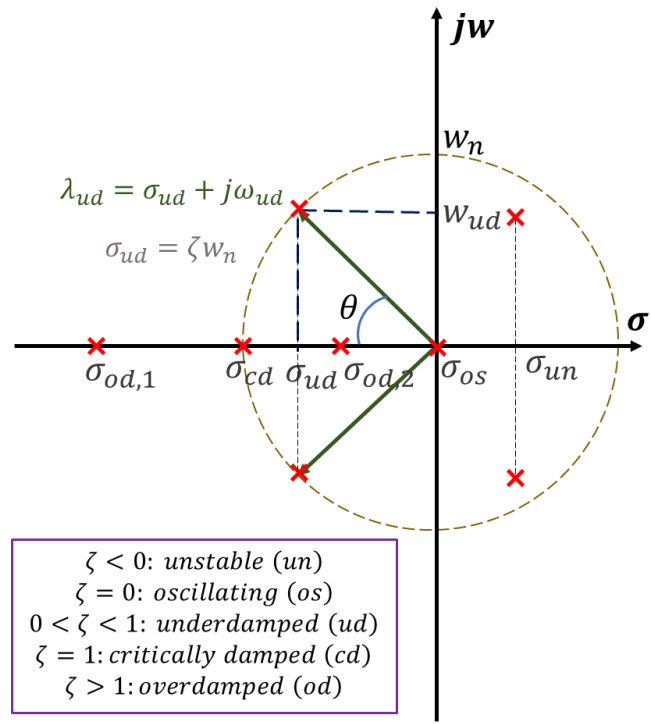

Figure 3.2: Stability profile from the plot of system damping ratio.

## 3.5 Preparing the ML Models

After the data is obtained from the Modelica through the python-dymola interface, the asymmetrical eigen data needs to be imported for ML classifiers. The following steps are to be executed before training of the ML models is started [137].

- Converting the categorical data to numerical values: there are various ways categorical values could be converted to numerical values; each way comes along with relevant trade-offs and impacts on the feature set. Two vastly used conversion methods are one-hot encoding and label encoding. In label encoding, the categorical values in a column are converted into normal in a simple manner. The shortcomings of the label encoder lie in the aspect that in this process, the produced numerical data sometimes seemed to form hierarchical orders in them and got misinterpreted by the algorithms. One way to address the hierarchical tendency is to use another method, one-hot encoding. In the one-hot encoding technique, the categorical values are first converted into a new column, having assigned 1 or 0 for the column as a notation for true and false.

- Compensating skewed dataset: In an ideal normal distribution curve, the data distribution to the left and right are symmetrical, and as a result, the value of asymmetry or skewness is zero. However, practical datasets are not symmetrical and got some skewness in them. The skewness could be divided in terms of right (positive) and left (negative). The asymmetry in the dataset causes the algorithms to be trained with rare and extreme values. As a result, in the case of a dataset of right or positive skewness, the algorithms would give better predictions with lower values than the higher values and vice versa. In statistical models, such as regression and classification, skewness directly affects the prediction accuracy and degrades the model performance. Thus, to ensure proper ML functioning, the skewed dataset needs to be transformed into a dataset with normal distribution. The most widely used method of transformation of skewed data set is the log transformation. However, the method or procedures to be applied to convert skewed data directly would depend on the characteristics of the data and the weight of the given data values.

- Initiating cost-function: Training and boosting the performance of an ML model is an iterative process. A cost (or loss) function is usually provided to measure the improvement in the model's performance. The cost function usually considers the predicted outcomes and actual values at any given iteration and provides the difference or error between them. The error or difference is then fed to the ML model, and the model tries to tune or adjust its parameters to reduce the error and improve the prediction. The process repeats till the error becomes closer to or at zero; at this stage, the ML stops learning. This technique of reaching the zero or closer to zero error is referred to as the search for local or global minima by moving towards the steepest descent till the cost function becomes null. The lost function refers to the error in the prediction of a single example data, whereas the cost function is the average of all the errors in the entire training dataset. Oftentimes, the cost function is derivated to produce the gradient descent value that helps to measure the steps with which the model searches for the local or global minima.

- Splitting data into training/testing and executing ML algorithms: The processed eigenvalue dataset would be imported into the Python library. In this research, the *scikit-learn* and *tensor-flow* would be used for ML classifier design. The dataset would be divided to train and test data with 75% and 25% ratios. After that, reduced training or testing dataset is obtained for the ML model. Traditional ML classifiers such as logistic regression, support vector machines, k-nearest neighbors, decision trees, and naïve Bayes would also be designed for comparative analysis. After that, a total of five layer-based neural networks (node pattern: 2×100×100×100×5) would be considered to improve stability classification accuracy. The accuracy, time of prediction, precision, and overall score of all the modeled classifiers would be compared to outline the actual accomplishment of the proposed model. The amount of misclassification and its impact on the stability analysis would also be performed.

The traditional and neural network models for stability analysis are to be studied in the next section.

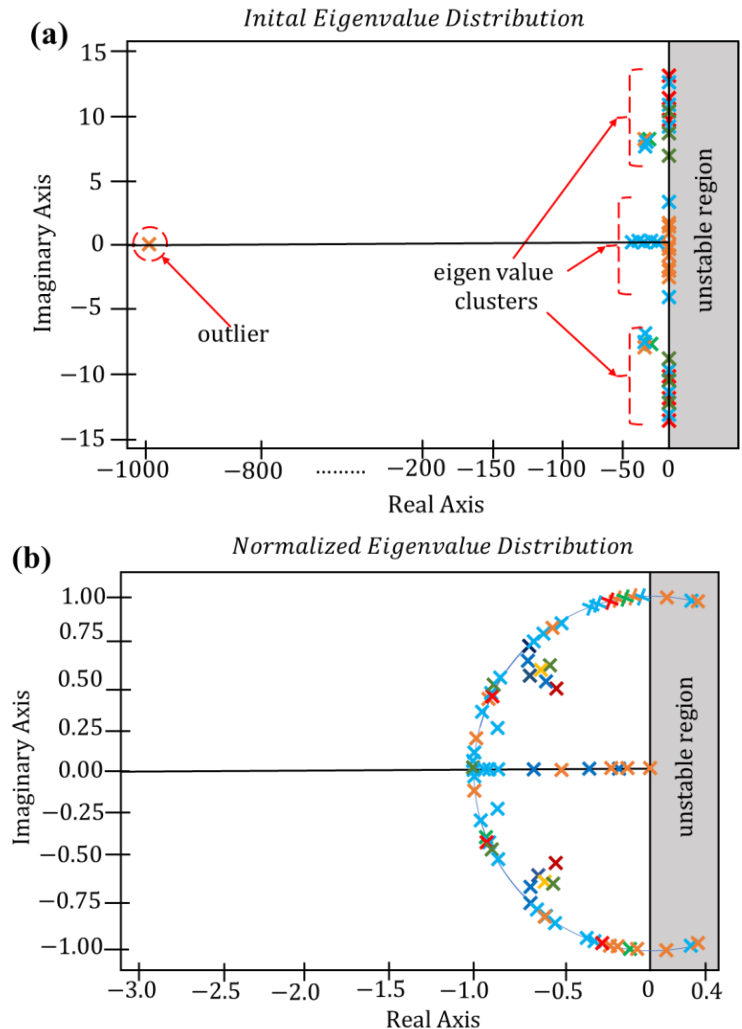

Figure 3.3: Eigenvalue distribution, (a) initial eigenvalues, and (b) normalized eigenvalues.

### 3.5.1 Logistic Regression

Logistic or nominal regression is used in binomial or multinomial models as a statistical technique for classifying categorical data. In this regression technique, a set of equations are formed that provides the probability of an event based on the input data. For each output probability category, an equation is formed that relates all the relevant input field categories. After the model is generated, it could be used to predict an outcome when new data is provided. The highest probability category during the prediction process is selected as the resultant category by the process. Due to the probabilistic nature of outcomes, the dependent variable is bounded in binary 0 (false) and 1 (true). Logistic regression is defined by logit transformation of the odds, also known as log odds or natural algorithm of odds. Accordingly, the probability is measured by dividing the probability of success by the probability of failure and expressed as [139]:

$$Logit(P_i) = \frac{1}{1 + e^{-P_i}} \tag{45}$$

$$\ln\left(\frac{P_i}{1 - P_i}\right) = \beta_0 + \beta_1 X_1 + \cdots + \beta_k X_k \tag{46}$$

In the above relationship, the independent variable $X$ is related to the response or dependent variable $logit(P_i)$ through the $\beta$ coefficient, which is calculated by maximum likelihood estimation. The log-likelihood function is used for the value of beta to get tuned and adjusted in each iteration to get the best fit of the log odds. The number of beta coefficients is one larger than the number of independent variables. Once the optimized coefficients are found, the conditional probability for each sample is calculated and logged. The resultant values for all the observations are summed together to yield a predicted probability. Types of classification play an important role. For instance, for binary classification, when the probability hits lower than the middle between 0 and 1, the outcome is forced to 0, and when the probability is larger than 0.5, the result is pushed to 1. It is a good practice to evaluate the model performance and score once the model is properly developed. The most popular assessment tool for logistic regression model fit is the Hosmer-Lemeshow test. The cost function for logistic regression is often coined as the log loss since it simply provides the negative average of the log of properly predicted probability. The cost function could be expressed for logistic regression as follow [139]:

$$J_{LR} = -\frac{1}{N}\sum_{i=1}^{N}(y_i\log\left(p(y_i)\right) + (1 - y_i)\log\left(1 - p(y_i)\right) \tag{47}$$

### 3.5.2 Linear Support Vector Machines

SVM is one of the widely used ML algorithms for regression and classification. It started as the most robust algorithm out there with the help of AT&T Bell labs and Vapnik. SVM could increase the margin of separation between two unique classes and helps in the operation of nonlinear functions by exploiting parallel kernels [37]. SVM creates an optimal hyperplane of the input values and classifies the data with an optimal margin of separation that leads to lower error and minimizes the empirical risk. The structural risk minimization principle is used in SVM to reduce the generalization error [37]. The empirical risk refers to the average loss of estimation while it deals with a finite data chunk. The SVM performance in defining the boundaries between the clusters is then increased by dispatching an empirical risk minimizer function [37]. Primarily SVM could be classified into two types:

- Linear SVM: The linear SVM creates a line of margin for linearly separable data of two defined classes. The classification is then assessed in terms of data in a false position.
- Non-linear SVM: When the dataset contains data that are non-linear in nature, then a non-linear type of SVM is used. In this case, instead of a single straight line, a hyperbolic plane is defined with a broader margin to properly cluster the non-linear data.

In order to understand the SVM operation, first, consider a given train dataset, $\{(x1, y1), (x1, y2) \dots (xi, yi)\}; x_i \in R^n; y_i \in R$, where $(x, y)$ is the input and output pair for $n$ number of features. SVM target to map a plane, $f(x): R^n \to R \, f(x)$, which fits the nearest data values and provides a margin between the groups of data clusters. The plane is formulated as [140]:

$$f(x) = W^T \cdot X + b \qquad (48)$$

Where $b$ and $W$ refer to the bias and coefficient of the input-output relation. The function, $f(x)$ provides the margin path which requires the highest magnitude of deviations, σ, from the nearest target $y_i$. The transformation function, $\phi: R^n \to R^s \, (s > n)$, is applied to all the input data points to handle the nonlinear relationship between the input and output; however, hyperplanes that result in deviation smaller than σ are ignored. Thus the input space is mapped to formulate the linear feature space through the following transformation [140]:

$$f(s) = W^T \cdot \phi(X) + b \qquad (49)$$

The transformation also helps in reducing the dimensionality of the input vectors. SVM solutions follow the typical convex optimization procedures. At the end of the computational step, each training point is assigned a pair of Lagrange multipliers (LP). Moreover, when the SVM is trained, and the optimal plane is obtained, the support vectors are considered to be points outside of the outlier σ band and characterized by having at least one zero LP [38].

### 3.5.3 K-Nearest Neighbors

The nearest neighbor principle of classification considers a predefined number of sample points nearest to the point of interest for prediction. The choice of the number of samples could be user-defined; for instance, the number of samples to be considered is k amount in k-nearest neighbor, or the number of samples is based on the location density, like the samples within a defined radius as in radius-based neighbor learning. The distance between the samples could be expressed in a general metric such as standard Euclidean distance. The majority of the training data are analyzed in neighbor-based methods, which makes them slower to predict. However, the nearest-neighbor method could result in very accurate classification and regression of practical applications, especially with irregularly defined decision boundaries, like recognizing handwritten digits and analyzing satellite image scenes. Unlike the other classification techniques, such as support vector machine or logistic regression, the k-NN does not have a loss function that needs to be minimized during training.

In the nearest neighbor methods, the data points that are closer to the query point in consideration are found by first calculating the mutual distance between the two. This distance also helps in defining the decision boundaries and separates the query points into different regions. The commonly used methods of distance calculation include euclidian, manhattan (for continuous), Minkowski, and hamming distance (for categorical). The distance metrics are described below:

- Euclidean distance ($p = 2$): The most common place distance metric for real vectors is the euclidean distance. In this, the distance between the query point $y_i$ and the closest point $x_i$ is simply the square root of the differences and is expressed as [141]:

$$d(x, y) = \sqrt{\sum_{i=1}^{k} (y_i - x_i)^2} \tag{50}$$

- Manhattan distance ($p = 1$): In this metric, the absolute value between the query point and the closest neighbor is calculated. The Manhattan distance metric is also referred to as the taxicab distance or city block distance as it is best to be visualized in a grid to navigate from one address to another. The metric is expressed by [141]:

$$d(x, y) = \sum_{i=1}^{k} |y_i - x_i| \tag{51}$$

- Minkowski distance: The Euclidean and Manhattan distance metrics are exploited to formulate Minkowski distance. In this, a new parameter $p$ is defined. The value of $p$ engenders other distance metrics. For instance, $p = 1$ is for manhattan distance, and $p = 2$ for the euclidean metric. The generalized formula looks like this [141]:

$$d(x, y) = \sum_{i=1}^{k} (y_i - x_i)^{\frac{1}{p}} \tag{52}$$

- Hamming distance: When the data vectors are of Boolean or string type, the identification of points becomes difficult since the vector is usually mismatched. The hamming distance metric is used in such situations as an overlap metric and is expressed as [141]:

$$d_H(x, y) = \begin{cases} 1, & x = y \\ 0, & x \neq y \end{cases} \tag{3.30}$$

### 3.5.4 Decision Trees

Decision trees are a supervised machine learning algorithm used for regression and classification purposes. This non-parametric model predicts the outcome of a target value based on a decision query which is dispatched one after one till the required value is found. The decision rules are formed from the features of the data on hand, and the structure follows a flowchart-like tree. The internal nodes and branches of the tree denote the test attribute and outcome of a test, respectively. Moreover, the terminal node refers to the class label, and the whole process mimics a piecewise constant approximation procedure. First, the trees employ a divide-and-conquer strategy by dispatching a greedy search to get the optimal split points. In a top-down recursive manner, the process is repeated till all the major labels are classified under specific labels. The classification process directly depends on the data size and the complexity of the tree. The splitting stops when it renders no future value to the predictions. Since the process is non-parametric, the decision tree is appropriate for knowledge search purposes and can handle high-dimensional data with good accuracy.

Information gain and Gini impurity are two of the most popular splitting criteria to select the best possible attribute for each node. They evaluate the quality of the test condition and predict the performance of the test condition to classify samples. Gain and entropy are two essential concepts to be considered. From the information theory context, the inherent impurity of the sample values is expressed through entropy and could be expressed as [142]:

$$Entropy(S) = -\sum_c p(c) \log_2 p(c)$$ (3.31)

Where $S$, $c$, and $p(c)$ refer to the data set for entropy calculation, classes in the data set, and the proportion of data points in class $c$ to all points in set $S$. The typical entropy value remains between 0 and 1. When the samples in $S$ are of the same class, the entropy or the impurity of the sample is null. When the total samples in $S$ are of two classes of similar size, then the entropy reaches the highest value of 1. Before splitting the dataset, it is required to find the best feature to split. This could happen when the dataset with the smallest entropy is used to split and find the optimal decision tree.

Information gain is fundamentally the mismatch or change in entropy of a dataset before and after a split on a given feature. The best split gives the highest information gain, and the classification of the training data provides lower error or mismatch. Information gain is directly related to entropy and could be expressed through the following formula [142]:

$$Gain(S, a) = Entropy(S) - \sum_v \left| \frac{S_v}{S} \right| Entropy(S_v)$$ (53)

Where, $a$, $Entropy(S)$, and $Entropy(S_v)$ refers to a specific attribute, the entropy of dataset $S$, and the entropy of dataset $S_v$, respectively. The cost function for the decision tree algorithm for data size $m$ could be provided through the mean squared error (MSE) [142]:

$$J_{DT} = MSE = \frac{1}{2m} \sum_{i=1}^{m} (\hat{y} - y)^2$$ (3.33)

### 3.5.5 Naïve-Bayes

Naïve Bayes works on simple probabilistic terms in that a test point is classified based on the probability of the point being of that class rather than what is defined as the label of the test point. It is one of the most common supervised ML algorithms and works on the Bays theorem. The underlying working process follows very basic Bayesian network models, and only after the kernel density estimation strategy is embedded does it hits higher accuracy. Unlike others, the Naïve Bayes is only suited for classification strategy. While predicting the likelihood of a test point being of a certain class, this algorithm makes assumptions that are virtually not feasible in practical data; that is the reason for being naïve. The conditional probability is exploited to measure the likelihood of individual probability. In conditional probability, the presence or absence of a specific feature is independent of the presence or absence of any other feature.

According to Bayes' theorem, when any given class variable $y$ is independent of the feature vector $x_1$ through $x_n$, the probability becomes [143]:

$$P(y \mid x, \dots, x_n) = \frac{P(y)P(x_1, \dots, x_n \mid y)}{P(x_1, \dots, x_n)} \tag{54}$$

When the following naïve conditional probability is considered, Bayes' theorem is tuned to the model required for the Naïve Bayes algorithm [143]:

$$P(x_i \mid y, \dots, x_{i-1}, x_{i+1}, \dots, x_n) = P(x_i \mid y) \tag{55}$$

Thus, the relation could be simplified to the following, given that $P(y)$ and $P(x_i \mid y)$ are the class probability and conditional probability, respectively, as [143]:

$$P(y \mid x, \dots, x_n) = \frac{P(y) \coprod_{i=1}^{n} P(x_i \mid y)}{P(x_1, \dots, x_n)} \tag{56}$$

In the above relationship, the numerator and the denominator respectfully refer to the Bayes numerator and Bayes denominator. The calculation requires the result of the prior class probabilities, which usually is directly estimated through the joint inverse conditional attribute value probability of the data on a given class. In the Gaussian Naïve Bayes relation, the continuous values related to each feature are perceived to have a Gaussian distribution or normal distribution across the space. In this method of classification, the likelihood of a feature being of a class is assumed to be Gaussian, and the conditional probability is expressed through the following relation [143]:

$$P(x_i \mid y) = \frac{1}{\sqrt{2\pi\sigma_y^2}} \exp\left(-\frac{\left(x_i - \mu_y\right)^2}{2\sigma_y^2}\right) \tag{57}$$

Where the maximum likelihood is dispatched to attain the value of the terms $\sigma_y$ and $\mu_y$. The Naïve Bayes classifier model does not require optimizing any cost function.

### 3.5.6 Neural Network

Neural networks, or ANN, are a subset of ML algorithms and are widely viewed as the heat of deep learning algorithms. The NN mimics and is inspired by the biological neurons' signal transmission from one another in the human brain. Each ANN is comprised of two input-output node layers and one or more hidden intermittent layers. Each node is provided with weight and threshold. An output at any individual node is higher than the predefined threshold that nodes get activated and forward the data to the next layer of the ANN. When the output is lower than the threshold, no passing along to the next layer takes place. NN are continuously trained to recognize patterns more accurately and to predict with lower errors. The more diverse the data fed to the NN are, the higher and more accurate prediction could be observed. When the algorithm reaches maturity and is fined-tuned, it can efficiently handle any extrinsic or unknown data, and this makes ANN a powerful tool for building AI and ASI.

Considering the individual isolated node having a linear regression model of the input vector, output vector, weights, and bias, the expression to sum and output prediction becomes [144]:

$$\sum_{i=1}^{m} w_i x_i + bias = w_1 x_1 + w_2 x_2 + w_3 x_3 + bias \qquad (3.38)$$

$$output = f(x) = \begin{cases} 1, \sum w_1 x_1 + b \geq 0 \\ 0, \sum w_1 x_1 + b < 0 \end{cases} \qquad (3.39)$$

Weights are assigned after the input layer is defined, and they are tuned to improve the prediction accuracy in the training stage. The individual inputs are each multiplied with an associated weight and summed together to get the complete model of all data for output estimation. The output is compared to a threshold; if it is higher than the threshold, it activates the next node. In this way, the output of the previous node is fed as the input of the next node, and the process is defined as a feedforward network.

After each recursion, the neural network cross-validates the prediction and tries to minimize a cost function to fit the observation. It then adjusts the bias and weights accordingly. In some instances, this reinforcement learning helps boost the model performance and to reach local minima of convergence. In gradient descent search, the model proceeds to reduce errors or cost function and obtains minima. The cost function for the neural network could be expressed by MSE as [144]:

$$J_{NN} = MSE = \frac{1}{2m} \sum_{i=1}^{m} (\hat{y} - y)^2 \qquad (58)$$

In some cases, the feedforward direction from the input to the output layer of a given NN could be changed to create a path for backpropagation from the output layer directly to the input layer.

Backpropagation reduces the estimation time and errors associated with each neuron in the NN and helps to better tune the model parameters.

The number neurons in each layer could be considered through the following criterion [144]:
- The number of neuron in the input layer is equal to the number of features column.
- The number of neuron in hidden layers could be calculated using the following relation: for one hidden layer only,

$$N_h = \frac{Ns}{\alpha \times (N_i + N_o) \times 10},$$
(59)

where,

$N_h$ = number of neurons in hidden layers,

$N_s$ = number of samples in training set,

$N_i$ = number of neurons in input layer,

$N_o$ = number of neurons in output layer, and

$\alpha$ = design factor, usually between 2 to 10.

- The number of neurons in output layer, for softmax activation, is equal to the number of class label.

### 3.6 System Flow Diagram

The concepts of power system control are first used to investigate the time-domain and frequency-domain response of the AVR model's gain parameters and time constants in the presence of external controller sets such as PID, LQR, and LQG. After that, the research shifts to modeling and assessing a neural network-based ML model to recognize power system stability, more precisely, the stable system poles from the unstable ones. The dataset required for the ML model is generated in Dymola and System Modelica. The Poisson's probability function is used to filter out the most probable contingency combination, and the Monte Carlo optimization technique is used to model linearization faster with a lesser computational burden. The generated dataset is then fed to the ML models with various proposed and traditional algorithms. A scoring system is used to properly understand the worth of the proposed neural network compared to other methods.

In summary, the overall stability analysis procedure can be visualized from the flow diagram shown in Fig. 3.4.

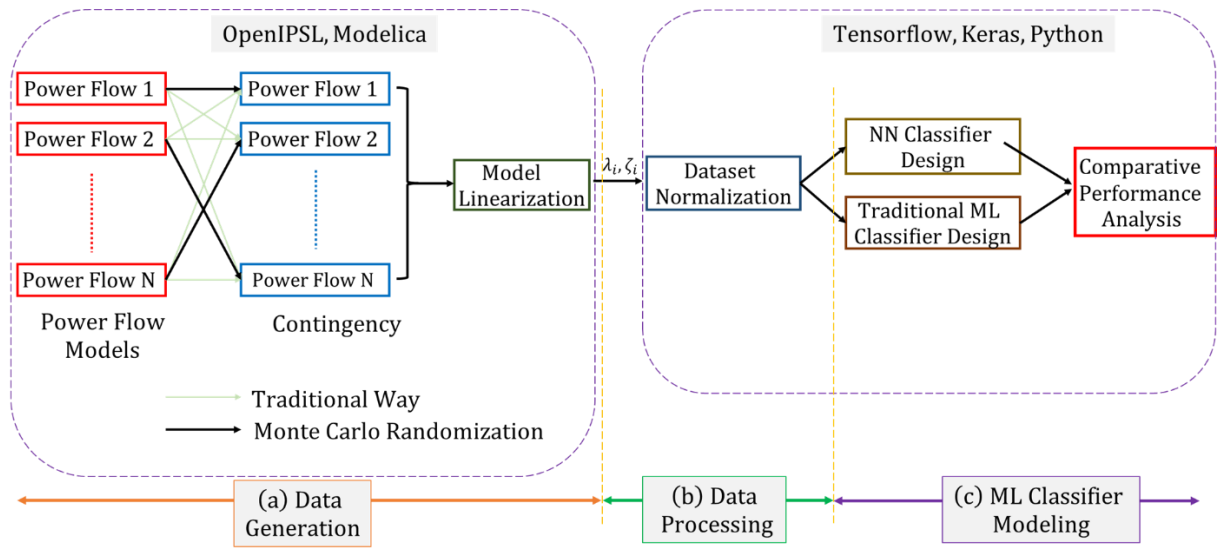

Figure 3.4: Flow diagram of the proposed research method.

## 3.7 Simulation Tools

This investigation is totally simulation-focused. The investigation of the AVR system, modeling, controller parameter optimization, and analysis of step responses are performed in the Matlab-Simulink software package and uses a control box toolbox. For data generation purposes, the System Modelica (OpenIPSL) and Dymola software were used. Both environment vastly uses the Modelica programming language. The Anaconda environment is used for Python programming, and Sci-kit learn with the tensor-flow framework is used for ML model training.

- **OpenIPSL library:** OpenIPSL is a power system modeling library forked from the iTesla power system library and maintained by the SmarTS Lab research group. This is an interesting library because it holds a set of power system components to develop power networks, phasor modeling, and dispatching time-domain simulation. As a result, additional computational procedures such as linearizing system eigen value and eigen analysis could be performed. The OpenIPSL works on finding the solution of load flow equations for the power system model, where the initial values of the parameters are computed from the model and/or could be provided into the initial equation. The library has been cross-referenced with the performance of the prevalent power system simulation tools such as PSAT and PSS/E and observed to be better performing and scalable with a few uniquely defined work procedures for power network simulation.

- **Dymola:** Dymola or System Dymola is a commercial modeling and simulation platform that extensively uses Modelica language for system development and simulation. Dymola allows the building of complex and large systems composed of numerous components, models, and variables. The mathematical modeling is primarily programming based and sufficiently encircles the dynamic system behavior under various external conditions. Dymola platform allows multi-physics and multi-engineering models consisting of components from diverse engineering domains, such as structural, mechanical, and electrical engineering. The inter-connected subsystem is expressed in a mathematical relationship, and each additional connection engenders an additional equation. The process could be made automatic if the GUI is used instead of the programming, where the drag and drop of branches automatically creates the required programming sections. When the entire system with all the possible components is provided, Dymola starts processing the entire system behavior. Alternatively, sub-systems could be simulated for testing and troubleshooting purpose. The domain-specific knowledge is usually attached to the Modelica libraries, which provide an extensive toolbox of electrical, mechanical, control, thermal, thermodynamics, and many more commercial libraries with standard IEEE system layouts. Moreover, dymola facilitates encryption of the system and hides crucial information from the model section. These features made the feasibility of using Modelica from an application like aerospace modeling to electrical system analysis.

- **Anaconda IDE:** Anaconda is an integrated development environment specifically designed for scientific computation in Python and R programming. The application includes but is not limited to data science, ML applications, signal and image processing, and statistical modeling and computation of big data. Anaconda is embedded with a simplified package manager that makes it easy to import the python toolbox and deploy. The IDE helps to use data-science packages across various operating systems.

- **Matlab/Simulink:** Matlab is computer programming used for scientific programming, modeling, and simulation. Simulink is the graphical programming environment designed for Matlab. The Simulink interface makes use of a graphical block, diverse built-in toolbox, and customizable block libraries. The Simulink could be worked with integration to the core Matlab script development interface or could be used independently. Simulink is widely used in engineering processes, such as automatic control, digital image/signal procession, power system analysis, multi-domain simulation, and model-based designing.

## 3.8 Summary

This chapter details the methods used for power system stability analysis using neural networks. The process is broken down into data generation, data preprocessing, and data training sections. Data is generated in the Dymola simulation tool for IEEE 14 bus. The preprocessing and machine learning algorithms are modeled within the Scikit learn framework using the Python programming language. The comparative analysis of the stability analysis using neural networks and other algorithms, such as k-NN, regression, etc., are also performed. The proposed method hits the highest score compared to the other traditional methods in precise stable/unstable system pole identification.

# CHAPTER IV

# Results and Discussion

## 4.1 Introduction

This thesis focuses on the control of traditional power system voltage regulators and investigates the classification efficiency of ML algorithms in terms of stable, marginally stable, critically stable, and unstable system states. Among the diverse ML algorithms, k-NN, Naive Baise, random forest, and 4-stage neural networks have been considered. This chapter of the thesis report would primarily focus on the result obtained in terms of (i) power system AVR model control with PID, LQR, and LQG controllers, (ii) use of the System Modelica tool to generate system states for an IEEE-14 bus system, and (iii) utilization of machine learning models to properly classify system stability. Moreover, a comparative analysis of the ML algorithms is attached to justify the usefulness of the proposed neural network models compared to traditional ML models.

## 4.2 Power System Stability with Close-loop Controller

### 4.2.1 Performance of an AVR without a controller

The characteristics of an AVR system in the absence of any controller are investigated. The basic linearized AVR plant consists of the generator, sensor, amplifier, and exciter subsystems. Each of the four components is presented with the associated transfer function, where the gain parameter and the time constant values are used. Table 4.1 represents the time constant and gain parameter values which are randomly considered to inspect the static performance of the AVR model with and without close-loop controllers when the subsystem parameters are pushed to the upper limits. The static After modeling in Matlab Simulink, a unit step function is fed to the system, and the observed time domain characteristics are measured. The step response showed that the response shows a very oscillatory nature with a large settling time. The root locus plot of the AVR system is shown in Fig. 4.1(a) [32]. The presence of pure imaginary complex conjugate pole pair is responsible for showing underdamped oscillation. Moreover, both the negative poles stay very closer to the imaginary axis, making the system stability margin narrower. In Fig. 4.1(b), the bode plot is attached to further investigate the reason behind the poor response [32]. This frequency domain analysis provides that the phase crossover frequency (PCF) for the AVR system is small than the gain crossover frequency (GCF); the measured two frequencies are 4.4050 rads⁻¹ and 6.1238 rads⁻¹, respectively. Furthermore, the gain margin (GM) and phase margin (PM) observed is 1.9250 dB and 18.593º, respectively, which is quite smaller. The GM and PM are derived from the following equations [32]:

$$GM = 20log\left(\frac{1}{M_{pc}}\right)dB = -20log\ Mpc\ dB \tag{60}$$

$$PM = 180° + \varphi_{pc} \tag{61}$$

In the above equations, $M_{pc}$ and $\varphi_{Pc}$ refers to the gain magnitude at the PCF and phase angle at GCF. The performance of an AVR system could be improved by embedding the closed-loop controller dynamics into the system, as is provided in the next section.

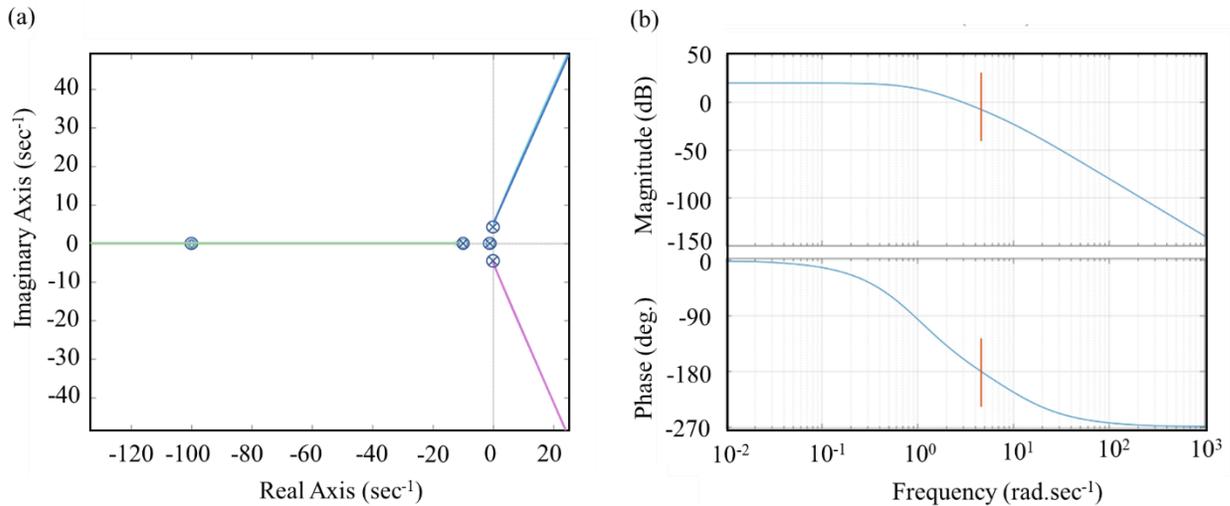

Figure 4.1: (a) Root locus plot and (b) Bode plot of a linearized AVR system in the absence of any closed-loop controller [32].

### 4.2.2 Comparative Analysis of the Controllers

The salient oscillatory nature of the AVR system could be improved by using a particularly defined closed-loop controller. In this part, PID, LQR, and LQG controllers are to be embedded in the basic AVR system model, and comparative performance analysis between these three would be driven. Moreover, the effect of the system time constant and gain parameters on the step response would be shown. The observed closed-loop poles with the three controller arrangements are attached in Table 4.2. In the presence of the controller, the closed loop poles have been pushed towards a further negative region, making a large separation between their position and the imaginary axis. The more negative the closed loop poles are, the more stability is shown by the system. Accordingly, the most stable controller system is obtained for the LQG, followed by the LQR and then PID. Moreover, for the LQR and LQG, the number of closed-loop poles reduces.

The step response observed for the AVR with PID, LQR, and LQG dynamics is shown in Fig. 4.2 [32]. The PID controller is the simplest of the three. With PID, the system needs white lengthier time to settle down. The step characteristics are better with the LQR and best with the LQG controllers. Table 4.3 summarizes the important step characteristics parameters, rise time, peak time, settling time, and steady-state error, for the three controllers under consideration.

Table 4.1: The standard and considered time constant and gain parameter values for a linearized AVR model [32].

|  | Gain parameter | The time constant (s) | Value considered |
|---|---|---|---|
| Amplifier | $10 \leq K_A \leq 40$ | $0.02 \leq T_A \leq 0.1$ | $K_A = 10$, $T_A = 0.1$ s |
| Exciter | $1.0 \leq K_E \leq 10$ | $0.5 \leq T_E \leq 1.0$ | $K_E = 1.0$, $T_E = 0.5$ s |
| Generator | $0.7 \leq K_G \leq 1.0$ | $1.0 \leq T_G \leq 2.0$ | $K_G = 1.0$, $T_G = 1.0$ s |
| Sensor | $1.0 \leq K_S \leq 2.0$ | $0.001 \leq T_S \leq 0.06$ | $K_S = 1.0$, $T_S = 0.01$ s |

Table 4.2: The observed closed-loop poles of an AVR system with PID, LQR, and LQG controllers.

| Without controller | PID controller | LQR controller | LQG controller |
|---|---|---|---|
| -99.9 + 0.0i | -0.5 + 0.0i | -99.9 + 0.0i | -59.5 + 102.8i |
| -12.4 + 0.0i | -100.1 + 0.0i | -19.5 + 0.0i | -59.5 − 102.8i |
| -10.0 + 0.0i | -10.0 + 0.0i | -8.62 +14.2i | -118.9 + 0.0i |
| -0.52 + 4.7i | -4.59+ 4.95i | -8.62 -14.2i | -100.1 + 0.0i |
| -0.52 - 4.7i | -4.59 − 4.95i | -10.0 + 0.00i | -10.0 + 0.0i |
| -2.5+ 0.0i | -2.71 + 0.0i | -2.5 + 0.00i | -2.50 + 0.0i |
| -1.0 + 0.0i | -2.50 + 0.0i | -1.0 + 0.00i | -1.00 + 0.0i |
|  | -1.09 + 0.0i |  |  |
|  | -1.00 + 0.0i |  |  |

Among the three, the observed smallest steady-state error (∼ 0.0000064%) results for the LQG controller, and the highest is for PID (∼ 0.015%). The percent overshoot is also the highest for the PID controller than LQR and LQG. The rise time is in a closer range for the PID and LQR. The rise time for the LQG is shorter than the other two. During the investigation, we dispatched a trial and error based on manual gain parameters tuning without any additional algorithm. The optimization usually takes place by minimizing a cost function. Thus the use of advanced controller gain optimization techniques, such as particle swarm optimization, kidney-search algorithms, etc., could boost the controller performance in dynamic system conditions. It is to be noted that the control weighing matrix is provided a large value to get the fast response for the LQG controller.

Next, the comparative performance of the PID, LQR, and LQG controllers is carried out by changing the time constants of the AVR subsystem blocks. During the process, the controller gains are kept unaltered. The time constant of the exciter, sensor, generator, and amplifier is set to the highest steady-state limit (as shown in Table 4.1), and then step responses for the three controllers are carried out (Fig. 4.3) [32]. The step response observed with the PID controller suffers greater with the change in the time constant, and the system becomes unstable with the PID controller. The instability appears due to the shift of the closed loop poles to the right half side of the complex s-plane. One way to solve such a misfit is to dynamically optimize the PID controller parameters. This could be achieved by dispatching well-known ZN methods. Moreover, heuristical or meta-heuristical optimization algorithms could also be used to update the PID to gain a dynamic system with or without external parameters such as noise or disturbances through a closed-loop recursive computation.

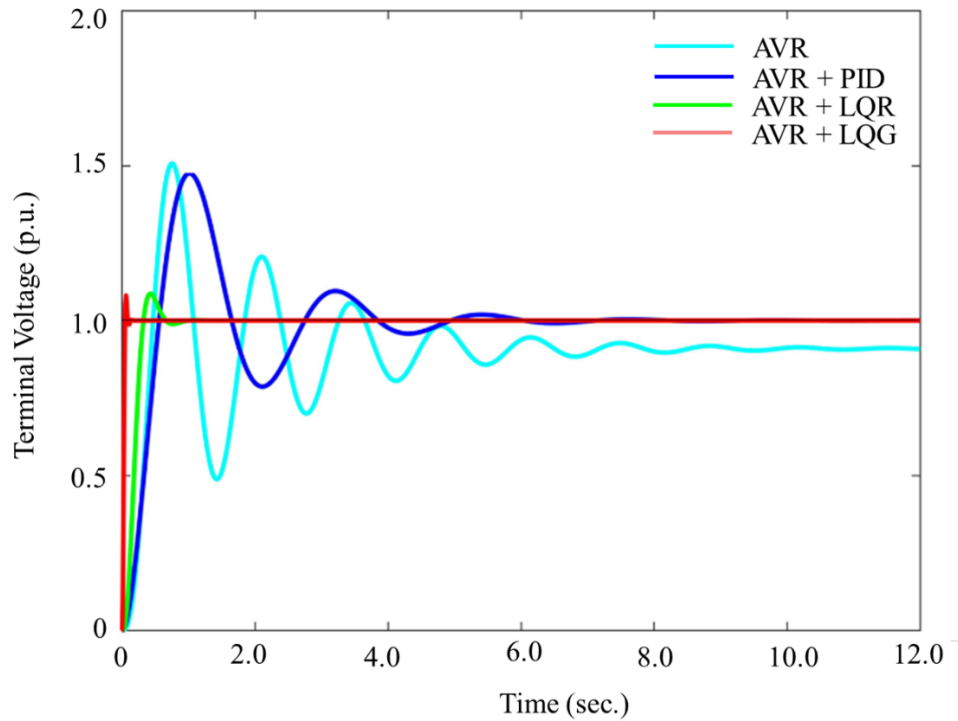

Figure 4.2: Observed step-response of a linearized AVR system with closed-loop PID, LQR, and LQG controllers [32].

The most important block of the AVR that is very much susceptible to environmental noise and disturbances is the sensor block. In this section, the sensor feedback gain is increased, and the performance of the AVR system terminal voltage control with the PID, LQR, and LQG controllers is observed. The step response provided in Fig. 4.4 represents that increased sensor gain drastically affects the PID controller performance with settling time, oscillation, and steady-state error, all reaching very bad values [32]. The LQR and LQG controllers, however, are only slightly affected by the change in the sensor gain. Thus, LQR and LQG controllers are better suited to handle the sudden change in the sensor gain due to disturbance, but PID controller requires extensive gain optimization to handle the situation. Finally, in Fig. 4.5, the step performance of the AVR system with is shown for the three controllers when the amplifier gain is increased in a step of 10, from 10 to 40 [32]. It is found that higher amplifier gain results in larger error magnification of the system, and this results in improved performance by the controllers to handle the error. For all three controllers, lower amplifier gain results in the lowest performance, higher steady-state error and settling time, overshoot, and rise time.

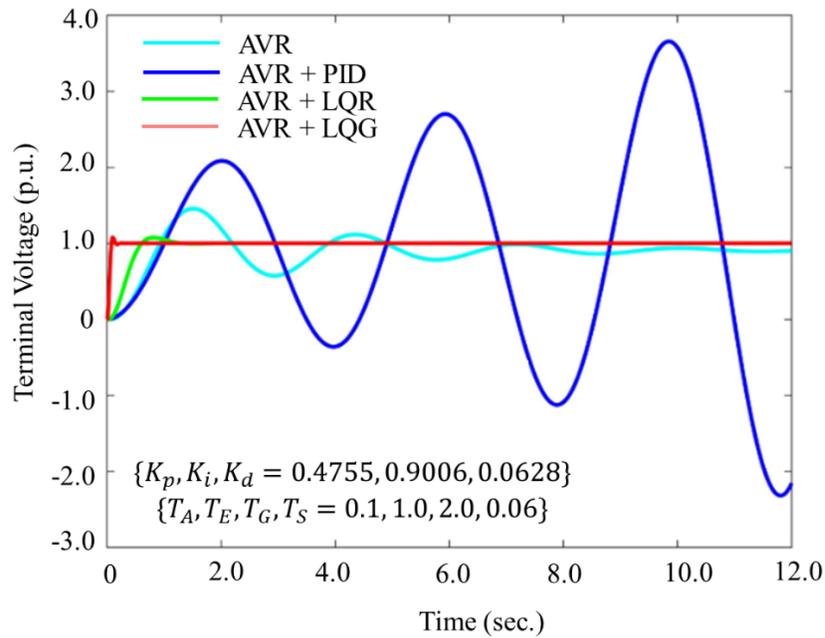

Figure 4.3: The observed step response of a linearized AVR model with the subsystem time constant set to a maximum steady-state value with PID, LQR, and LQG controllers gain kept static [32].

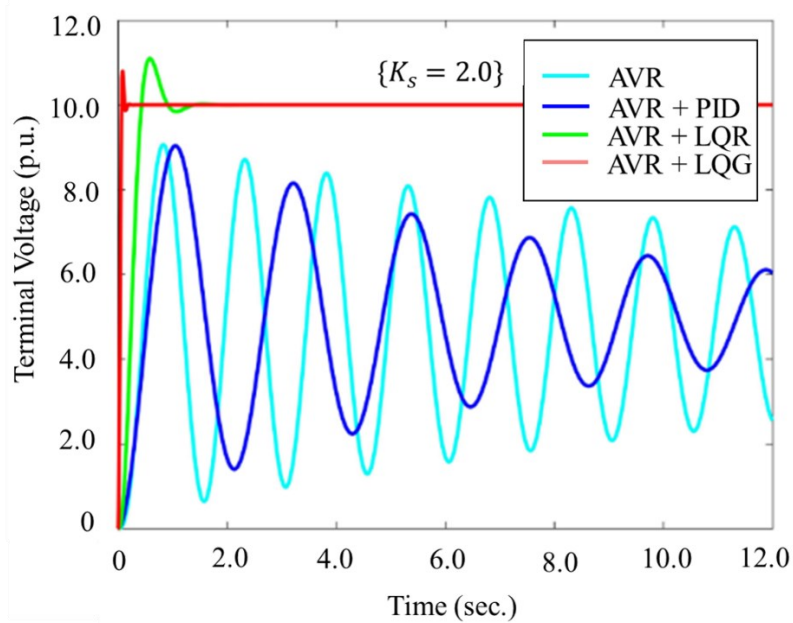

Figure 4.4: The step response of an AVR system with PID, LQR, and LQG controllers when the sensor gain is set at 2.0 [32].

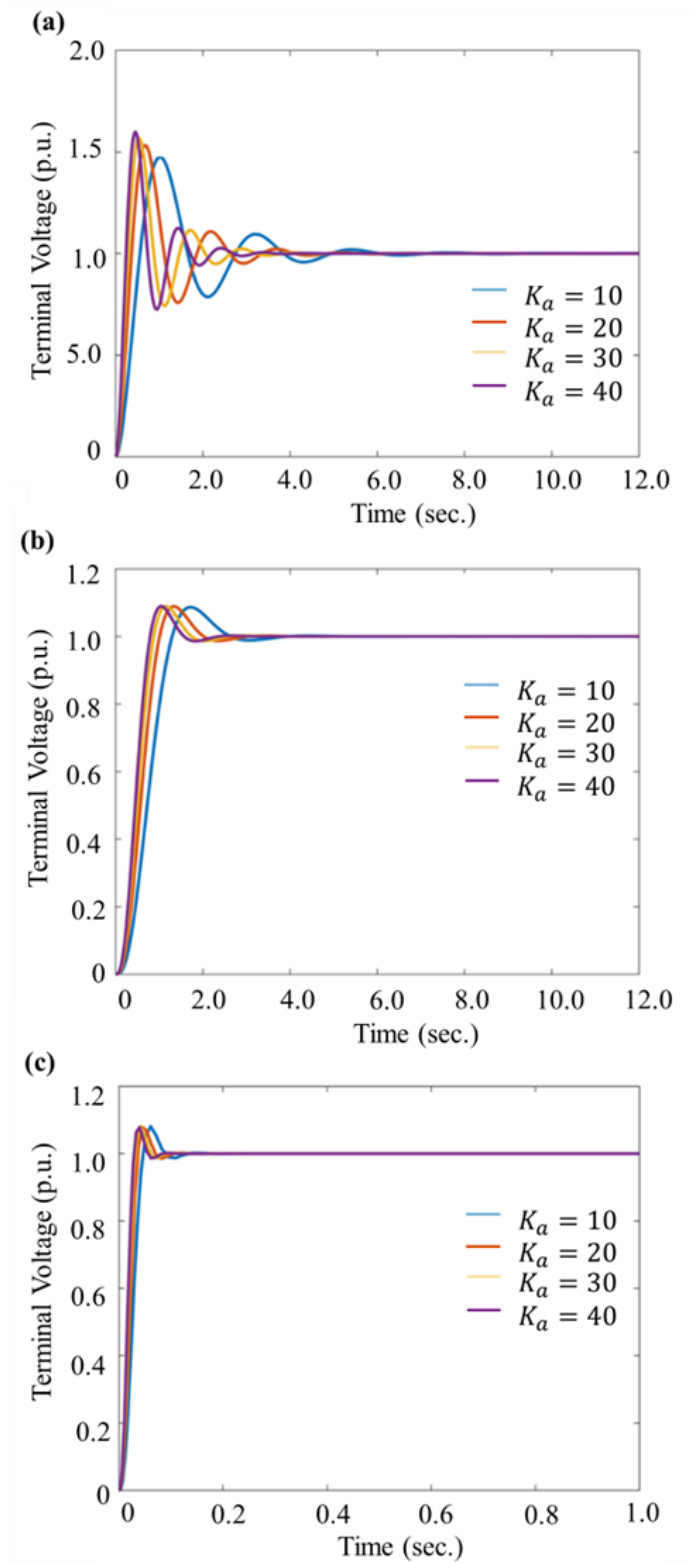

Figure 4.5: The step response of an AVR system with (a) PID, (b) LQR, and (c) LQG controllers when the amplifier gain is set at 10, 20,30, and 40 [32].

Table 4.3: Time domain parameters were observed from the step response of an AVR system with PID, LQR, and LQG controller.

| | *Without Controller* | *PID* | *LQR* | *LQG* |
|---|---|---|---|---|
| Rise Time (s) | 0.2607 | 0.3841 | 0.1979 | 0.0282 |
| Settling Time (s) | 6.9865 | 4.7135 | 0.5974 | 0.0809 |
| Peak Time (s) | 1.5066 | 1.0016 | 0.4293 | 0.0611 |
| Peak (p. u.) | 0.7522 | 1.4744 | 1.0864 | 1.0805 |
| Overshoot (%) | 65.7226 | 47.4385 | 8.6426 | 8.0523 |
| Steady-state Error (%) | 9.18 | 0.014809 | 0.000015 | 0.0000064 |

### 4.3 Power System Stability with Traditional ML Algorithms

- **Logistic regression:**

Logistic regression is a supervised ML algorithm that works on the principle of the log odds of a probabilistic event. It considers the probability of an outcome based on the presence of the independent variables in a dataset. After pre-processing the dataset, the logistic regression algorithm is performed for the system stability states of the IEEE 14 bus generated in the System Modelica environment. Figure 4.6 demonstrates the output obtained with the logistic regression model. The stability eigenvalues are highlighted with dark green (satisfactory), light green (good), green (acceptable), orange (critical), and red (unstable). The prediction time and accuracy with logistic regression are obtained to be as small as 0.0156 seconds and 78.04%. The precision and recall value for logistic regression is 68.08% and 82.99%, respectively (Table 4.4). It is clear that the eigenvalues nearest to the stability boundary (vertical dotted line) are not properly defined by the logistic regressions; 'critical' eigenvalues are misclassified as 'irrelevant' (pink color). Moreover, eigenvalues in the closer range of the real negative axis are considered critical, whereas, in truth, they are 'good'. This could burden the system to make a change in order to make eigenvalue acceptable, where they already exist as 'good'. The F1 score obtained with logistic regression stays near 70%.

- **Linear support vector machines:**

SVM is one of the most commonly used robust ML algorithms for regression and classification purposes. SVM provides two unique features that separate it from the rest of the ML classifiers. First, it tries to maximize the separation boundary between two data clusters or classes, and it can predict data of non-linear characteristics by converting the data to an n-dimensional hyperplane where the classification accuracy could be increased. After pre-processing the dataset, the linear support vector machine algorithm is performed for the system stability states of the IEEE 14 bus generated in the System Modelica environment. Figure 4.7 demonstrates the output obtained with a linear support vector machine. The stability eigenvalues are highlighted with dark green (satisfactory), light green (good), green (acceptable), orange (critical), and red (unstable). The prediction time and accuracy with the linear support vector machine are obtained to be as small as 0.376 seconds and 71.78%. The precision and recall value for the linear support vector machine is 43.12% and 42.55%, respectively (Table 4.4). The F1 score for LSVM is 37.64%. LSVM renders the worst classification accuracy and shows misclassification of important 'acceptable' and 'critical' eigenvalues as 'irrelevant'. The stability margin is not strict and dispersed across 'irrelevant', 'unstable', and 'critical', which is very uneconomical.

- **K-nearest neighbors:**

The k-NN is a non-parametric ML algorithm that makes a classification decision based on the k-number of data nearest to the point of query. The nearest neighbor could be euclidean distance or could be within a defined radius boundary. Apart from other ML, for each decision, all the neighbors' information is also considered; this makes the classification process slower and increases computational time. However, the accuracy of the k-NN is very high, and this makes it very feasible to apply with a smaller dataset to predict with good accuracy. After pre-processing the dataset, the k-NN algorithm is performed for the system stability states of the IEEE 14 bus generated in the System Modelica environment. Figure 4.8 demonstrates the output obtained with k-NN. The stability eigenvalues are highlighted with dark green (satisfactory), light green (good), green (acceptable), orange (critical), and red (unstable). The prediction time and accuracy with k-nearest neighbors are obtained as 11.414 seconds and 99.89%. The precision, recall, and F1 value for k-nearest neighbors is 99.66%, 99.98%, and 99.93%, respectively (Table 4.4). The k-NN provides the highest classification accuracy. However, the required is very large, nearly 300 times than NN, and the prediction time makes the k-NN very costly to consider.

- **Decision trees:**

Decision trees are another non-parametric supervised ML algorithm used to predict a decision based on the decision rules inferred from the relationship of the features in a given dataset. The process mimics a piecewise constant approximation where a top-down divide and conquer procedure is exploited in a greedy search method to obtain the best optimum to split the tree. After pre-processing the dataset, the decision trees algorithm is performed for the system stability states of the IEEE 14 bus, generated in the System Modelica environment. Figure 4.9 demonstrates the output obtained with decision trees. The stability eigenvalues are highlighted with dark green (satisfactory), light green (good), green (acceptable), orange (critical), and red (unstable). The prediction time and accuracy with decision trees are obtained as 0.0317 seconds and 98.82%. The precision, recall, and F1 value for decision trees is 93.44%, 96.19%, and 93.79%, respectively (Table 4.4). In its overall performance, the decision tree provides quite a good prediction. The only classification error was obtained in the form of misclassifying some 'critical' eigenvalues as 'satisfactory'.

- **Naïve Bayes:**

Naïve Bayes is a supervised ML algorithm used for classification purposes. The algorithm works on the pseudo assumption that the features are mutually independent and works on the Bayesian theory. This probabilistic algorithm is mostly wide with a categorical dataset. After pre-processing the dataset, the naïve Bayes algorithm is performed for the system stability states of the IEEE 14 bus generated in the System Modelica environment. Figure 4.10 demonstrates the output obtained with decision trees. The stability eigenvalues are highlighted with dark green (satisfactory), light green (good), green (acceptable), orange (critical), and red (unstable). The prediction time and accuracy with naïve Bayes are obtained as 0.122 seconds and 97.08%. The precision, recall, and F1 score value for naïve Bayes is 86.21%, 93.92%, and 88.25%, respectively (Table 4.4). Naïve Bayes fails to properly follow the stability margin and considers a few 'critical' eigenvalues as 'unstable'. Moreover, this algorithm results in misclassifying some 'good' and 'acceptable' eigenvalues as 'critical' and could place an unnecessary burden on the controllers.

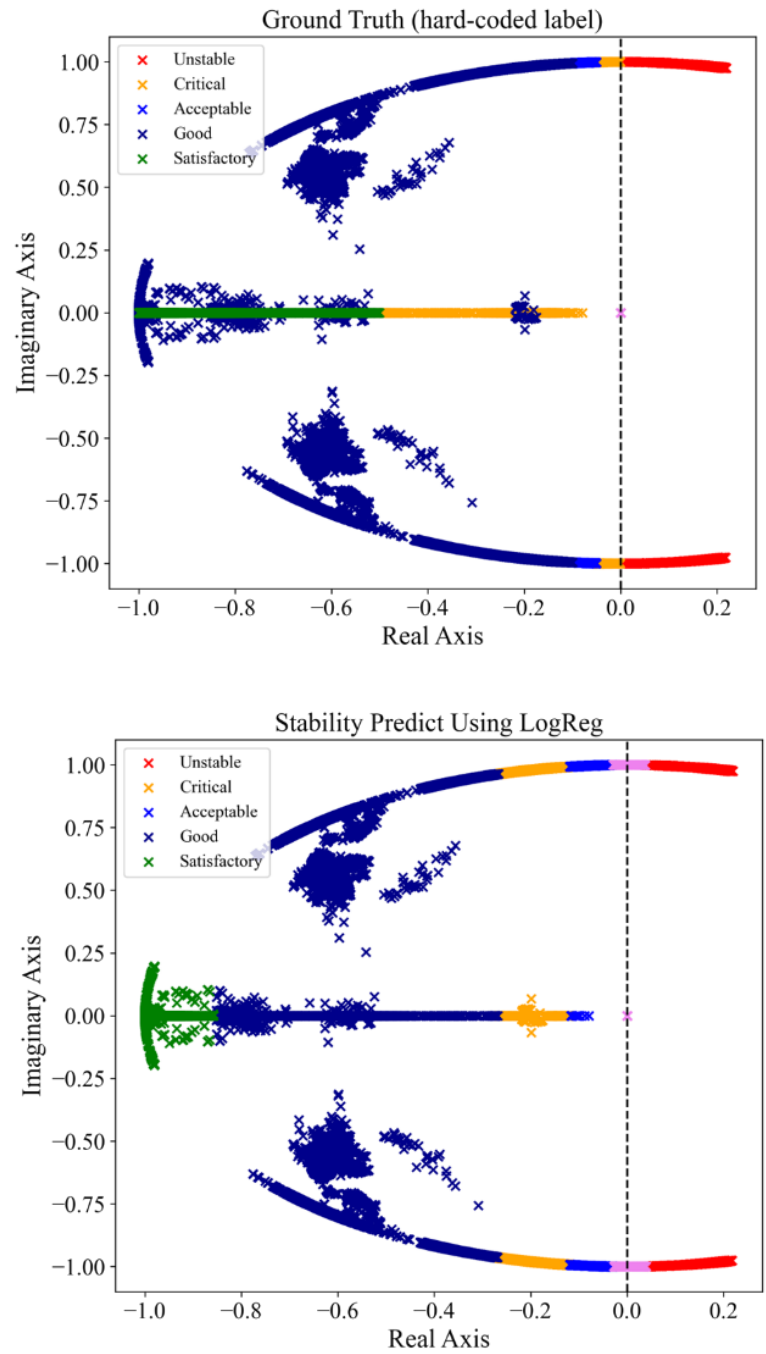

Figure 4.6: Observed stability classification with logistic regression (LogReg) for IEEE 14 bus bar system.

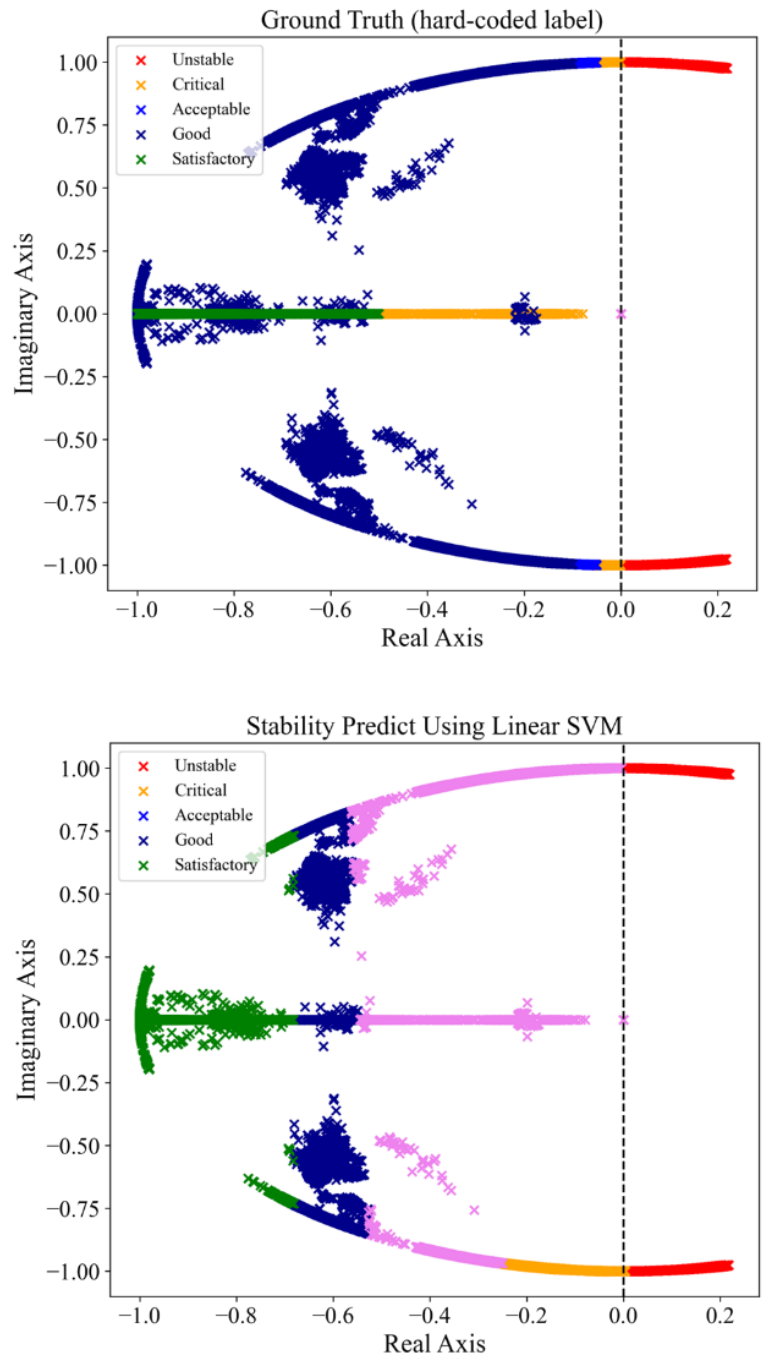

Figure 4.7: Observed stability classification with linear support vector machine for IEEE 14 bus bar system.

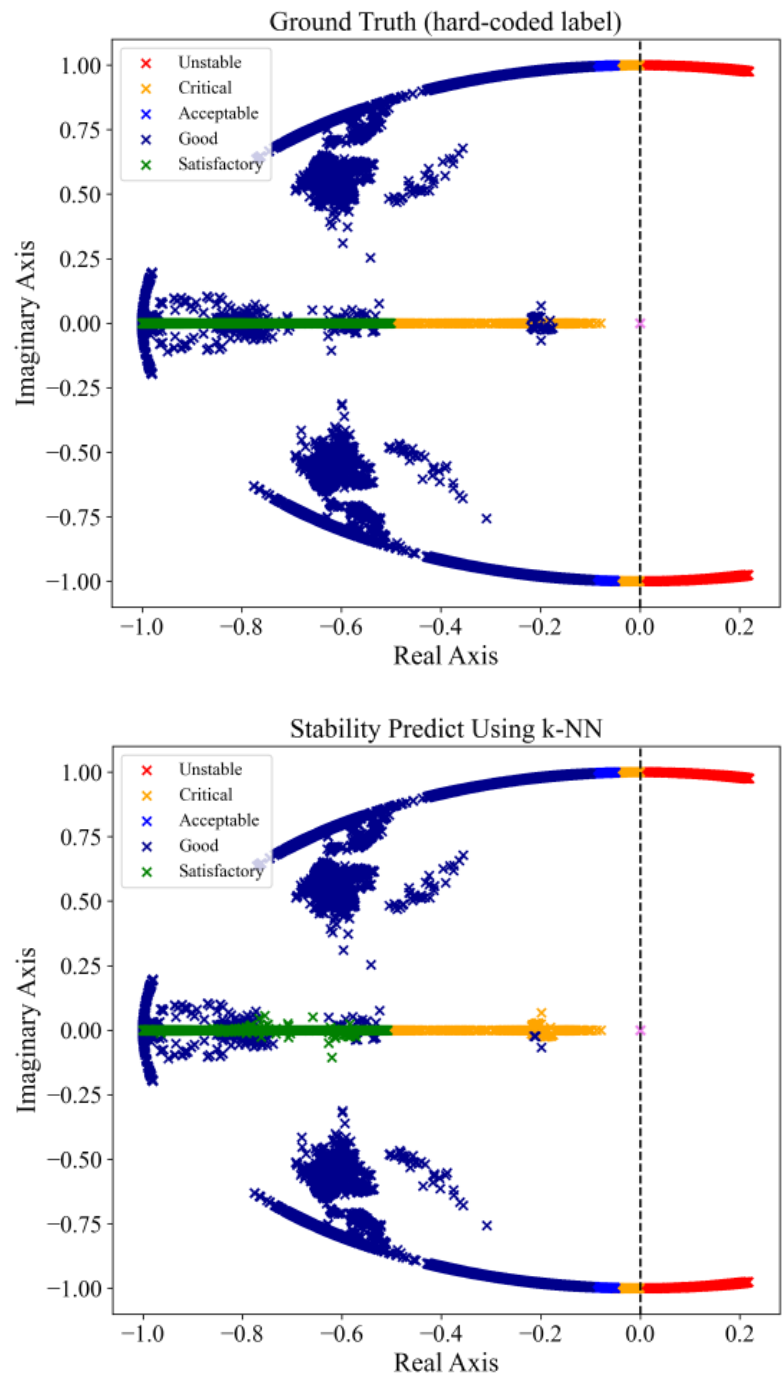

Figure 4.8: Observed stability classification with the k-nearest neighbor for IEEE 14 bus bar system.

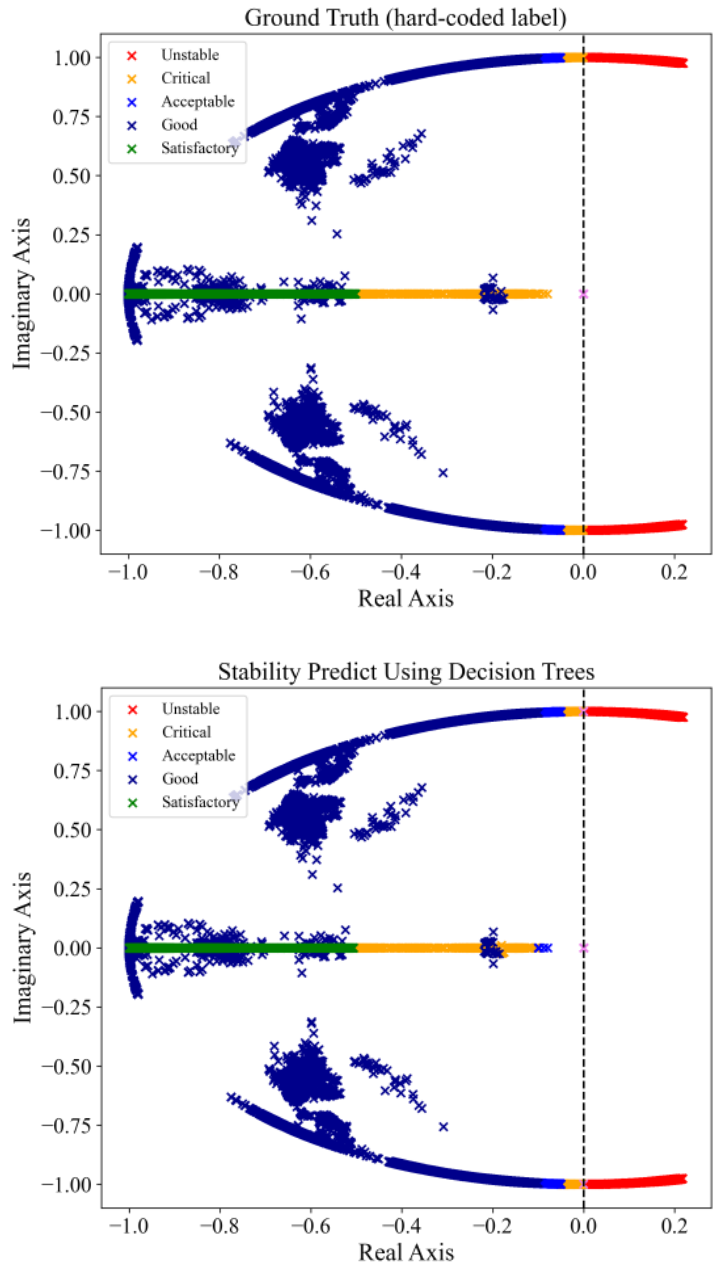

Figure 4.9: Observed stability classification with decision trees for IEEE 14 bus bar system.

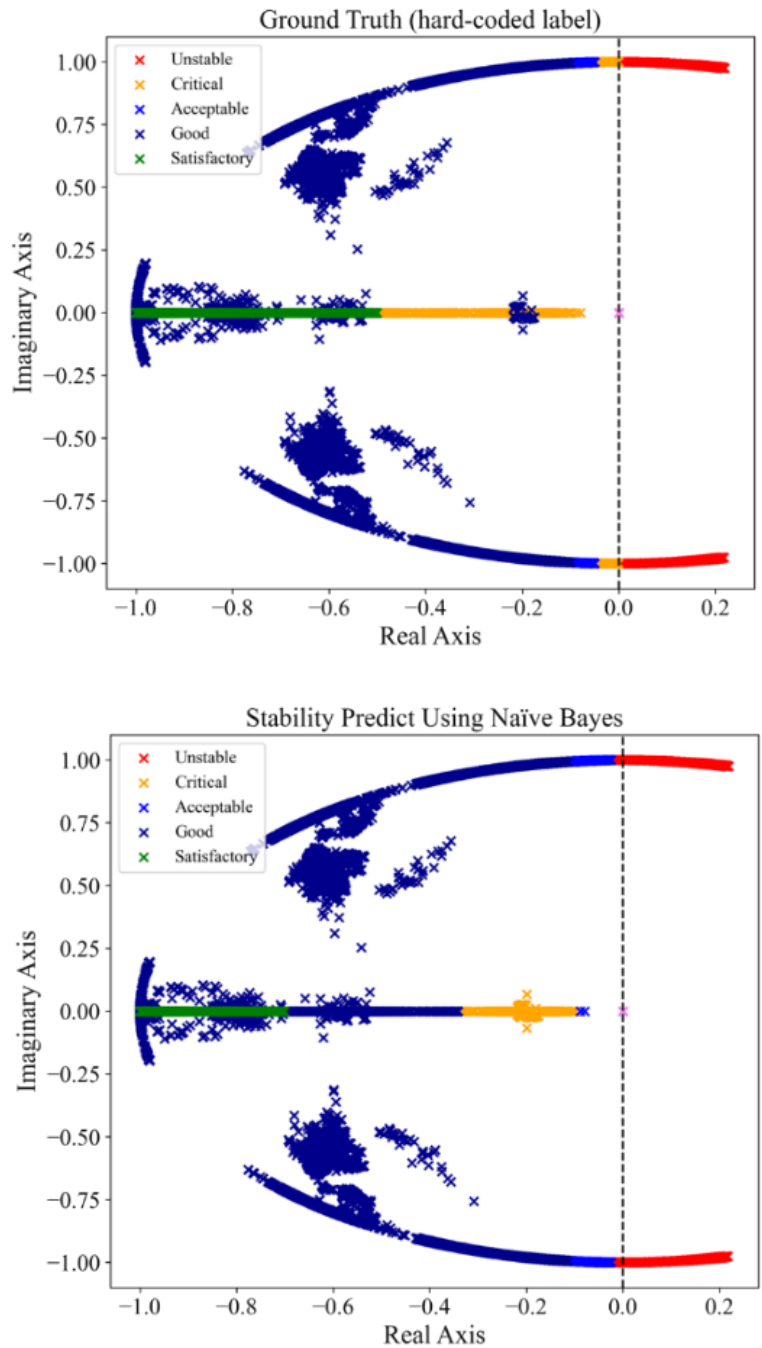

Figure 4.10: Observed stability classification with Naïve Bayes for IEEE 14 bus bar system.

## 4.4 Power System Stability with Neural Network Classifiers

NN or ANN is a subset of ML that mimics the operation of human brain neurons. Each NN consists of an input layer, one or a few hidden intermediate layer(s), and an output layer. The structure could be made as feed-forward or back propagated depending on the necessity of such use. Each node is given a weight and threshold; the threshold defines whether the data should be passed to the next node of the network. The output of the previous node is fed as the input of the next node. After the prediction is made, cross-validation is carried out, and the performance data, such as accuracy and error, are stored. The NN tries to adjust the weight of the nodes to increase the model efficacy. The wider the dataset is, the more time it takes the NN to attain good accuracy. However, once the training stage is completed, the NN can predict very accurately; even if completely new data is fed, the efficiency is retained to a tolerable value.

Three important NN node activation methods are used as a trial and error culture with four NN architectures to formulate the best NN model. First, the input and hidden layers are considered with ReLU, SeLU, and Sigmoid functions only, and the output layer is activated with the Softmax function. Softmax is well suited for multiclass classification. Moreover, a few dropout layers are connected to the NN architecture to prevent overfitting. The number of dropout layers is kept equal to the number of dropout layers. The observed performance metrics for the combination of activation methods and layer structures are presented in Fig. 4.11. Accordingly, the SeLU activation methods provide comparatively higher accuracy scores than the rest. A combination of SeLU with NN architecture of 100×100×100×6 provides nearly 98% precision, 92% accuracy, and 96% recall score, with a considerably low prediction time (~ 0.03 seconds). With the ReLU activation method, the result is slightly lower than the SeLU. Sigmoid activation provides the worst performance. In addition, all four NN architectures are then considered with a combination of ReLU and SeLU activation methods (Fig. 4.12). The input and the hidden layers are activated with either the same or a combination of these two functions. The Softmax regression is kept for output layer activation. It is observed that choosing the SeLU function for both the input and hidden layer activations results in the comparatively best model. The performance of SeLU+SeLU is followed by ReLU+SeLU and then SeLU+ReLU. Thus, the NN model of 100×100×100×6 with the SeLU activation method for input and hidden layers and Softmax activation method for output layers is considered for further investigation of stability classification.

For the six-layered NN model, input feature matrix $X$, input layer matrix $I$, dropout layer matrix $D$, hidden layer $H$, probability-distributed output layer matrix $Y$, probability selection matrix $S$, weight $W$, and bias $b$ are selected. The probability-distributed output layer matrix is formed by enumerating the probability of $m$th input and keeping it at $m$th column. Each $m$th vector of the probability selection matrix pulls the probability of the highest possible outcome to zero and nullifies all other values. For a NN having an architecture of and m input training samples, the mathematical modeling with the activation units could be realized as follows [137]:

$$I^1 = SeLU(XW^1 + b^1)$$
$$D^1 = dropout\ (I^1)$$
$$H^2 = SeLU(D^1W^2 + b^2)$$
$$D^2 = dropout\ (H^2)$$
$$H^3 = SeLU(D^2W^3 + b^3)$$
$$Y = \sigma_M(H^3W^4 + b^4)$$
$$S = argmax(Y)$$

(62)

The considered NN model with three different numbers of hidden layers and dropout layers of architecture, respectively 100×100×6, 100×100×100×6, and 100×100×100×100×6, are considered for performance investigation, shown in Fig 4.13. The NN architectures are employed for the system stability states of the IEEE 14 bus generated in the System Modelica environment. The stability eigenvalues are highlighted with dark green (satisfactory), light green (good), green (acceptable), orange (critical), and red (unstable). The prediction time and accuracy with the neural network classifier (1, 2, 3) hidden layer architectures are obtained as (0.00673, 0.00709, 0.00772) seconds and (98.38, 98.53, 98.53)% (Fig 4.14). The precision, recall, and F1 score value were obtained as (90.72, 91.26, 91.26)%, (95.75, 95.79, 95.79)%, and (92.09, 92.39, 92.39)%, respectively (Table 4.4). It is to be noted that increasing the number of NN layer architecture more than fix-layers does not increases the accuracy or precision, or recall value. However, the prediction time only increased. Thus, for this investigation, six-layered architecture is the most suitable one.

The effect of the dataset size on the considered stability classification is dispatched. Thedataset generated by the ALSETLab is further investigated in this part of the thesis to realize theperformance of classifiers with IEEE 14 bus bar data generated in the Dymola for stability classification. All the classifiers were run on Inter 8-core processor, 16 GB of GPU, and 16 GB of RAM. The performance metrics obtained with five classical supervised classifiers and three topologies of NN, at 100%, 75%, 50%, and 25% dataset size, are all summarized in Table 4.4. To keep the computation lighter, only four dataset segment is considered to observe the trend in model performance with change in dataset size and hidden layers. It is observed that the number of hidden layers significantly impacts the prediction time and accuracy for the ANN. It could be perceived that increasing the hidden layer from one to two shows a distinguishable change in performance metrics; however, increasing it to three hidden layers only results in a negligible change but increases the prediction time. Whenever the hidden layers are improved by more than five, the model score starts to decrease. For the 25% dataset, the small data become overfitted to a denser neural network, and thus F1 score decreases. The time of prediction, precision, recall, and the F1 score are all considered for finding the best model. The performance of the ANN stays within a considerable limit (~95%) with all the various dataset sizes, whereas the performance of other ML algorithms is affected by the smaller dataset. At 25% dataset size, the precision of the LogReg and LSVM falls to ~70% and ~30%, respectively. The metrics could be calculated using the following equation:

$$accuracy = \frac{true_{pos} + true_{neg}}{true_{pos} + true_{neg} + false_{pos} + false_{neg}}$$

(63)

$$precision = \frac{true_{pos}}{true_{pos} + false_{pos}} \qquad (64)$$

$$recall = \frac{true_{pos}}{true_{pos} + true_{neg}} \qquad (65)$$

$$F1_{score} = \frac{2 \times precision \times recall}{precision + recall} \qquad (66)$$

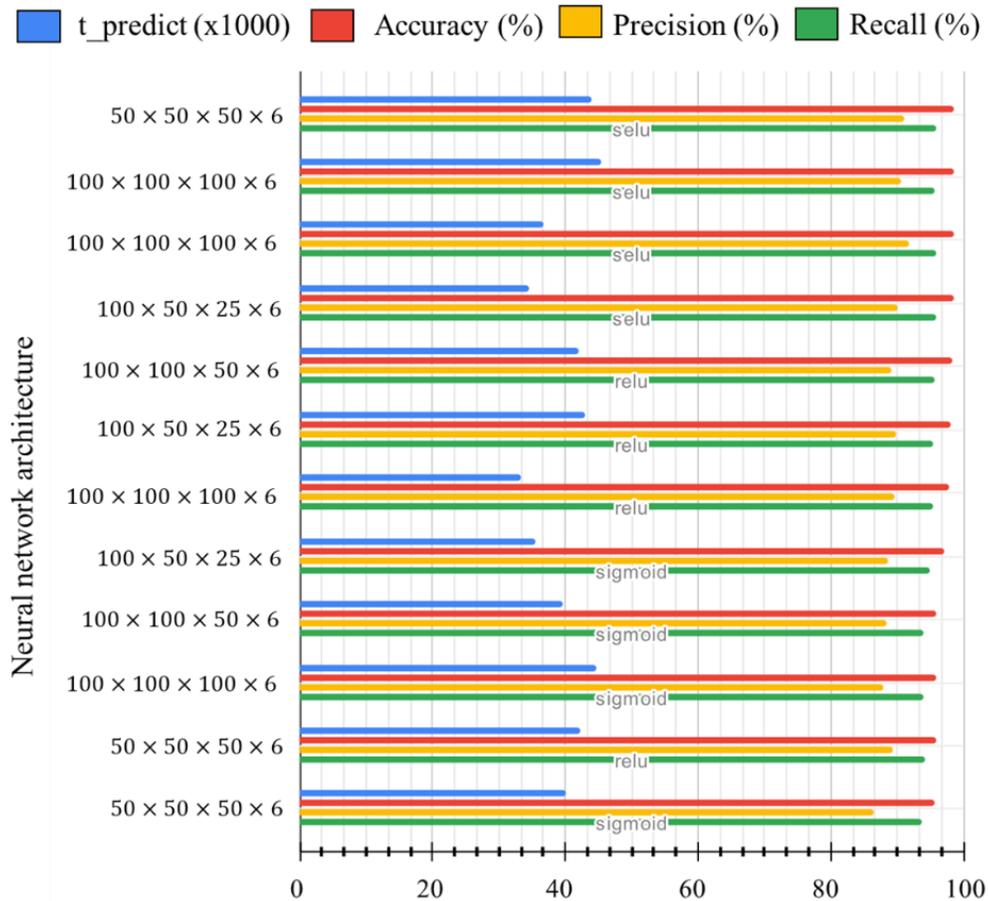

Figure 4.11: Performance metrics of neural networks with ReLU, SeLU, and Sigmoid activation function for the input and hidden layer with four different architectures.

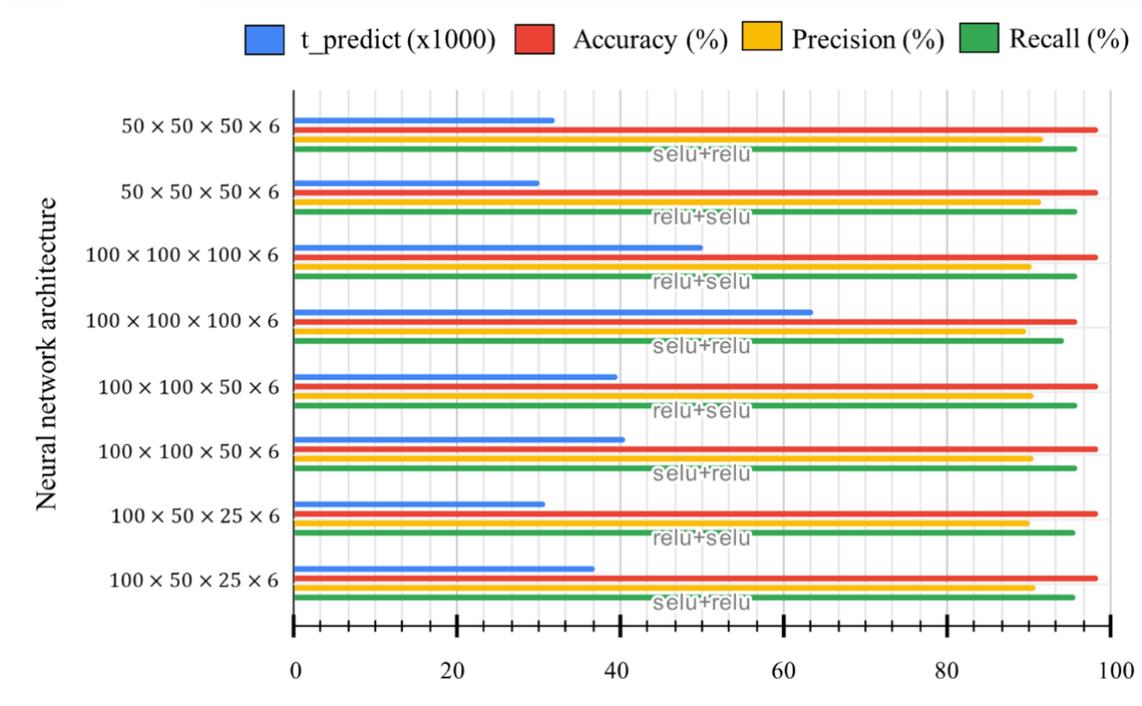

Figure 4.12: Performance metrics of neural networks with combinations of ReLU and SeLU activation functions for the input and hidden layer with four different architectures.

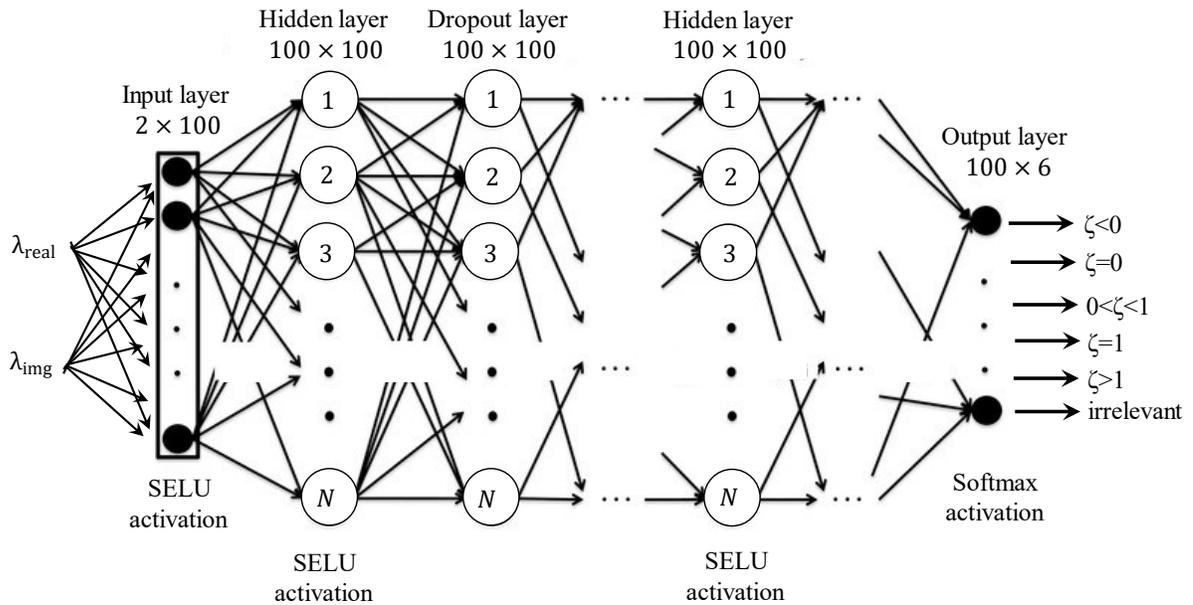

Figure 4.13: Architecture of the considered neural networks with hidden and dropout layers.

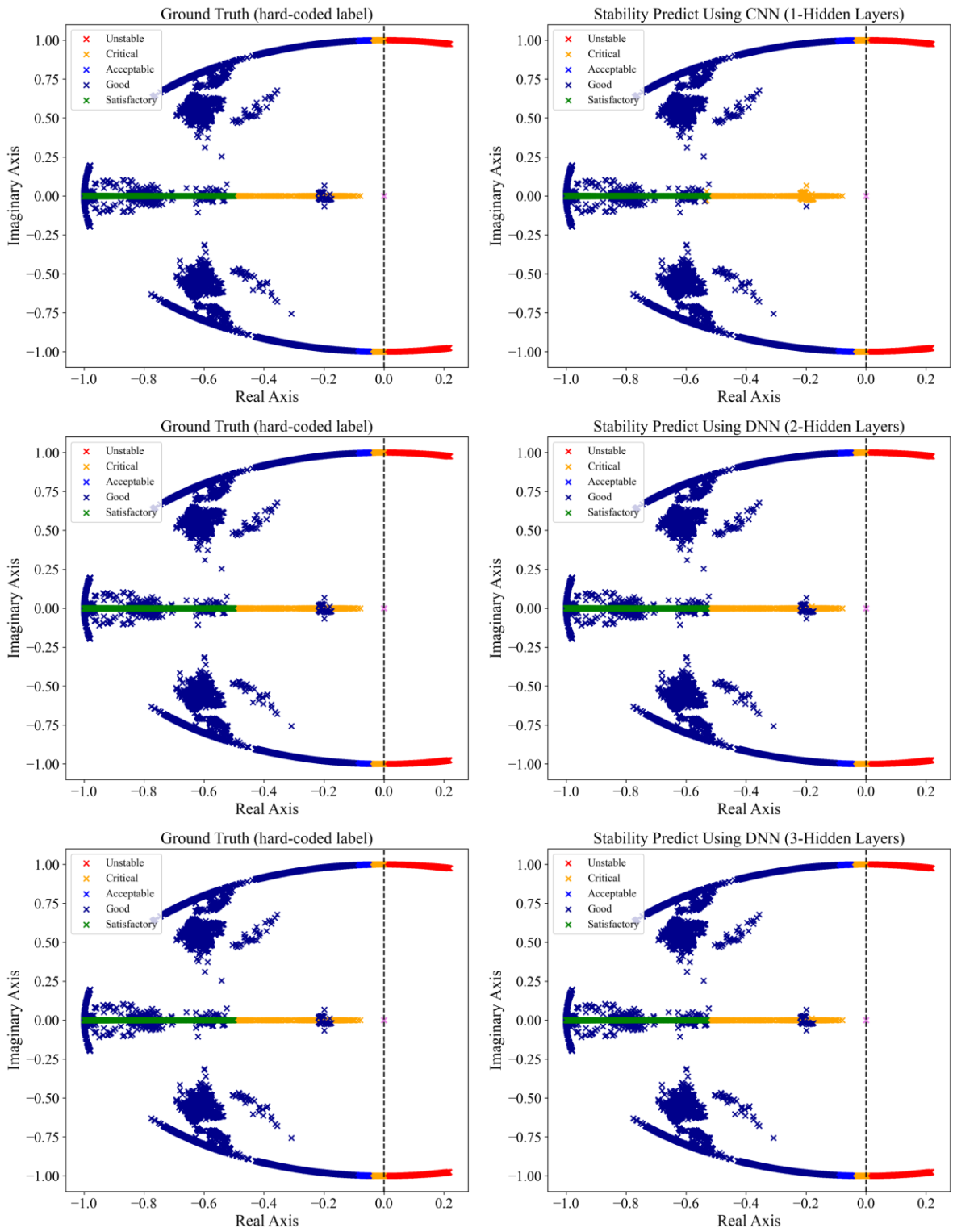

Figure 4.14: Observed stability classification with neural network classifier for IEEE 14 bus bar system.

Table 4.4: Comparative performance of the considered ML and NN classifiers for the power system stability classification of IEEE 14 bus system data at various dataset sizes.

| Dataset Size | Type of Algorithm | | Time of prediction | Accuracy (%) | Precision (%) | Recall (%) | F1_score (%) |
|---|---|---|---|---|---|---|---|
| 100% | Naïve Bayes | | 0.12168 | 97.08 | 86.20 | 93.92 | 88.25 |
| | Linear SVM | | 0.37640 | 71.79 | 43.12 | 42.55 | 37.64 |
| | k-Nearest N | | 11.4142 | 99.90 | 99.67 | 99.94 | 99.80 |
| | Logistic Regression | | 0.01562 | 78.04 | 68.08 | 82.99 | 68.99 |
| | Decision Trees | | 0.03175 | 98.83 | 93.45 | 96.20 | 93.80 |
| | Neural Networks | 3-layers | 0.00673 | 98.38 | 90.72 | 95.75 | 92.09 |
| | | 5-layers | 0.00709 | 98.53 | 91.26 | 95.79 | **92.39** |
| | | 7-layers | 0.00772 | 98.53 | 91.26 | 95.79 | 92.39 |
| | | 11-layers | 0.00835 | 87.43 | 93.98 | 91.34 | 89.32 |
| | | 15-layers | 0.01093 | 87.43 | 93.98 | 91.34 | 89.32 |
| | | 21-layers | 0.01033 | 87.43 | 93.98 | 91.34 | 89.32 |
| 75% | Naïve Bayes | | 0.06383 | 97.09 | 86.24 | 93.94 | 88.31 |
| | Linear SVM | | 0.33415 | 66.34 | 33.24 | 42.50 | 33.41 |
| | k-Nearest N | | 6.73377 | 99.87 | 99.87 | 99.93 | 99.72 |
| | Logistic Regression | | 0.01194 | 77.40 | 77.40 | 77.40 | 77.40 |
| | Decision Trees | | 0.01197 | 98.84 | 93.50 | 96.25 | 93.88 |
| | Neural Networks | 3-layers | 0.00488 | 98.51 | 91.99 | 95.76 | 92.41 |
| | | 5-layers | 0.00410 | 98.59 | 91.17 | 95.89 | **92.81** |
| | | 7-layers | 0.00454 | 98.59 | 91.17 | 91.17 | 92.81 |
| | | 11-layers | 0.00720 | 95.93 | 89.86 | 94.24 | 90.52 |
| | | 15-layers | 0.00923 | 95.93 | 89.86 | 94.24 | 90.52 |
| | | 21-layers | 0.01154 | 95.93 | 89.86 | 94.24 | 90.52 |
| 50% | Naïve Bayes | | 0.04730 | 97.12 | 86.57 | 93.97 | 88.55 |
| | Linear SVM | | 0.00793 | 71.23 | 40.53 | 44.99 | 38.45 |
| | k-Nearest N | | 3.96739 | 99.76 | 99.22 | 99.88 | 99.54 |
| | Logistic Regression | | 0.00897 | 78.09 | 68.28 | 83.06 | 83.06 |
| | Decision Trees | | 0.01197 | 98.83 | 93.39 | 96.24 | 93.81 |
| | Neural Networks | 3-layers | 0.00482 | 98.41 | 90.48 | 95.71 | 91.90 |
| | | 5-layers | 0.00569 | 84.93 | 91.04 | 95.75 | **92.28** |
| | | 7-layers | 0.00683 | 98.44 | 98.44 | 95.75 | 92.28 |
| | | 11-layers | 0.00870 | 96.11 | 89.14 | 94.25 | 89.49 |
| | | 15-layers | 0.00977 | 96.11 | 89.14 | 94.25 | 89.49 |
| | | 21-layers | 0.01262 | 96.11 | 89.14 | 94.25 | 89.49 |
| 25% | Naïve Bayes | | 0.01595 | 97.07 | 85.59 | 93.79 | 87.70 |
| | Linear SVM | | 0.00498 | 64.56 | 27.45 | 34.67 | 30.08 |
| | k-Nearest N | | 1.69949 | 99.64 | 98.03 | 99.83 | 98.89 |
| | Logistic Regression | | 0.00399 | 77.79 | 67.29 | 82.85 | 68.18 |
| | Decision Trees | | 0.00399 | 98.76 | 93.10 | 96.08 | 93.53 |
| | Neural Networks | 3-layers | 0.00140 | 97.38 | 88.88 | 95.01 | **90.59** |
| | | 5-layers | 0.00200 | 95.70 | 88.19 | 94.20 | 89.85 |
| | | 7-layers | 0.00380 | 95.70 | 88.19 | 94.20 | 89.85 |
| | | 11-layers | 0.00486 | 95.68 | 87.86 | 93.66 | 87.60 |
| | | 15-layers | 0.00769 | 95.68 | 87.86 | 93.66 | 87.60 |
| | | 21-layers | 0.01189 | 95.68 | 87.86 | 93.66 | 87.60 |

## 4.6 Summary

In this chapter, the simulation result of the AVR system with closed loop PID, LQR, and LQG controller is attached with a relative comparison among the three controllers. Moreover, the prediction of the system stability of the IEEE 14 bus system, in the presence of any contingency scenario, is investigated with a few traditional ML algorithms. After that, the performance with the neural network having a five to seven layers configuration is observed. The performance metrics of dataset size and the number of NN layers are also observed. The overall score of the proposed neural network is found to be better than all five other investigated traditional ML algorithms. It is concluded that increasing NN architecture to more than six layers does not increase any performance metrics but only increases the prediction time.

# CHAPTER V

# Conclusions and Recommendations

## 5.1 Conclusion

The research work attached throughout this thesis report is two-folded. The first part focuses on the close-loop advanced power system control of an automatic voltage regulator system with PID, LQR, and LQG controllers considered. Second, the classification and prediction strategies of a power system eigenvalues are investigated using ML algorithms and classifiers. The dataset required for training the ML models could be generated from the System Modelica and Dymola environments by simulating the possible contingencies. The number of simultaneous contingency scenarios could be first reduced by employing Poisson's probability distribution of having more than one contingency at a time (Monte-Carlo optimization).

The resultant dataset obtained through the Python-Dymola interface is pre-processed and normalized using the Skewed principle before feeding to the ML models. During the comparative study of the ML classifiers, it has been observed that for the same dataset and simulation tools (hardware and software configurations), the use of a neural network provides the best performance. In terms of prediction processing time, a neural network takes lesser time than any other considered classifiers. It is also observed that in terms of accuracy, precision, and recall, the k-nearest neighbor gives the best possible outcomes. However, the prediction time with k-NN is tremendously higher than other classifiers. The effect of the number of hidden layers and the amount of data in the dataset has also been investigated. It is observed that the neural network does not provide better prediction and accuracy when the hidden layers are increased beyond three for 100%, 75%, 50%, and 25% data size. For 100x1000x100x6 neural network architecture, when the number of the hidden layers increases from one to two, the time of prediction with the neural network changes from 0.00673 seconds to 0.007091 seconds, and the amount of accuracy, precision, recall, and F1 score changes from 98.38% to 98.53%, 90.72% to 91.26%, 94.75% to 95.79%, and 92.09% to 92.39%, respectively. With data sets ranging from 25%, 50%, and 75%, the F1 score observed for three and five-layers neural networks are 92.09%, 92.41%, 91.90%, 90.59%, and 90.39%, 92.81%, 92.28%, 89.85%, respectively.

Apart from that, power system control was realized by analyzing the step response of a grid-isolated automatic voltage regulator unit with and without a closed-loop controller. The main focus was given to the influence of the AVR subsystem models parameter, such as time constant and gain parameter variations, on the close loop response of the AVR model with PID, LQG, and LQR controller. Moreover, a comparative analysis of the controller was summarized. It is obtained that the LQG controller performs best compared to the PID and LQR. The PID controller performance is very much susceptible to parameter variation and requires to be embedded with a metaheuristic or parameter optimization algorithm to automatically tune the PID gain parameter values in the dynamic environment. The LQR and LQG can maintain stability in the presence of external disturbance or overshooting of the amplifier gain and change in the sensor gain parameters. This analysis helps in clearly outlining the expected behavior of a system at various system parameter values when static gain parameters of PID, LQR, and LQG controllers are set and are not updated autonomously using any optimization algorithm. The research on the AVR model would help predict the change in AVR parameters and the relevant change in the system step response, which could directly lead to better utilizing any control algorithms or the design of newer close loop controllers for the system. In the future, the three controllers and the machine learning algorithms would be further investigated to justify their suitability and to improve their operating characteristics in dynamic or static environments for distributed renewable sources for futuristic smart grid or microgrid implementation.

## 5.2 Future Outlook

The two primary research aspects carried out in this thesis, power system controller design and power system analysis with the artificial neural network, are of sublime impact on the concurrent trend of investigating futuristic distributed generations, micro-grid, and smart grid technologies. The ML classifiers and models proposed and employed during the IEEE test-grid stability profile classification could be extended for practical grids. The incorporation of ML algorithms could prevent disruption of the power grid from instantaneous breakdown, and time could be saved in dealing with the stability identifications. Recently, the stability information rendered through the ML models is being fed to the close loop controllers to alter or modify the system operation, machine speed, and excitation and to handle load distribution more precisely and swiftly. Dymola is a system modeling platform that works on the basis of Modelica language and is very much suited for muti-engineering complex system design, simulation, and practical realization. The platform gives direct access to a few powerful IEEE standardized test systems to work with automation, aerospace, robotics, vehicle manufacturing processes, and power system investigation. It is becoming a fascinating tool for companies such as Audi, BMW, Ford, and Toyota for efficient system design. Also, power plant providers, such as ABB, EDF, and Siemens, use Modelica, as well as many other companies. The dataset required to model the neural network proposed in this thesis is generated from the Dymola environment. The effect of the hidden layer and data size analyzed herein in this dissertation could directly help in realizing the data quality and implementation aspects of the data generated via Dymola.

# PUBLICATIONS

1. **Md. Rayid Hasan Mojumder** and Naruttam Kumar Roy, "PID, LQR, and LQG Controllers to Maintain the Stability of an AVR System at Varied Model Parameters," 2021 5th International Conference on Electrical Engineering and Information Communication Technology (ICEEICT), 2021, pp. 1-6, DOI: 10.1109/ICEEICT53905.2021.9667897.

2. **Md. Rayid Hasan Mojumder** and Naruttam Kumar Roy, "Effect of Dataset Size and Hidden Layers on the Stability Classification of IEEE-14 Bus System Using Deep Neural Network," 2022 International Conference on Energy and Power Engineering (ICEPE), 2022. (Best Paper Award – 2nd Place).

3. Naruttam Kumar Roy and **Md. Rayid Hasan Mojumder**, "Review of meta-heuristic optimization algorithms to tune the PID controller parameters for automatic voltage regulators," Computers and Electrical Engineering, Elsevier (submitted).